\documentclass{article}

\usepackage[margin=1in]{geometry}

\usepackage{authblk}
\usepackage{hyperref}
\usepackage{amsmath}
\usepackage{mathabx}
\usepackage{tensor}

\usepackage{graphicx}
\usepackage{caption}
\usepackage{subfig}

\usepackage[export]{adjustbox}

\usepackage{floatrow}

\usepackage{multirow}

\usepackage{siunitx}

\usepackage{soul}
\usepackage[table]{xcolor}
\definecolor{mypurple}{HTML}{800080}
\definecolor{mygreen}{HTML}{008000}
\definecolor{mylb}{HTML}{c9daf8}

\usepackage{tabularx}

\usepackage{makecell}
\usepackage{pbox}
\usepackage{arydshln}

\begin{document}

\title{Modeling multi-scale data via a network of networks}
\author[1]{Shawn Gu}
\author[1]{Meng Jiang}
\author[2]{Pietro Hiram Guzzi}
\author[1, *]{Tijana Milenkovi\'{c}}
\affil[1]{Department of Computer Science and Engineering, University of Notre Dame, Notre Dame, IN 46556, USA}
\affil[2]{Department of Surgical and Medical Sciences, University Magna Graecia of Catanzaro, Catanzaro, IT}
\affil[*]{To whom correspondence should be addressed (email: tmilenko@nd.edu).}
\date{}

\maketitle

\abstract{Prediction of node and graph labels are prominent network science tasks. Data analyzed in these tasks are sometimes related: entities represented by nodes in a higher-level (higher-scale) network can themselves be modeled as networks at a lower level. We argue that systems involving such entities should be integrated with a ``network of networks'' (NoN) representation. Then, we ask whether entity label prediction using multi-level NoN data via our proposed approaches is more accurate than using each of single-level node and graph data alone, i.e., than traditional node label prediction on the higher-level network and graph label prediction on the lower-level networks. To obtain data, we develop the first synthetic NoN generator and construct a real biological NoN. We evaluate accuracy of considered approaches when predicting artificial labels from the synthetic NoNs and proteins' functions from the biological NoN. For the synthetic NoNs, our NoN approaches outperform or are as good as node- and network-level ones depending on the NoN properties. For the biological NoN, our NoN approaches outperform the single-level approaches for just under half of the protein functions, and for 30\% of the functions, only our NoN approaches make meaningful predictions, while node- and network-level ones achieve random accuracy. So, NoN-based data integration is important.}

\section{Introduction} \label{sec:intro}
Networks can be used in many domains to model entities and the complex systems involving them. For example, in biological networks, nodes are biological entities (such as genes or their protein products, tissues, etc.) and edges are interactions between them; in social networks, nodes are generally individuals and edges are social interactions between them; and more. By modeling systems as networks, the important relationships can be studied, which can lead to deeper insights compared to analyzing each entity on its own. 

Two important tasks in network science are node label prediction \cite{bhagat2011node} and graph label prediction \cite{nikolentzos2017matching}. In the former, given a single network, the goal is to predict labels of its \textit{nodes}. In the latter, given multiple networks, the goal is to predict labels of those \textit{networks}. For example, the former can be applied to a protein-protein interaction (PPI) network (PIN), where nodes are proteins and edges are PPIs, to predict proteins' functions; to a social network to uncover individuals' demographics, hobbies, etc.; and more. The latter can be applied to multiple proteins' structure networks (PSNs), where nodes are amino acids and edges join those that are close in the 3D crystal structure, also to predict proteins' functions; to multiple chemicals' molecule networks, where nodes are atoms and edges are bonds, to predict their properties; and more.

\begin{figure*}[ht]
    \centering
    \includegraphics[width=0.95\textwidth]{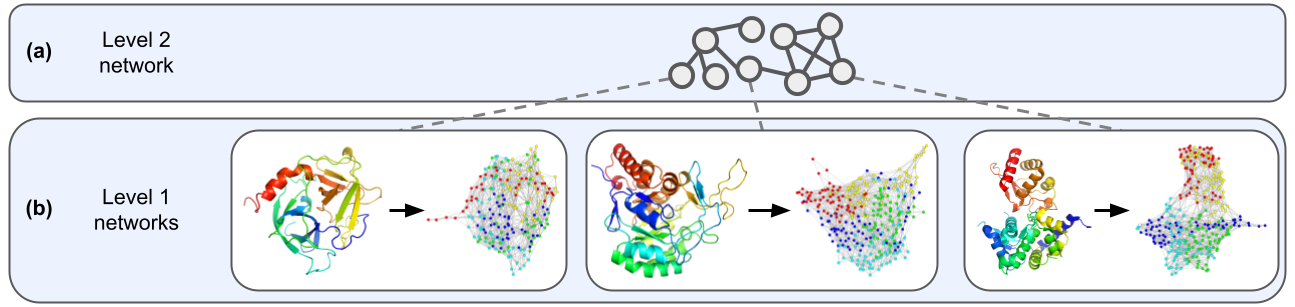}
    \caption{\label{fig:pin-psn-non} Illustration of a two-level biological NoN. Level 2 nodes (proteins) in \textbf{(a)} the level 2 network (PIN) are joined to their corresponding \textbf{(b)} level 1 networks (PSNs) by dotted lines. Only three level 1 networks are shown for simplicity, but generally every level 2 node can have a corresponding level 1 network. Nodes in the PSNs are colored based on their corresponding amino acids in the ribbon diagram and are not indicative of node labels.}
\end{figure*}

So, sometimes, the entities (e.g., proteins) that are represented by nodes in a network (e.g., a PIN) can themselves be modeled as networks (e.g., PSNs).
We argue that the systems involving such entities should be integrated into a ``network of networks'' (NoN), where nodes in a network at a higher level (i.e., higher scale) are themselves networks at a lower level (Fig. \ref{fig:pin-psn-non}). More specifically, we refer to the higher level of the NoN as the level 2 network (Fig. \ref{fig:pin-psn-non}(a)), which contains level 2 nodes and level 2 edges. Each level 2 node has a corresponding level 1 network at the lower level of the NoN (Fig. \ref{fig:pin-psn-non}(b)), which contains level 1 nodes and level 1 edges. We number levels in this way with the idea that lower-level networks are the building blocks of higher-level networks. However, we tend to discuss level 2 networks first, as doing so is often more convenient for developing intuition.
Even though we analyze two-level NoNs in this study, NoNs can encompass more: proteins interact with each other to carry out cellular functioning, cells interact with each other to form tissues, and so on, up the levels of biological organization. We hope to extend our work to encompass more than two levels in the future.

Then, we ask whether entity label prediction using information from multiple levels as an NoN is more accurate than entity label prediction using only a single level, i.e., node label prediction using only the level 2 network (Fig. \ref{fig:pin-psn-non}(a)) and graph label prediction using only the level 1 networks (Fig. \ref{fig:pin-psn-non}(b)). 

The novelty of this work is that we construct and provide a new source of NoN data, we develop novel approaches for NoN label prediction, and we are the first to test whether using NoN data in label prediction is more accurate than using only a single level. Note that existing studies have combined protein structural data with PPI data \cite{peng2014improving, zhang2019deepfunc} for various tasks. However, they generally do so by incorporating more basic \textit{non-network} structural properties, such as proteins' domains and families, with PPI data. On the other hand, our approaches combine PSN representations of detailed 3D protein structural properties with PIN data through the NoN representation. Importantly, we then evaluate whether NoN label prediction is actually more accurate than each of node- and network-level alone. So, our entire study, which is already comprehensive, studies \textit{network-based} data integration. A comparison with other, \textit{non-network-based} data integration schemes is outside the scope of the current study and the subject of future work.


Before continuing, note that there exist some other network models of higher-order data. These include: multiplex, multimodal, multilevel, and interdependent networks \cite{roth2017emergence, morone2017model, chen2018identifying, li2018evidential, dong2020network, perich2020rethinking}, which are sometimes used interchangeably and sometimes also referred to as ``networks of networks''; hierarchical networks \cite{clauset2008hierarchical}; higher-order networks \cite{xu2016representing}; hypergraphs \cite{berge1973graphs}; and simplicial complexes \cite{munkres2018elements}. However, these all model different types of data compared to NoNs as we define them, so we cannot consider these other network types in our study.


There are also studies that do model data as NoNs. However, they differ from our proposed work in terms of data analyzed, application domain, and/or network science task. With respect to data, besides synthetic NoNs, we analyze a PIN-PSN biological NoN. However, these other studies analyze NoNs where the level 2 network is a disease-disease similarity network and the level 1 networks are disease specific PINs \cite{ni2016disease}, where the level 2 network is a social network and the level 1 networks are individuals' brain networks \cite{falk2017brain, parkinson2018similar, bassett2017network}, or where the level 2 network is a chemical-chemical interaction network and level 1 networks are molecule networks \cite{wang2020gognn}. With respect to application domain, while we aim to predict protein function, these other studies aim to identify disease causing genes \cite{ni2016disease}, answer sociologically motivated questions like whether similarities between friends mean they have similar ways of thinking \cite{parkinson2018similar}, or predict new chemical-chemical interactions \cite{wang2020gognn}. With respect to network science task, while we aim to predict entities' labels, these other studies aim to identify important entities (level 1 nodes) \cite{ni2016disease}, predict links between entities (level 2 nodes) \cite{wang2020gognn}, or embed multiple networks at the same level into a common low dimensional space, using an NoN as an intermediate step \cite{du2019mrmine}. While it \emph{might} be possible to extend some of these existing studies to ours or vice versa, doing so could require considerable effort, as it would mean developing new methods, and code is not publicly available for all of the existing methods. All of this makes any potential extensions hard. As such, we cannot compare against these existing NoN-like methods.

We need labeled NoNs for our task. Since to our knowledge such data is limited, we first develop an NoN generator that can create a variety of synthetic NoNs (Section \ref{data:synthetic-non}). Intuitively, given any set of random graph models, such as geometric \cite{penrose2003random} or scale-free \cite{barabasi1999emergence}, our generator combines instances of these models at each level. In this way, we can label each entity (level 2 node and its level 1 network) based on which combination of graph models it is involved in at the two levels. 
Our generator can control a variety of network structural parameters (Section \ref{data:synthetic-non}).
Second, we construct a biological NoN, consisting of a PIN from BioGRID \cite{stark2006biogrid} at the second level and PSNs for proteins from Protein Data Bank (PDB) \cite{berman2000protein} at the first level. Proteins are labeled based on their functions via Gene Ontology (GO) annotation data \cite{ashburner2000gene} (Section \ref{data:pin-psn-non}). For each of the 131 GO terms considered, the goal is to predict whether or not each protein is annotated by that GO term. We expect the NoN data resulting from our study to become a useful resource for future network science research.

Label prediction approaches generally extract features of the entities and then perform supervised classification, i.e., prediction of the entities' labels based on their features. So, for our study, there are three types of approaches to consider: (i) those that extract node-level features (i.e., level 2 only), (ii) those that extract network-level features (i.e., level 1 only), or (iii) those that extract NoN features (i.e., integrated level 1 and level 2). To our knowledge, approaches of type (iii) do not exist yet, so we create NoN features in two ways: by combining existing node- and network-level features and by applying the novel graph neural network (GNN) \cite{wu2020comprehensive} approach that we propose for analyzing NoNs. We aim to demonstrate that approaches of type (iii) outperform those of types (i) and (ii), therefore providing evidence that NoN-based data integration is useful for label prediction. To determine which approach types are the best, we evaluate them in terms of accuracy for synthetic NoNs, as class sizes are balanced, and in terms of the area under the precision recall curve (AUPR), precision, recall, and F-score for the biological NoN, as class sizes are unbalanced.

For synthetic NoNs, we find that our NoN approaches outperform single-level node and network ones for those NoNs where the majority of nodes are not densely interconnected (i.e., where nodes do not tend to group into densely connected modules). For NoNs where there are groups of densely interconnected nodes (i.e., where there is clustering structure), an existing single-level approach performs as well as NoN approaches.
For the biological NoN, we find that our NoN approaches outperform the single-level ones in a little under half of the GO terms considered. Furthermore, for 30\% of the GO terms considered, only our NoN approaches make meaningful predictions, while node- and network-level ones achieve random accuracy. Also, while deep learning does not perform the best \textit{overall}, it seems to be useful for otherwise difficult-to-predict protein functions. As such, NoN-based data integration is an important and exciting direction for future research. 

\section{Materials and methods}

\subsection{NoN definition} \label{methods:non-def}

We define an NoN with $l$ levels as follows. Let $G^{(l)} = (V^{(l)}, E^{(l)})$ be the level $l$ network with node set $V^{(l)}$ and edge set $E^{(l)} \subseteq V^{(l)} \times V^{(l)}$. Each ``level $l$ node'' $v^{(l)}_i \in V^{(l)}$ itself corresponds to a ``level $l-1$ network'' $G^{(l-1)}_i = (V^{(l-1)}_i, E^{(l-1)}_i)$.
Note that we allow each level 1 network to contain no nodes (and thus no edges). That is, $G^{(l-1)}_i$ can be an order-zero graph, signifying that $v^{(l)}_i$ has no corresponding level $l-1$ network. We assume that nodes from different level $l-1$ networks do not overlap -- for example, amino acids (nodes) from different PSNs do not represent the same physical entities, even if the types of the amino acids are the same. That is, $V^{(l-1)}_1 \cap ... \cap V^{(l-1)}_{|V^{(l)}|} = \emptyset$.
Each level $l-1$ node $v^{(l-1)}_{i_j} \in V^{(l-1)}_{i}$ in each level $l-1$ network $G^{(l-1)}_{i} \in V^{(l)}$ itself corresponds to a level $l-2$ network $G^{(l-2)}_{i_j} = (V^{(l-2)}_{i_j}, E^{(l-2)}_{i_j})$. This recursion continues until level 1. 
We illustrate a two-level NoN in Fig. \ref{fig:pin-psn-non}. 

\subsection{Problem statement} \label{methods:task}
Given a two level NoN $\{G^{(2)} = (V^{(2)}, E^{(2)})$ and $\{G^{(1)}_1,...,G^{(1)}_{|V^{(2)}|}\}\}$ and labels for each entity (level 2 node, or equivalently, its corresponding level 1 network),
our framework extracts node-level, network-level, or NoN features of the entities and performs label prediction. That is, our framework hides features and labels of some of the entities, trains a classifier to learn patterns between features and labels of the remaining entities, and then applies the classifier to try to uncover labels of the hidden entities based on their features.
Again, if we were to only consider $G^{(2)}$, we would be doing node classification (predicting node labels), and if we were to only consider $\{G^{(1)}_1,...,G^{(1)}_{|V^{(2)}|}\}$, we would be doing graph classification (predicting graph labels). 
As such, we can fairly compare the accuracy of single-level approaches to our NoN approaches. 
Following, we describe data (Section \ref{data:data}), label prediction approaches (Section \ref{methods:approaches}), and evaluation (Section \ref{methods:eval-framework}).

\subsection{Data} \label{data:data}

\subsubsection{Our synthetic NoN generator} \label{data:synthetic-non}

We develop a generator that can create synthetic NoNs with a variety of parameters and multiple levels. In this study, we focus on two levels. While analyzing NoNs of three or more levels would be interesting, doing so would be difficult in the context of our study, especially since available real-world NoN data of so many levels is scarce. Our goal is to test whether NoN-based integration is worth it. With two levels, there exist very clearly defined and fairly comparable tasks: NoN vs. node-level vs. graph-level label prediction. With more levels, this is no longer the case. So, we leave such investigation of NoNs with more than two levels for future work.

Our goal is to generate a two-level NoN with labeled entities (i.e., level 2 nodes, or equivalently, level 1 networks) such that only an approach that uses information from both levels should be able to attain high entity classification accuracy. To do so, our generator combines instances of random graph models that are known to have different network topologies, such as geometric (GEO) \cite{penrose2003random} and scale-free (SF) \cite{barabasi1999emergence}, at level 1 and level 2. Intuitively, if combined level 1 and level 2 network topologies of two entities differ from one another (either only level 1 differs, only level 2 differs, or both differ), then only an approach that uses information from both levels should be able to detect the difference. Hence, this combination can be used to label entities' nodes. For example, consider an NoN with two regions: one where the level 2 network topology is GEO-like and the corresponding level 1 network of each level 2 node in that region is of type GEO, and another where the level 2 network topology is SF-like but the corresponding level 1 network of each level 2 node in that region is of type SF. Clearly, every entity in the former NoN region has a different combination of level 1 and level 2 network topologies compared to every entity in the latter NoN region. As such, the entities can be labeled based on the (level 1, level 2) random graph model pair involving them, i.e., (GEO, GEO) in the former and (SF, GEO) in the latter.

\begin{figure}[ht]
    \centering
    \includegraphics[width=0.4\textwidth]{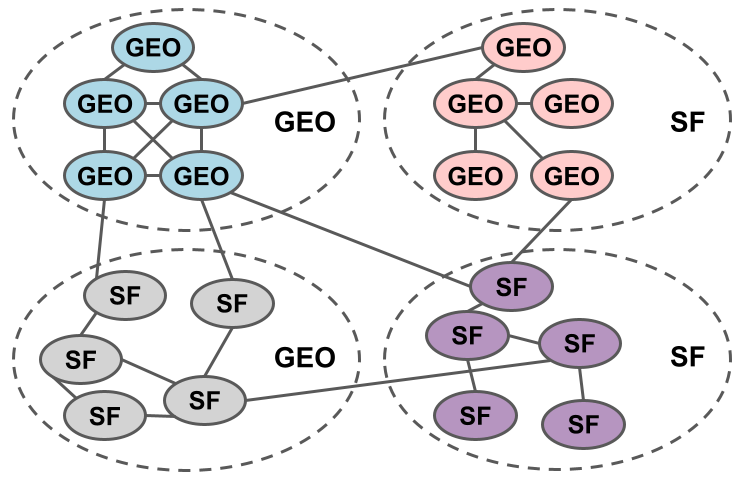}
    \caption{\label{fig:synthetic-non} A toy synthetic NoN generated from two graph models. Large dotted circles represent node groups (originating from isolated NoN regions) whose level 2 nodes are connected in a random geometric (GEO) or scale-free-like (SF) fashion. Small solid circles represent level 2 nodes whose level 1 networks are of the random graph type indicated. Level 1 nodes and edges are not shown. Level 2 nodes are colored based on their label, i.e., their combination of level 1 and level 2 network topology (\{(GEO, GEO), (GEO, SF), (SF, GEO), and (SF, SF)\}).}
\end{figure}

Our generator starts by creating several isolated NoN regions and then combining them. In our study, we generate four types of isolated NoN regions -- corresponding to the four possible combinations of GEO and SF at level 1 and level 2: \{(GEO, GEO), (GEO, SF), (SF, GEO), (SF, SF)\} -- with multiple instances of each type. We generate the isolated NoN regions so that, when combined into a single connected NoN (as described below), the resulting NoN's level 2 network has roughly 15,000 nodes and 300,000 edges to approximate the size of the human PIN, and so that each level 1 network has 200 nodes and 800 edges to approximate the average size of the PSNs.
The resulting NoN has four labels, with an equal number of entities having each label, so balanced multiclass classification is performed. We visualize a toy NoN having a single instance of each of the four random graph model combinations in Fig. \ref{fig:synthetic-non}.

To form one connected NoN, we join the isolated NoN regions at the second level; we refer to such regions as level 2 node groups in the connected NoN. Specifically, we randomly remove edges within level 2 node groups and randomly add the same number of edges across level 2 nodes groups (\textit{across-edge} amount). That is, we repeat the following process $a\% \times$ 300,000 times: (i) randomly select a level 2 node group, (ii) randomly select an edge in that node group, (iii) delete that edge, (iv) randomly select two level 2 nodes from different node groups, and (v) add an edge between the selected nodes.
We start with $a=5$ to retain most of the level 2 node groups' original GEO- and SF-like network topologies, and we vary $a$ to be 25, 50, 75, and 95 to test the effect of breaking the network topologies down. This also means that at $a=5$ there is significant clustering (each level 2 node group consists of densely interconnected nodes), while at $a=95$ there is very little clustering.

We also introduce random rewiring to test each approach's robustness to data noise (\textit{rewire-noise} amount). Specifically, for $r\%$ rewire-noise, for each level 1 network and each level 2 node group, we randomly remove $r\%$ of the total edges and randomly add the same number of edges back.
We vary $r$ to be 0 (no noise), 10, 25, 50, 75, and 100 (completely random). Combining the $a$ and $r$ parameters, we generate a total of $5 \times 6 = 30$ synthetic NoNs. For a formal description of the NoN generation process and the parameters we vary, see Supplementary Section \ref{supp:methods-data-synthetic}.

In our study, we report results for two-model NoNs, i.e., for $\{$GEO, SF$\}$. Note that we also analyzed three-model NoNs, adding the Erd\H{o}s-R\'{e}nyi (ER) model \cite{erdHos1960evolution}, i.e., for $\{$GEO, SF, ER$\}$. Because results are qualitatively and quantitatively similar, we do not discuss this analysis in the paper due to space constraints.


\subsubsection{Biological NoN} \label{data:pin-psn-non}
We also investigate whether integration is useful in the applied task of protein functional prediction. We construct a biological NoN using the human PIN and the proteins' associated PSNs (see also Supplementary Section \ref{supp:methods-data-bio}). We construct a PIN using human PPI data from BioGrid \cite{stark2006biogrid} version 4.1.190; this PIN has 18,708 nodes and 434,527 edges. Then, we map each protein ID to its corresponding PDB chain, resulting in 4,776 PDB chains.
Finally, we construct PSNs from these chains using an established process: nodes represent amino acids and edges join two amino acids if the distance between any of their heavy atoms (carbon, nitrogen, oxygen, or sulfur) is within 6 \r{A} \cite{faisal2017grafene}. The obtained biological NoN has 18,708 proteins at level 2, of which 4,776 have PSNs at level 1.

To obtain label information, we rely on protein-GO term annotation data (accessed in October 2020) \cite{ashburner2000gene}. Of all protein-GO term annotations, we focus on biological process (BP) GO terms in which the annotations were experimentally inferred (EXP, IDA, IPI, IMP, IGI, IEP). From those, we keep only GO terms annotating the 4,776 proteins that have PSNs, which results in 131 unique GO terms, i.e., classification labels. For each label $g$, proteins annotated by $g$ constitute positive data instances. 
While we could consider negative data instances to be all proteins not annotated by $g$, this could add bias for proteins that are not annotated by $g$ but are by GO terms related to $g$ and would also create an extreme positive/negative imbalance. Instead, we define negative data instances to be proteins that are not currently annotated by any BP GO term, reducing the bias and resulting in more balanced classes. 
Ultimately, each label has between 20 and 277 positive data instances and 61 negative data instances; as there are 131 labels total, we perform binary classification 131 times (Section \ref{methods:eval-framework}). Note that not all proteins have labels. Regardless, when extracting information from the level 2 network, we consider all 18,708 nodes and 434,527 edges. However, for each label, we only perform classification on the positive and negative data instances.

\subsection{Approaches for label prediction} \label{methods:approaches}

We consider graph theoretic approaches that are based on graphlets \cite{milenkovic2008uncovering}, and graph learning approaches, namely, SIGN \cite{rossi2020sign} and DiffPool \cite{ying2018hierarchical}. 

Graphlets are small subgraphs (a path, triangle, square, etc.) that can be considered the building blocks of networks, and they can be used to extract features of both nodes and networks (Supplementary Section \ref{supp:methods-existing}). The graphlet-based feature of a node in a general network is called its \textit{graphlet degree vector} (GDV), and GDVs of all nodes in a network can be collected into the network's \textit{GDV matrix} (GDVM) feature.
One drawback of GDVM is that its dimensions depend on the size of the network -- if performing graph classification of different sized networks using GDVM features, issues can arise. Thus, we also consider a transformation of GDVM, the graphlet correlation matrix (GCM) \cite{yaverouglu2014revealing}, which always has the same dimensions regardless of network size.

Given these definitions of graphlet features for nodes in a general network or for the entire general network itself, we now explain which features we use for nodes in a level 2 network and which features we use for level 1 networks. For the former, we extract each level 2 node's GDV (L2 GDV). For the latter, we extract each level 1 network's GDVM and GCM (L1 GDVM and L1 GCM). We use L1 GDVM when analyzing synthetic NoNs since we found that it outperformed L1 GCM. For the biological NoN, L1 GCM is the only viable feature since level 1 networks (PSNs) have different numbers of nodes (amino acids).

Then, to obtain NoN graphlet features, we concatenate level 2 nodes' L2 GDVs with their networks' L1 GDVMs or L1 GCMs. 
This results in five graphlet-based features: those for level 1 networks (L1 GDVM and L1 GCM) that are used for graph label prediction, those for nodes in a level 2 network (L2 GDV) that are used for node label prediction, and those for the entire NoN (L1 GDVM + L2 GDV and L1 GCM + L2 GDV) that are used for entity label prediction. In order to perform classification, for each graphlet-based feature, we train a logistic regression classifier (Supplementary Section \ref{supp:methods-eval}). So for example, when we say L2 GDV, we mean the L2 GDV feature under logistic regression.

SIGN aims to perform node classification (Supplementary Section \ref{supp:methods-eval}). It first computes adjacency matrix-based features and then uses them in a neural network classifier.
Mathematically, SIGN can be thought of as an ensemble of shallow graph convolutional network (GCN) classifiers, which is why it is a graph learning approach. In this study, when we say L2 SIGN, we mean its adjacency matrix-based features paired with its own classifier for node classification using only a level 2 network.

DiffPool aims to perform graph classification (Supplementary Section \ref{supp:methods-existing}). For each input network, DiffPool's GNN summarizes nodes' initial features into a hidden feature for the entire network. Then, given hidden features corresponding to the input networks, the GNN is trained on these hidden features to perform graph classification.
When we say L1 DiffPool, we mean its GNN with the initial features chosen (Supplementary Section \ref{supp:methods-eval}), for graph classification using only level 1 networks.

As SIGN and DiffPool are single-level graph learning approaches, we also combine them into an NoN graph learning approach. Given each level 2 node's SIGN feature, we concatenate it with the level 2 node's corresponding level 1 network's hidden feature computed by DiffPool's GNN. The GNN is then trained on these concatenated features to perform classification (any general purpose feature can be incorporated into DiffPool like this).
When we say L1 DiffPool + L2 SIGN, we mean entity label prediction using the process described above, incorporating SIGN's feature into DiffPool's GNN.
So, we use three graph learning-based approaches: L1 DiffPool, L2 SIGN, and L1 DiffPool + L2 SIGN.

We also combine L1 GDVM + L2 GDV or L1 GCM + L2 GDV with L1 DiffPool + L2 SIGN to test whether integrating information across the graph theoretic and graph learning domains improves upon either alone. Graphlet-based features can be incorporated into DiffPool using the process described previously.

In summary, thus far, we have described five single-level approaches and five NoN approaches that we use (Table \ref{tab:approach-cats}).

\begin{table}[ht!]
\centering
\begin{tabular}{lll}
\hline
  \multicolumn{2}{c}{Single-level approaches \vspace{0.5mm}} & \multicolumn{1}{c}{NoN approaches} \\
  Node-level & Network-level &      \\
\hline
    L2 GDV       & L1 GDVM &  L1 GDVM + L2 GDV  \\
     & L1 GCM  &   L1 GCM + L2 GDV    \\
    L2 SIGN & L1 DiffPool & L1 DiffPool + L2 SIGN \\
    & & Combined all (L1 GDVM) \\
    & & Combined all (L1 GCM) \\
\hline
\end{tabular}
\caption{Existing approaches that we consider and their generalized NoN counterparts. ``Combined all (L1 GDVM)'' refers to L1 GDVM + L2 GDV + L1 DiffPool + L2 SIGN; ``Combined all (L1 GCM)'' is similarly named.}
\label{tab:approach-cats}
\end{table}


\newcommand{\presuper}[3]{\tensor*[^{#1}]{#2}{#3}}
Next, we describe our integrative GCN-based approach. We focus on GCNs for two reasons: (i) recent work has suggested that other GNN architectures do not offer very much benefit over GCNs \cite{shchur2018pitfalls, wu2019simplifying, rossi2020sign}, making such methods more complex for little gain and (ii) the extension of GCNs to NoNs is intuitive.

The basic unit of a GCN is a graph convolutional layer. Graph convolution layers allow each node to see information about its neighbors.
So, we generalize graph convolution layers to NoNs so that each node receives information not only from its neighbors (in the same level), but also from its corresponding network at a lower level or from the network it is a part of at a higher level. This would be in line with intuition that, for example, the feature of a protein should contain information about how it interacts with other proteins (i.e., its topology in the level 2 network) and structural properties of the protein itself that allow for such interactions (topology of level 1 nodes in its level 1 network). Then, we can stack multiple NoN graph convolutional layers (with intermediate layers in between) to form an NoN-GCN (Supplementary Section \ref{supp:methods-gcn}). We refer to an NoN-GCN approach using $\lambda$ layers as ``GCN-$\lambda$''.


\subsection{Evaluation} \label{methods:eval-framework}

For a given NoN $\{G^{(2)} = (V^{(2)}, E^{(2)})$ and $\{G^{(1)}_1,...,G^{(1)}_{|V^{(2)}|}\}\}$, its label set $Y = {y_1,...,y_c}$, and a function that maps level 2 nodes (and thus their corresponding level 1 networks) to their labels $f_{true}: V^{(2)} \rightarrow Y$, the goal is to learn a predictive function $f_{pred}: V^{(2)} \rightarrow Y$. We do this by first splitting the data into three disjoint sets: training ($V^{(2)}_{tr}$), validation ($V^{(2)}_{val}$), and testing ($V^{(2)}_{te}$). Then, we train a classifier on the training set that aims to minimize the cross-entropy loss between $f_{true}(V^{(2)}_{tr})$ and $f_{pred}(V^{(2)}_{tr})$. We use $V^{(2)}_{val}$ to optimize hyperparameters and finally report the classifier's performance on $V^{(2)}_{te}$, an independent set never seen in the training process and not used for determining hyperparameters. As typically done, we form these disjoint sets using stratified sampling, repeating multiple times and averaging the results to reduce bias from the randomness of the sampling. For details on hyperparameter optimization and sampling, see Supplementary Section \ref{supp:methods-eval}.


Regarding how we measure classification performance of an approach, for synthetic NoNs, we report classification accuracy (Supplementary Section \ref{supp:methods-eval}) since class sizes are balanced. For the real-world NoNs, we report area under precision-recall (AUPR), precision@k, recall@k, and F-score@k (Supplementary Section \ref{supp:methods-eval}), since class sizes are not balanced. As commonly done, we also perform statistical tests to see whether each approach's performance is significantly better than random, i.e., is ``significant'' (Supplementary Section \ref{supp:methods-eval}).

\section{Results and discussion}

\subsection{Synthetic NoNs} \label{sec:results-accuracy-synthetic}

\begin{figure*}

\subfloat[]{\includegraphics[width=0.3\textwidth]{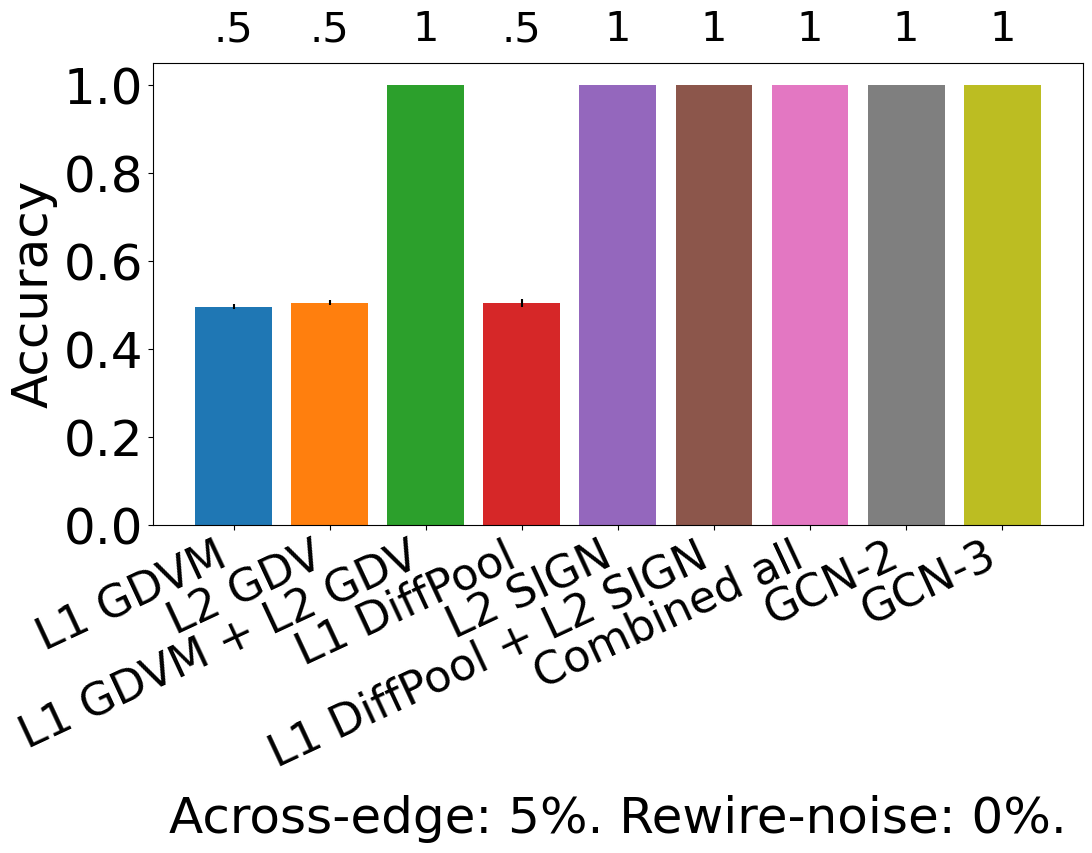}} 
\subfloat[]{\includegraphics[width=0.3\textwidth]{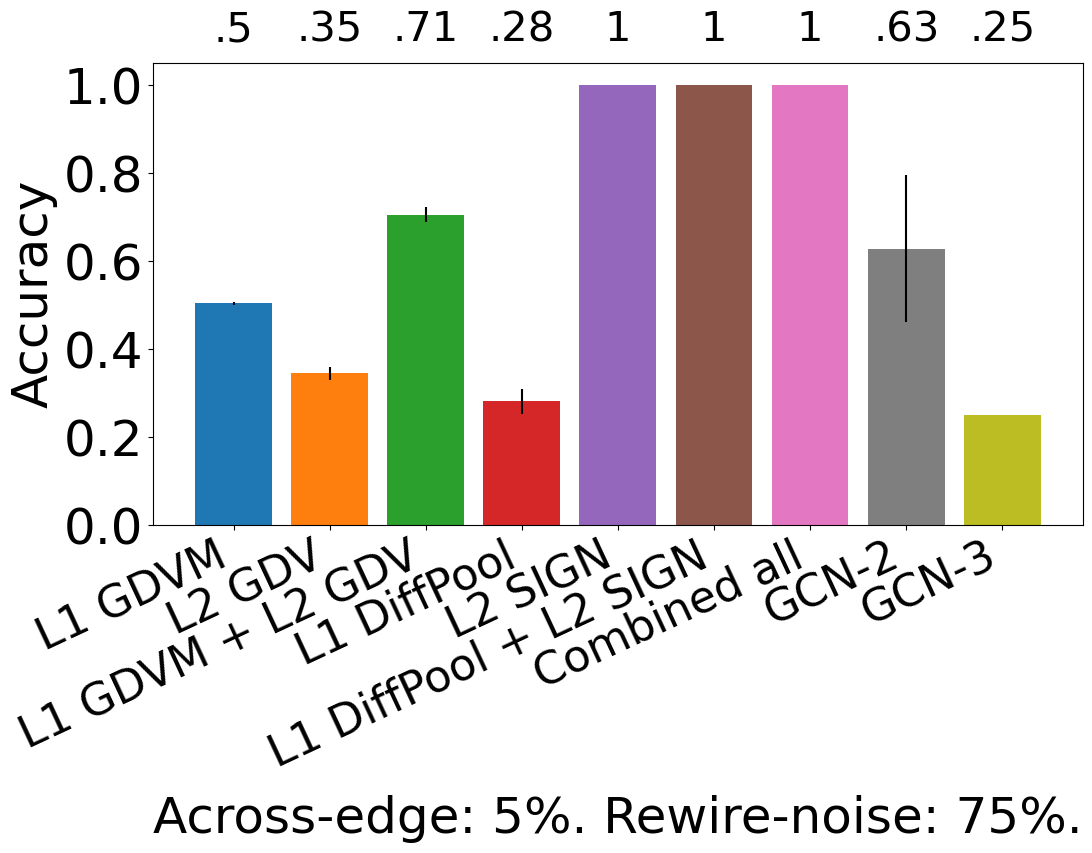}} 

\subfloat[]{\includegraphics[width=0.3\textwidth]{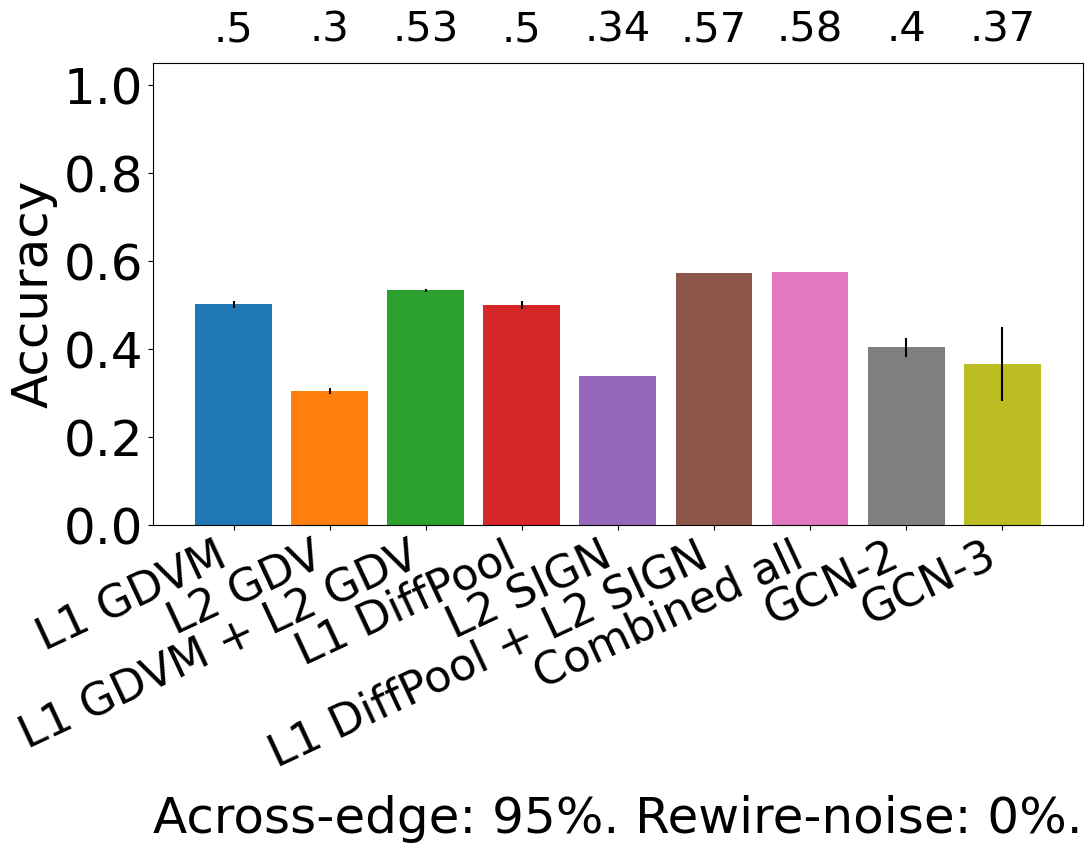}} 
\subfloat[]{\includegraphics[width=0.3\textwidth]{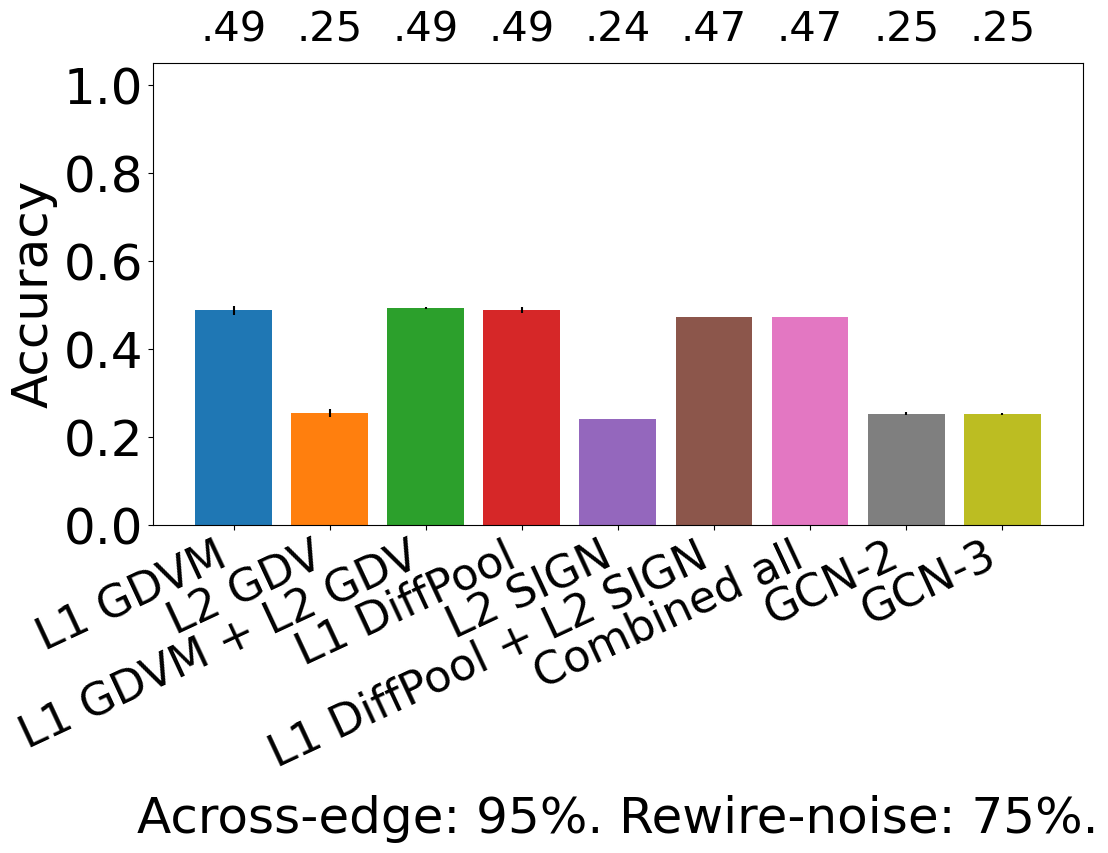}}
          
\caption{Comparison of the nine considered approaches in the task of label prediction for synthetic NoNs with the following parameters: \textbf{(a)} 5\% across-edge and 0\% rewire-noise amount, \textbf{(b)} 5\% across-edge and 75\% rewire-noise amount, \textbf{(c)} 95\% across-edge and 0\% rewire-noise amount, and \textbf{(d)} 95\% across-edge and 75\% rewire-noise amount. ``Combined all'' refers to L1 GDVM + L2 GDV + L1 DiffPool + L2 SIGN. Accuracy is shown above the bars. We expect an approach that only uses a single level and does not capture clustering information to have around $\frac{\textrm{\# of models}}{\textrm{\# of labels}}$, or 0.5, accuracy when both across-edge and rewire-noise amount are low (Supplementary Section \ref{supp:synthetic-results}). 
Results for other parameter combinations are shown in Supplementary Figs. \ref{suppfig:synthetic-5-across-edge}-\ref{suppfig:synthetic-95-across-edge}.}
  \label{fig:synthetic-non-results}
\end{figure*}

We expect NoN approaches to outperform single-level ones. We find that at least one NoN approach (L1 GDVM + L2 GDV, L1 DiffPool + L2 SIGN, L1 GDVM + L2 GDV + L1 DiffPool + L2 SIGN, GCN-2, or GCN-3) outperforms or ties (is within 1\% of) all single-level approaches (L1 GDVM, L2 GDV, L1 DiffPool, and L2 SIGN) for 30 out of the 30 synthetic NoNs (Fig. \ref{fig:synthetic-non-results} and Supplementary Figs. \ref{suppfig:synthetic-5-across-edge}-\ref{suppfig:synthetic-95-across-edge}). Specifically, at least one NoN approach outperforms all single-level approaches for 9 out of the 30 NoNs, and at least one NoN approach is tied with L2 SIGN for 21 out of the 30 NoNs. L2 SIGN is the only single-level approach that ties NoN approaches. However, before we discuss why L2 SIGN performs as well as NoN approaches, we need to understand the effects of both across-edge amount and rewire-noise amount.


Recall that when we increase across-edge amount, level 2 node groups' original GEO- and SF-like network topologies are increasingly broken down and eventually become entirely random. When across-edge amount is high, most edges will exist across level 2 node groups, not within (and there will be very little, if any, clustering structure in the level 2 network). Thus, approaches using only level 2 information (L2 GDV and L2 SIGN) will be making predictions on random data, and approaches that combine level 1 and level 2 information (all NoN approaches) will be making predictions on partially random data (level 1 networks are unaffected by across-edge amount). So, for the former approaches, we expect that as across-edge amount increases, prediction accuracy will drop to 0.25 (since there are four labels and class sizes are balanced, random performance is $\frac{1}{4}$). For the latter approaches, we expect that prediction accuracy will drop to 0.5, for the following reason. The only signal NoN approaches can pick up when across-edge amount is high is from level 1 networks, essentially turning NoN approaches into a single-level approaches (level 1 only). And, the maximum expected accuracy of any single level approach is $\frac{\textrm{\# of models}}{\textrm{\# of labels}}$, or 0.5 (Supplementary Section \ref{supp:synthetic-results}).
Indeed, we observe these drops in accuracy for all approaches (Fig. \ref{fig:synthetic-non-results}(c, d) and Supplementary Figs. \ref{suppfig:synthetic-5-across-edge}-\ref{suppfig:synthetic-95-across-edge}).

Recall that we increase rewire-noise amount to investigate approaches' robustness to increasing data noise. When rewire-noise amount is high, both the level 2 node groups' and level 1 networks' original GEO- and SF-like network topologies will now be random (note, however, that clustering structure will not be affected since rewire-noise occurs within node groups, not across). So, all types of approaches will be making predictions on random data.
As such, we expect that as rewire-noise amount increases, prediction accuracy will decrease.
We observe these drops in accuracy for all approaches except L2 SIGN (Fig. \ref{fig:synthetic-non-results}(b, d) and Supplementary Figs. \ref{suppfig:synthetic-5-across-edge}-\ref{suppfig:synthetic-95-across-edge}), which we discuss below.

To summarize, high across-edge amount leads to a random level 2 network with very little clustering structure, and high rewire-noise amount leads to a random level 2 network with clustering structure (in addition to random level 1 networks). Since L2 SIGN performs poorly for the former (Fig. \ref{fig:synthetic-non-results}(c, d)) but well for the latter (Fig. \ref{fig:synthetic-non-results}(b)), despite the level 2 networks having random network topology in both situations, we hypothesize that L2 SIGN 
is able to capture the clustering structure in the level 2 network, i.e., it is able to detect the existence of densely interconnected level 2 node groups. So, L2 SIGN is able to perform as well as NoN approaches for 21 out of the 30 NoNs simply because there exists clustering structure in the level 2 networks of those 21 NoNs.
L2 SIGN's ability to capture clustering structure is also likely why at low across-edge amounts, regardless of rewire-noise amount, NoN approaches incorporating L2 SIGN perform as well as they do. This also suggests that when one expects clustering structure in the data, incorporating SIGN could help. 

Above, we analyze single-level approaches versus NoN approaches as well as trends regarding across-edge amount and rewire-noise amount. However, recall that the approaches we consider come from either the graph theoretic or graph learning domain. So, we also compare the two domains. For simplicity, we focus on the NoN approaches, i.e., L1 GDVM + L2 GDV from the graph theoretic domain and L1 DiffPool + L2 SIGN from the graph learning domain, as we already know that they outperform or tie single-level approaches. We find that L1 DiffPool + L2 SIGN outperforms L1 GDVM + L2 GDV for 20 out of the 30 NoNs, is tied for 9 out of the 30 NoNs, and is worse for 1 out of the 30 NoNs. However, as discussed above, for NoNs where across-edge amount is low and rewire-noise amount is high, L1 DiffPool + L2 SIGN's performance likely comes from L2 SIGN. We also investigate whether combining research knowledge from the graph theoretic and graph learning domains improves upon each domain individually. This does not appear to be the case on the synthetic data, as L1 DiffPool + L2 SIGN is as good as L1 GDVM + L2 GDV + L1 DiffPool + L2 SIGN for 29 out of the 30 NoNs and is worse for only one NoN (Fig. \ref{fig:synthetic-non-results} and Supplementary Figs. \ref{suppfig:synthetic-5-across-edge}-\ref{suppfig:synthetic-95-across-edge}).

Finally, recall that our extensions of existing node/graph label prediction approaches to their NoN counterparts (L1 GDVM + L2 GDV, L1 DiffPool + L2 SIGN, L1 GDVM + L2 GDV + L1 DiffPool + L2 SIGN) are concatenation-based, which is why we developed integrative NoN-GCN approaches (GCN-2 and GCN-3) as well. Regarding the NoN-GCN approaches themselves, we expect that GCN-3 will outperform GCN-2, as the former is a deeper model. However, this is not the case, as GCN-3 only outperforms GCN-2 for 2 out of the 30 NoNs, ties for 21 out of the 30, and is worse for 7 out of the 30 (Fig. \ref{fig:synthetic-non-results} and Supplementary Figs. \ref{suppfig:synthetic-5-across-edge}-\ref{suppfig:synthetic-95-across-edge}). This, combined with the fact that GCN-3 takes more time than GCN-2 (Section \ref{sec:results-running-time}), is why we did not consider GCN-3 for the biological NoN. Still, we expect that the integrative NoN-GCN approaches will outperform the concatenation-based ones. We find that while the NoN-GCN approaches do perform well for low across-edge amounts and low rewire-noise amounts, they are not as robust to changes in those parameters compared to the concatenation-based ones. Specifically, NoN-GCN approaches perform as well as concatenation-based ones for 7 out of the 30 NoNs and are worse for 23 out of the 30 NoNs (Fig. \ref{fig:synthetic-non-results} and Supplementary Figs. \ref{suppfig:synthetic-5-across-edge}-\ref{suppfig:synthetic-95-across-edge}).
These findings suggest that deep learning might not offer an advantage on this kind of synthetic data, or that more complex models are needed.

\subsection{Biological NoN} \label{res:pin-psn}
\begin{figure}[ht]
     \centering
         \includegraphics[width=0.55\textwidth]{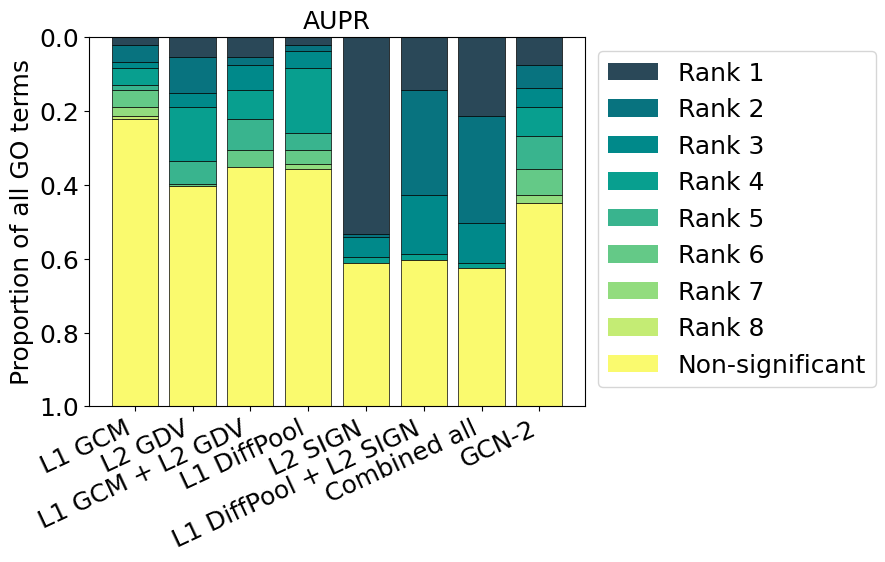}
    \caption{Summarized results of the eight considered approaches (as GCN-3 is not used for the biological NoN) in the task of protein functional prediction in terms of AUPR. For each GO term (out of the 131 total), we rank the eight approaches' from best (rank 1) to worst (rank 8). 
    Then, we calculate the proportion of GO terms each approach achieves each rank. ``Combined all'' refers to L1 GDVM + L2 GDV + L1 DiffPool + L2 SIGN. Results for other evaluation measures are shown in Supplementary Fig. \ref{suppfig:bio-overall-rankings}}
    \label{fig:pin-psn-overall-rankings}
\end{figure}

Again, we expect NoN approaches to improve upon single-level ones. Since we consider 131 GO terms and parsing raw results for every single one would be difficult, we instead present summarized results over the 131. Specifically, given the eight considered approaches (L1 GCM, L2 GDV, L1 GCM + L2 GDV, L1 DiffPool, L2 SIGN, L1 DiffPool + L2 SIGN, L1 GCM + L2 GDV + L1 DiffPool + L2 SIGN, and GCN-2), for each of AUPR, precision, recall, and F-score, for each GO term, we do the following. We rank each of the eight approaches that is significant (Section \ref{methods:eval-framework}) from 1st best (rank 1) to 8th best (rank 8), considering any approaches within 1\% of each other to be tied. Then, for each approach, we count how many times (i.e., for how many GO terms) it has rank 1, 2, etc. We find that NoN approaches have rank 1 for 49 out of the 131 GO terms with respect to AUPR, 37 out of 131 for precision, 35 out of 131 for recall, 33 out of 131 for F-score, and 69 out of 131 for at least one of the four evaluation measures (Fig. \ref{fig:pin-psn-overall-rankings} and Supplementary Fig. \ref{suppfig:bio-overall-rankings}). We examine in more detail why NoN approaches work better than single-level approaches for some but not all GO terms, as follows.

First, we investigate whether the GO terms for which NoN approaches have rank 1 are different than the GO terms for which L2 SIGN, the best approach overall, has rank 1. If not, then NoN approaches would be redundant to L2 SIGN. To do so, for each NoN approach, we measure the overlap between the set of GO terms for which the given NoN approach has rank 1 and the set of GO terms for which L2 SIGN has rank 1. We find that NoN approaches have rank 1 for mostly different GO terms compared to L2 SIGN, with a maximum overlap of around 6\% (Supplementary Fig. \ref{suppfig:sign-overlaps}). This suggests that NoN approaches are not redundant to L2 SIGN. 

So, it makes sense to continue analyzing NoN approaches in comparison to single-level approaches. To better understand for which kinds of GO terms NoN approaches have rank 1 versus for which kinds of GO terms single-level approaches have rank 1, we do the following. For each evaluation measure, we split the 131 GO terms into six groups based on how single-level approaches perform in relation to NoN approaches, with ``S'' referring to single-level approaches and ``C'' referring to combined-level (i.e., NoN) approaches, as outlined in Table \ref{tab:group-desc}. As an example, for AUPR, ``S < C'' indicates that the performance of single-level approaches (``S'') is worse than (``<'') the performance of NoN approaches (``C''). In other words, the group ``S < C'' contains all GO terms for which at least one NoN approach has rank 1 (multiple NoN approaches can be tied with each other for rank 1), and all single-level approaches have rank 2 or worse, with respect to AUPR. Note that for a GO term in the above scenario, if no single-level approaches are significant, that GO term would be in the ``C only'' group instead, corresponding to those GO terms for which only NoN approaches are significant.

\begin{table}[ht]
\centering
\begin{tabularx}{0.99\textwidth}{l|X}

\hline
S only           & At least one ``S'' approach is significant; \newline no ``C'' approaches are significant. \\[0mm]
\hdashline
S $>$ C & At least one ``S'' approach is significant and has rank 1; \newline at least one ``C'' approach is significant but none have rank 1.      \\[0mm]
\hdashline
S = C   & At least one ``S'' approach is significant and has rank 1; \newline at least one ``C'' approach is significant and has rank 1.    \\[0mm]
\hdashline
\rowcolor{mylb} S $<$ C    & At least one ``S'' approach is significant but none have rank 1; \newline at least one ``C'' approach is significant and has rank 1.   \\[0mm]
\hdashline
\rowcolor{mylb} C only           & No ``S'' approaches are significant; \newline at least one ``C'' approach is significant.      \\[0mm]
\hdashline
No sig.          & No approaches are significant.     \\
\hline

\end{tabularx}
\caption{Description of the six GO term groups based on how single-level (S) and combined-level (C), i.e., NoN, approaches perform. Rows with blue backgrounds correspond to the groups where NoN approaches are the best.}
\label{tab:group-desc}
\end{table}

Given these groups, we investigate whether there are any GO terms where NoN approaches are necessary if one wants to make accurate predictions. We do so by looking at the number of GO terms for which at least one NoN approach has rank 1 and all single-level approaches are strictly worse, i.e., not tied for rank 1. This corresponds to the number of GO terms in the groups ``S $<$ C'' and ``C only''. We find that NoN approaches have rank 1 and are untied with any single-level approach for around 20\%-30\% of all GO terms, depending on evaluation measure (Table \ref{tab:goterm-counts}). Taking the union over all evaluation measures, we find that there are 33 (25\% of) GO terms in ``S $<$ C'' and 38 (29\% of) in ``C only'', i.e., a total of 60 (46\% of) GO term across the two groups. That is to say, there are 33 GO terms where NoN approaches outperform single-level approaches (but single-level approaches are still significant) for at least one evaluation measure and, importantly, 38 GO terms where only NoN approaches are able to perform significantly better than random for at least one evaluation measure. In other words, for those 38 GO terms, only NoN approaches make meaningful protein functional predictions, while single-level ones achieve random accuracy.
Taking the groups together, we find that there are 60 GO terms where NoN approaches have rank 1 and single-level approaches are strictly worse for at least one evaluation measure. These results suggest that NoN approaches are necessary, especially if one wants to make predictions for certain GO terms.

\newcolumntype{R}[1]{>{\raggedleft\let\newline\\\arraybackslash\hspace{0pt}}m{#1}}
\newcolumntype{L}[1]{>{\raggedright\let\newline\\\arraybackslash\hspace{0pt}}m{#1}}
\newcolumntype{C}[1]{>{\centering\let\newline\\\arraybackslash\hspace{0pt}}m{#1}}

\begin{table}
  \centering
  \renewcommand{\arraystretch}{1.2}
  \begin{tabular}{c|rr|rr|rr|rr|r}
    \hline
    & \multicolumn{9}{c}{Number of GO terms in each group for}  \\
     & \multicolumn{2}{c}{AUPR} & \multicolumn{2}{c}{Precision} & \multicolumn{2}{c}{Recall} & \multicolumn{2}{c}{F-score} & Union\\
    \hline
    S only           & 12 & 9\%  & 20 & 15\% & 33 & 25\% & 31 & 24\% & 46 \\
    S $>$ C & 63 & 48\% & 45 & 35\% & 22 & 17\% & 30 & 23\%  &  75 \\
    S = C            & 8 & 6\%   & 8 & 6\%   & 6 & 5\%   & 4 & 3\%   & 18   \\
    \rowcolor{mylb} S $<$ C    & 27 & 21\% & 8 & 6\%   & 8 & 6\%   & 9 & 7\%  & 33 \\
    \rowcolor{mylb} C only           & 14 & 11\% & 21 & 16\%   & 21 & 16\%   & 20 & 15\%  & 38  \\
    No sig.          & 7 & 5\%   & 29 & 22\% & 41 & 31\% & 37 & 28\% & 43  \\
    \hline
  \end{tabular}
  \caption{Number of GO terms in each of the six groups for AUPR, precision, recall, and F-score. For example, for AUPR, there are 14 GO terms in the group ``C only''. We also report the union of GO terms in a given group over all measures (Union). For example, there are 38 GO terms in the union of ``C only'' over all evaluation measures. Rows with blue backgrounds correspond to the groups where NoN approaches are the best. The IDs and names of GO terms in each group for each measure can be found in Supplementary Tables \ref{supptab:goinfo-aupr}-\ref{supptab:goinfo-f-score}.}
\label{tab:goterm-counts}
\end{table}

Since we now know that NoN approaches are important, we investigate which of them are the best.  Here, we comment on results for AUPR (Supplementary Fig. \ref{suppfig:combined-best-overlaps}(a)), only noting that results are qualitatively similar for other measures (Supplementary Fig. \ref{suppfig:combined-best-overlaps}-\ref{suppfig:f-score-grid}).
For ``S $<$ C'', L1 GCM + L2 GDV + L1 DiffPool + L2 SIGN, i.e., the combination of graph theoretic and graph learning approaches, is the best overall NoN approach. It has rank 1 for 19 GO terms, while all other NoN approaches have rank 1 for fewer than 19 GO terms (Supplementary Fig. \ref{suppfig:combined-best-overlaps}(a)). This suggests that 
integrating knowledge across domains is somewhat useful. For ``C only'', GCN-2 has rank 1 for 9 GO terms, while all other NoN approaches have rank 1 for fewer than 9 GO terms (Supplementary Fig. \ref{suppfig:combined-best-overlaps}(b)). In fact, for 7 out of the 9 GO terms, GCN-2 is the only approach that is significant (Supplementary Fig. \ref{suppfig:aupr-grid}). This suggests that deep learning could be useful for otherwise difficult-to-predict GO terms. 

Finally, note that we did analyze the properties of GO terms in each of the six GO term groups, in order to see whether different GO term groups contain different kinds of GO terms. Specifically, for each group, we computed the distribution of the depths of the GO terms in the GO tree and the distribution of class sizes (number of proteins annotated by each GO term, which ranges from 20 to 277), and compared groups' distributions to each other. We found that ``S < C'' and ``C only'' contain GO terms whose classes sizes are among the smallest, suggesting that NoN approaches may have some potential to make predictions for GO terms with limited training data. And while one might expect that GO terms with small class sizes correspond to those that are deep in the GO tree, we find that there is no significant difference between the six GO term groups with respect to GO term tree depth. 



\subsection{Running times}\label{sec:results-running-time}

Lastly, we analyze approaches' running times for the synthetic NoN with 5\% across-edge and 0\% rewire-noise amount as a representative; we choose a single NoN for two reasons. The first is that GCN-3 was only run on synthetic NoNs (Section \ref{sec:results-accuracy-synthetic}), so they are the only NoNs where we can analyze the trade-off between performance (Fig. \ref{fig:synthetic-non-results}) and running time. The second is simplicity: trends are qualitatively similar across all synthetic NoNs. For each approach, we record the time to extract all necessary features and the time for one epoch of training the associated classifier. For hardware details, see Supplementary Section \ref{supp:results-time}.


First, GCN-3, which we found does not have a clear advantage over GCN-2 in terms of accuracy (Section \ref{sec:results-accuracy-synthetic}), takes 4.25x longer to train. This poor tradeoff between accuracy and running time is why we did not consider GCN-3 for the biological NoN.

Second, recall that L1 DiffPool + L2 SIGN and L1 GDVM + L2 GDV + L1 DiffPool + L2 SIGN, the best approaches overall, are as good as each other in terms of accuracy, with the former being worse in only 1 out of the 30 NoNs. Thus, because L1 GDVM + L2 GDV + L1 DiffPool + L2 SIGN has longer feature extraction and training time than L1 DiffPool + L2 SIGN (Supplementary Table \ref{supptab:running-times}), L1 DiffPool + L2 SIGN would likely be the better approach to use for a general NoN when considering the trade-off between accuracy and running time. 
Also recall that L2 SIGN performs as well as L1 DiffPool + L2 SIGN and L1 GDVM + L2 GDV + L1 DiffPool + L2 SIGN for 21 out of the 30 NoNs, in those NoNs where there is significant clustering structure in the level 2 network. Thus, if one expects significant clustering structure in the level 2 network of a general NoN, L2 SIGN should be considered, as its feature extraction time is around 77x faster and its training time is around 1.5x faster than those of L1 DiffPool + L2 SIGN (Supplementary Table \ref{supptab:running-times}).
\section{Conclusion}

We present a comprehensive framework to test whether integrating network information into an NoN leads to more accurate label predictions than using information from a single level alone. We also develop the first synthetic NoN generator that can create NoNs with a variety of parameters for study, construct a biological NoN from PIN and PSN data, and propose a novel GCN-based model for label prediction on NoNs. We have shown that on synthetic data, NoN approaches are among the best, and that on a real-world biological NoN, NoN approaches are necessary to make predictions about certain protein functions. As such, research into NoN-based data integration is promising, and likely could be applied to a variety of other tasks, especially as such NoN data becomes readily available.

To our knowledge, our study is the first to investigate data integration for label prediction using NoNs. Recall that, for example, studies have combined protein sequence and protein structural data with PPI data \cite{zhang2009inferring, peng2014improving, zhang2019deepfunc}. So, another important direction is the comparison between different data integration schemes for various tasks. Also, our integrative NoN-GCN model is not significantly better than just combining features from the two levels. More sophisticated and scalable deep learning models for NoNs, perhaps taking inspiration from SIGN's precomputable neighborhood aggregators, are worth pursuing. Finally, we only analyze a two-level NoN in our study, so expanding in scale is an important future direction.

\section*{Acknowledgements}
We thank Khalique Newaz for creating visualizations in Fig. \ref{fig:pin-psn-non}(b). This work was funded by National Institutes of Health (1R01GM120733) and National Science Foundation (CAREER CCF-1452795).

\newpage

\newcommand{\beginsupplement}{%
        \renewcommand{\figurename}{Supplementary Figure}
        \renewcommand{\tablename}{Supplementary Table}
        \renewcommand{\thetable}{S\arabic{table}}%
        \renewcommand{\thefigure}{S\arabic{figure}}%
        \renewcommand{\thesection}{S\arabic{section}}%
        \setcounter{table}{0}
        \setcounter{figure}{0}
        \setcounter{section}{0}
        
}

\newcount\suppl
\suppl0
\def\Ref#1{\ifnum\suppl<1 \ref{#1} \else S\ref{#1}\fi}

\beginsupplement

\section*{Supplementary information for ``Modeling multi-scale data via a network of networks''}

\section{Materials and methods} \label{supp:methods}
\subsection{Data} \label{supp:methods-data}
\subsubsection{Our synthetic NoN generator} \label{supp:methods-data-synthetic}
Let $M$ be the set of random graph models we consider for generating the synthetic NoN. For random graph model $m \in M$, let $m(x, y)$ be a random graph of type $m$ with $x$ nodes and $y$ edges. Let $M^{(2)} = M \times M$ be the set of all possible combinations of the elements of $M$ with themselves. 
Let $|V^{(2)}|$ be the target number of nodes at level 2, $|E^{(2)}|$ be the target number of edges at level 2, $|V^{(1)}|$ be the target number of nodes for each network at level 1, and $|E^{(1)}|$ be the target number of edges for each network at level 1; these parameters allow us to generate synthetic NoNs that approximate the size of real-world NoNs. Note that in our synthetic NoN generation, we fix the size of the level 1 networks to eliminate any effect of level 1 network size; however, our model can easily generate level 1 networks of varying size.

For each $(m_1, m_2) \in M^{(2)}$, we generate $k$ isolated NoN regions where, for each region, the level 2 network is of type $m_2$ and every level 1 network is of type $m_1$. This results in $k|M^{(2)}|$ total isolated NoN regions. After combining all of them, the resulting NoN should have $|V^{(2)}|$ nodes and $|E^{(2)}|$ edges. As such, for each $(m_1, m_2) \in M^{(2)}$, we generate $k$ isolated NoN regions 
\begin{equation}
    \begin{split}
        \{G_{(m_1, m_2)}^{(2)} = m_2(\lfloor\frac{|V^{(2)}|}{k|M^{(2)}|}\rfloor, \lfloor\frac{|E^{(2)}|}{k|M^{(2)}|}\rfloor) \textrm{ and } \\
        \{G_{(m_1, m_2)_i}^{(1)} = m_1(|V^{(1)}|, |E^{(1)}|) \textrm{ for } i \in \{1,...,\lfloor\frac{|V^{(2)}|}{k|M^{(2)}|}\rfloor\}\}.
    \end{split}
\end{equation}

Because real-world systems are likely to have many groups of nodes, we set $k=5$ for our synthetic NoNs, corresponding to five instances of each of the four random graph model combinations.
Then, we connect these isolated NoN regions by randomly removing edges within level 2 node groups and randomly adding the same number of edges across level 2 nodes groups (\textit{across-edge} amount). Specifically, we repeat the following process $a\% \times |E^{(2)}|$ times: (i) randomly select a level 2 node group, (ii) randomly select an edge in that node group, (iii) delete that edge, (iv) randomly select two level 2 nodes from different node groups, and (v) add an edge between the selected nodes. If the resulting NoN is still disconnected, we redo the process with a different random seed. While we could impose a condition to guarantee connectedness, doing so would bias the generation. If a connected NoN can not be found after 10 tries, we just continue with the last one.
We start with $a=5$ to retain most of the level 2 node groups' original GEO- and SF-like network topologies, and we vary $a$ to be 25, 50, 75, and 95 to test the effect of breaking the network topologies down. This also means that at $a=5$ there is significant clustering (each level 2 node group consists of densely interconnected nodes), while at $a=95$ there is very little clustering. 


We also introduce random rewiring to test each method's robustness to data noise (\textit{rewire-noise} amount). Specifically, for $r\%$ rewire-noise, for each level 1 network, we randomly delete $\frac{r}{100} \times |E^{(1)}|$ edges and randomly add the same number back. For the level 2 network, for each node group, we randomly delete $r\%$ of $\lfloor\frac{|E^{(2)}|}{5|M^{(2)}|}\rfloor$ edges and randomly add the same number back. We vary $r$ to be 0 (no noise), 10, 25, 50, 75, 100 (completely random).

\subsubsection{PIN-PSN NoN} \label{supp:methods-data-bio}
We construct a biological NoN using the human PIN and the proteins' associated PSNs. We obtain human PPI data from BioGrid \cite{stark2006biogrid} version 4.1.190. We keep only physical interactions, remove selfloops and multiedges, and take the largest connected component. This results in a final size of 18,708 nodes and 434,527 edges.

We map proteins in our PIN to their corresponding PDB IDs as follows. Considering the proteins' BioGrid IDs, we use UniProt's \cite{uniprot2019uniprot} mapping service (version 2020\_06) to obtain BioGrid-to-UniProt mappings. Any mappings that are not reviewed (i.e., not Swiss-Prot) are discarded. Next, we remove any mapped data when more than one BioGrid ID is mapped to a UniProt ID and vice versa, leaving only one-to-one mappings between BioGrid IDs and UniProt IDs. Then, we repeat the process starting with the proteins' official symbol IDs. As such, for each protein, we have two UniProt IDs: one originating from its BioGrid ID and the other from its official symbol ID. To remove any ambiguity moving forward, we only keep proteins whose two UniProt IDs are equal. In total, we have 16,079 such UniProt IDs.

Given these UniProt IDs, we again use UniProt's mapping service, but this time to map UniProt IDs to PDB IDs. Then, we remove any PDB ID whose PDB structure has a resolution greater than or equal to 3.0\AA, as PDB considers these to be ``low resolution'' \cite{pdbresolution}. Next, to obtain a one-to-one mapping between Uniprot IDs and PDB IDs, we form one set out of every protein sequence associated with the UniProt IDs and another set out of every protein sequence associated with every PDB chain (each PDB ID can have multiple corresponding chains). We perform all-vs-all protein sequence comparison using BLASTP \cite{altschul1997gapped} between these two sets and take only reciprocal best hits as our final one-to-one UniProt-to-PDB mappings. After this step, we have 4,776 PDB chains.

Regarding GO term labels, we only consider those GO terms with 20 or more positive instances to ensure there is enough data to perform classification on.

\subsection{Existing approaches for label prediction} \label{supp:methods-existing}
Recall that we consider graph theoretic approaches based on graphlets and graph learning approaches, namely, SIGN and DiffPool. 

\textcolor{black}{Graphlets are small subgraphs (a path, triangle, square, etc.) that can be considered the building blocks of networks, and they can be used to extract features of both nodes and networks. 
For each node in a general network, for each automorphism orbit (intuitively, node symmetry group) in a graphlet, one can count the number of times the node is a part of a given graphlet orbit.
These counts are summarized into the node's feature, also called its \textit{graphlet degree vector} (GDV); when considering up to 4-node graphlets, GDVs will have length 15. 
Then, to extract features of the entire network, GDVs of all nodes can be collected into the network's \textit{GDV matrix} (GDVM) feature. One drawback of the GDVM feature is that its dimensions depend on the number of nodes in the network -- if performing graph classification of different sized networks using GDVM features, issues can arise. Thus, we also consider a transformation of the GDVM, the graphlet correlation matrix (GCM) \cite{yaverouglu2014revealing}, which always has the same dimensions regardless of network size.}

\textcolor{black}{Given these definitions of graphlet features for nodes in a general network or for the entire general network itself, we now explain which features we use for nodes in a level 2 network and which features we use for level 1 networks. For the former, we extract each level 2 node's GDV (L2 GDV). For the latter, we extract each level 1 network's GDVM and GCM (L1 GDVM and L1 GCM). We use L1 GDVM when analyzing synthetic NoNs since we found that it outperformed L1 GCM. For the biological NoN, L1 GCM is the only viable feature since level 1 networks (PSNs) have different numbers of nodes (amino acids).} 

\textcolor{black}{Then, to obtain NoN graphlet features, we concatenate level 2 nodes' L2 GDVs with their networks' L1 GDVMs or L1 GCMs. 
This results in five graphlet-based features: those for level 1 networks (L1 GDVM and L1 GCM) that are used for graph label prediction, those for nodes in a level 2 network (L2 GDV) that are used for node label prediction, and those for the entire NoN (L1 GDVM + L2 GDV and L1 GCM + L2 GDV) that are used for entity label prediction. In order to perform classification, for each graphlet-based feature, we train a logistic regression classifier (Supplementary Section \ref{supp:methods-eval}). So for example, when we say L2 GDV, we mean the L2 GDV feature under logistic regression.}

SIGN consists of two parts. First, it extracts different types of adjacency matrices from a network.
SIGN specifically considers the traditional adjacency matrix, the Personalized PageRank-based adjacency matrix \cite{klicpera2019diffusion}, the triangle-induced adjacency matrix \cite{monti2018motifnet}, and their powers (see Supplementary Section \ref{supp:methods-eval} for which powers are used); these matrices are concatenated row-wise. Second, they are given as input into a neural network classifier. Mathematically, SIGN overall is equivalent to an ensemble of multiple one-layer-deep (i.e., shallow) GCN classifiers, which is why it is considered a graph learning approach.

DiffPool aims to perform graph classification.
However, unlike graphlet-based approaches and SIGN, which extract ``general purpose'' features of nodes/networks that can be used in any downstream machine learning task (label prediction in our study), DiffPool does not extract general purpose features. Instead, for each input network, given initial features for each node, DiffPool uses a GNN to aggregate the nodes' initial features into a summary hidden feature for the entire network. Then, given hidden features corresponding to the input networks, the GNN is trained to perform graph classification. Since the GNN is trained over many iterations, the hidden feature is dependent on the training data and can only be used as a part of DiffPool's GNN. 
When we say L1 DiffPool, we mean its GNN with the initial features chosen (Supplementary Section \ref{supp:methods-eval}), for graph classification using only level 1 networks.

\textcolor{black}{As SIGN and DiffPool are single-level graph learning approaches, we also combine them into an NoN graph learning approach. Given each level 2 node's feature extracted by SIGN, we concatenate it with the level 2 node's corresponding level 1 network's hidden feature computed by DiffPool's GNN. The GNN is then trained on these concatenated features to perform classification (note that any general purpose feature can be incorporated into DiffPool like this).
When we say L1 DiffPool + L2 SIGN, we mean entity label prediction using the process described above, incorporating SIGN's extracted feature into DiffPool's GNN.
So, we use three graph learning-based approaches: L1 DiffPool, L2 SIGN, and L1 DiffPool + L2 SIGN. }

We also combine L1 GDVM + L2 GDV or L1 GCM + L2 GDV with L1 DiffPool + L2 SIGN to test whether integrating information across the graph theoretic and graph learning domains improves upon either alone. \textcolor{black}{Graphlet-based features can be incorporated into DiffPool using the process described previously.} 

In total, we have \textcolor{black}{five} single-level approaches: L1 GDVM, L1 GCM, L2 GDV, L1 DiffPool, and L2 SIGN; and \textcolor{black}{five} NoN approaches: L1 GDVM + L2 GDV, L1 GCM + L2 GDV, L1 DiffPool + L2 SIGN, L1 GDVM + L2 GDV + L1 DiffPool + L2 SIGN, and L1 GCM + L2 GDV + L1 DiffPool + L2 SIGN.

Finally, note that we did test node2vec \cite{grover2016node2vec}, a prominent random walk-based embedding method, as a graph learning approach. node2vec extracts general purpose features like graphlets and SIGN, so we used it with logistic regression. However, DiffPool outperformed node2vec in level 1 graph classification, SIGN outperformed node2vec in level 2 node classification, and L1 DiffPool + L2 SIGN outperformed any combination involving node2vec in level 2 node classification for the entire NoN.

\subsection{Our integrative GCN approach} \label{supp:methods-gcn}

Here, we describe how we generalize GCNs to apply to NoNs. First, we summarize basic GCNs. Second, we discuss our extensions. 

The important unit of a GCN is the graph convolutional layer, which works as follows. For each node in some network $G = (V, E)$, the node's features are aggregated with its neighbors' features and then these aggregated features are propagated to the next layer of the neural network. More formally, summarized from \cite{kipf2016semi}, let $A$ be the adjacency matrix of $G$ and $H$ be a $|V| \times d$ matrix of G's nodes' features at the current layer (the $i^{th}$ row corresponds to the feature of the $i^{th}$ node). Then, forward propagation is carried out through

\begin{gather}
    f(H, A) = \sigma(\tilde{D}^{-\frac{1}{2}} \tilde{A} \tilde{D}^{-\frac{1}{2}} H W),
\end{gather}

\noindent where $W$ is the trainable weight matrix for the current layer, $\sigma$ is an activation function, $\tilde{A} = A + I$ is the adjacency matrix with self-loops added (so that mathematically, the aggregation actually includes each node's features along with its neighbors features) and $\tilde{D}$ is the diagonal node degree matrix used for normalizing the adjacency matrix. 

\begin{figure}[ht]
    \centering
    \includegraphics[width=0.4\textwidth]{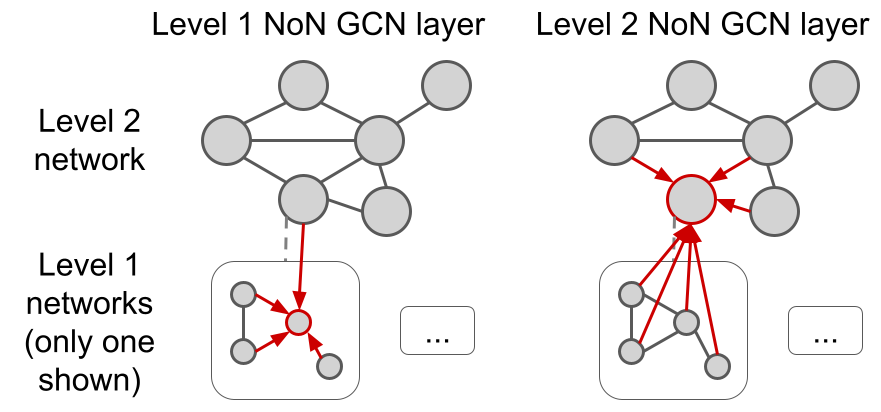}
    \caption{\label{suppfig:non-gcn-layer}   Illustration of the two part NoN-GCN layer. In the level 1 NoN-GCN layer, the level 1 node circled in red receives features from its neighbors in its level 1 network as well as features of the level 2 node its level 1 network corresponds to. This is done for every level 1 node in every level 1 network. In the level 2 NoN-GCN layer, the level 2 node circled in red received features from its neighbors in the level 2 network as well as features of each of the level 1 nodes in its level 1 network. This is done for every level 2 node in the level 2 network.}
\end{figure}

Essentially, graph convolutions allow each node to see information about its neighbors. So, what if we generalized graph convolutions to NoNs so that each node sees information not only about its neighbors (in the same level), but also about its corresponding network at a lower level or about the network it is a part of at a higher level? This would be in line with our intuition that the feature of a protein should contain information about how it interacts with other proteins (i.e., its topology in the level 2 network) and properties of the protein itself that allow for such interactions (topology of level 1 nodes in its level 1 network). So, we define a two part NoN-GCN layer consisting of one part that propagates level 2 nodes and another part that propagates level 1 nodes that attempts to do this (below and Fig. \ref{suppfig:non-gcn-layer}).

Let $\{G^{(2)} = (V^{(2)}, E^{(2)})$ and $\{G^{(1)}_1,...,G^{(1)}_{|V^{(2)}|}\}\}$ be an NoN. Let $A^{(2)}$ be the adjacency matrix of $G^{(2)}$. Let $\presuper{k}{H}{^{(2)}}$ be the $|V^{(2)}| \times d$ matrix of features for $G^{(2)}$ after the $k^{th}$ neural network layer; for $k=0$, this would correspond to the input feature matrix (for example, $G^{(2)}$'s GDVM or GCM). Let $\presuper{k}{h}{^{(2)}_i}$ be the feature vector of the $i^{th}$ node $v^{(2)}_i \in V^{(2)}$ (for example, $v^{(2)}_i$'s GDV). Let $A^{(1)}_i$ be the adjacency matrix of $v^{(2)}_i$'s level 1 network $G^{(1)}_i$. Let $\presuper{k}{H}{^{(1)}_i}$ be the $|V^{(1)}_i| \times d$ matrix of features for $G^{(1)}_i$ after the $k^{th}$ neural network layer. Let $\presuper{k}{h}{^{(1)}_{i_j}}$ be the feature vector of the $j^{th}$ node $v^{(1)}_{i_j}$ of $G^{(1)}_i$.

Propagation at level 2 works as follows. For each node $v^{(2)}_i$, for the feature matrix $\presuper{k}{H}{^{(1)}_i}$ of its corresponding level 1 network $G^{(1)}_i$, we take the average over all of $\presuper{k}{H}{^{(1)}_i}$'s rows to obtain a $1 \times d$ vector as a ``summary'' feature vector of $G^{(1)}_i$. Then, we combine these resulting vectors over all level 1 networks into a $|V^{(2)}| \times d$ matrix $\presuper{k}{\bar{H}}{^{(2)}}$, where each row corresponds to a level 1 network. In other words, $\presuper{k}{\bar{H}}{^{(2)}}$ can be thought of as the feature matrix of the level 2 network based on each node's level 1 network (whereas $\presuper{k}{H}{^{(2)}}$ is the feature matrix of the level 2 network based on each level 2 node). Then, our level 2 NoN-GCN layer forward propagation is carried out through

\begin{equation}
\begin{split}
    \presuper{k+1}{H}{^{(2)}} = \presuper{k+1}{f}{_{l2}}(\presuper{k}{H}{^{(2)}}, \presuper{k}{\bar{H}}{^{(2)}}, A^{(2)}) = \\ \sigma(\tilde{D^{(2)}}^{-\frac{1}{2}} \tilde{A^{(2)}} \tilde{D^{(2)}}^{-\frac{1}{2}} \\ (\presuper{k}{H}{^{(2)}} + \presuper{k}{\bar{H}}{^{(2)}}) \presuper{k+1}{W}{^{(2)}}),
\end{split}
\end{equation}

\noindent where $\presuper{k+1}{W}{^{(2)}}$ is the trainable weight matrix for the level 2 NoN-GCN layer, $\sigma$ is an activation function, $\tilde{A} = A + I$ is the adjacency matrix with self-loops added and $\tilde{D}$ is the diagonal node degree matrix used for normalizing the adjacency matrix. 

Propagation at level 1 works as follows. For each node $v^{(1)}_{i_j}$ in each level 1 network $G^{(1)}_i = (V^{(1)}_i, E^{(1)}_i)$, we sum its feature with all of its neighbors' features as well as the feature of $G^{(1)}_i$'s corresponding level 2 node. Mathematically, this corresponds to the following for each level 1 network. Let $\presuper{k}{\bar{H}}{^{(1)}_i}$ be a $|V^{(1)}_i| \times d$ matrix consisting of $\presuper{k}{h}{^{(2)}_i}$ repeated $|V^{(1)}_i|$ times. This can be thought of as the (naive) feature matrix of the level 1 network based on its corresponding level 2 node. Importantly $\presuper{k}{\bar{H}}{^{(1)}_i}$ has the same dimensions as $\presuper{k}{H}{^{(1)}_i}$. Then, level 1 NoN-GCN layer forward propagation is carried out for one level 1 network through

\begin{equation}
\begin{split}
    \presuper{k+1}{H}{^{(1)}_i} = \presuper{k+1}{f}{_{l1_i}}(\presuper{k}{\bar{H}}{^{(1)}_i}, \presuper{k}{H}{^{(1)}_i}, A^{(1)}_i) \\ = \sigma(\tilde{D^{(1)}}^{-\frac{1}{2}}_i \tilde{A^{(1)}}_i \tilde{D^{(1)}}^{-\frac{1}{2}}_i \\ (\presuper{k}{H}{^{(1)}_i} + \presuper{k}{\bar{H}}{^{(1)}_i}) \presuper{k+1}{W}{^{(1)}_i}),
\end{split}
\end{equation}

\noindent where $\presuper{k+1}{W}{^{(1)}_i}$ is the trainable weight matrix for the $i^{th}$ level 1 network for the current level 1 NoN-GCN layer, $\sigma$ is an activation function, $\tilde{A} = A + I$ is the adjacency matrix with self-loops added and $\tilde{D}$ is the diagonal node degree matrix used for normalizing the adjacency matrix. 

So, one full NoN-GCN layer takes as input 
\begin{equation}
    \begin{split}
        \presuper{k}{H}{^{(2)}}, \presuper{k}{\bar{H}}{^{(2)}},  A^{(2)}, \\  \{\presuper{k}{\bar{H}}{^{(1)}_1},\allowbreak...,\allowbreak \presuper{k}{\bar{H}}{^{(1)}_{|V^{(2)}|}}\}, \\ \{\presuper{k}{H}{^{(1)}_1},\allowbreak...,\allowbreak \presuper{k}{H}{^{(1)}_{|V^{(2)}|}}\}, \\ \textrm{ and } \{A^{(1)}_1,\allowbreak...,\allowbreak A^{(1)}_{|V^{(2)}|}\},
    \end{split}
\end{equation}
\noindent and returns $\presuper{k+1}{H}{^{(2)}}$ and $\{\presuper{k+1}{H}{^{(1)}_1},...,\presuper{k+1}{H}{^{(1)}_{|V^{(2)}|}}\}$. These outputs can then be fed as inputs (along with $\presuper{k+1}{\bar{H}}{^{(2)}}$, which can be constructed from $\{\presuper{k+1}{H}{^{(1)}_1},...,\presuper{k+1}{H}{^{(1)}_{|V^{(2)}|}}\}$, and each $\presuper{k+1}{\bar{H}}{^{(1)}_i}$, which can be constructed from its corresponding $\presuper{k+1}{h}{^{(2)}_i}$) into another NoN-GCN layer, thus allowing these layers to be chained.
        
%
We refer to a GCN approach using $\lambda$ layers as ``GCN-$\lambda$''.

Note that our implementation of the above is based on the \texttt{spektral} GNN library \cite{grattarola2020graph}.

\subsection{Evaluation} \label{supp:methods-eval}

For a given NoN $\{G^{(2)} = (V^{(2)}, E^{(2)})$ and $\{G^{(1)}_1,...,G^{(1)}_{|V^{(2)}|}\}\}$, its label set $Y = {y_1,...,y_c}$, and a function that maps level 2 nodes to their true labels $f_{true}: V^{(2)} \rightarrow Y$, the goal is to learn a predictive function $f_{pred}: V^{(2)} \rightarrow Y$. We do this by first splitting the data into three disjoint sets: training ($V^{(2)}_{tr}$), validation ($V^{(2)}_{val}$), and testing ($V^{(2)}_{te}$). Then, we train a classifier on the training set that aims to minimize the cross-entropy loss between $f_{true}(V^{(2)}_{tr})$ and $f_{pred}(V^{(2)}_{tr})$. We using $V^{(2)}_{val}$ to optimize hyperparameters and finally report the classifier's performance on $V^{(2)}_{te}$. Details are as follows.


Denote $Y = {y_1,...,y_c}$ to be the set of possible level 2 node labels (recall for synthetic NoNs, given $m$ random graph models, multiclass classification is done on $m \times m$ labels; for the real-world NoN, for each of the 131 ground truth datasets, binary classification is done on whether proteins have the corresponding label or not) and $f_{true}: V^{(2)} \rightarrow Y$ to be a function that maps level 2 nodes to their true labels. We split the set of level 2 nodes $V^{(2)}$ into three disjoint subsets as follows. $p\%$ of the data is randomly removed from $V^{(2)}$ and put into the training set $V^{(2)}_{tr}$. Half of the data remaining from $V^{(2)}$ is randomly removed and put into the validation set $V^{(2)}_{val}$. The remaining data is put into the testing set $V^{(2)}_{te}$. This results in three sets with size ratio $p$:$\frac{1-p}{2}$:$\frac{1-p}{2}$. Importantly, this splitting is done with the constraint that the distribution of node labels in each of the three sets matches the original label distribution of $V^{(2)}$ as closely as possible (i.e., stratified sampling). We train the classifier on $V^{(2)}_{tr}$, optimize hyperparameters using $V^{(2)}_{val}$, and report results on $V^{(2)}_{te}$. We repeat the random data splitting 3 different times and perform classification for each, reporting the average results over them. We do this 3 times so that 1) the effect of randomness from sampling reduced and 2) running the the approaches is still computationally feasible. For synthetic NoNs, we choose $p = 0.8$ (corresponding to a 8:1:1 data ratio), as this is a common split amount when data is not scarce. For real-world NoNs, we choose $p = 1/3$ (corresponding to a 1:1:1 data ratio). Because some of the ground truth sets have as few as 20 positive instances, larger values of $p$ would result in the validation/testing sets having too few of them.

Below, we describe classifier details. For graphlet-based approaches, we use each of L1 GDVM, L1 GCM, L2 GDV, L1 GDVM + L2 GDV, and L1 GCM + L2 GDV in logistic regression. For L2 SIGN, we use its features in own classifier. We refer to these as ``regular classification''. For approaches involving DiffPool, we run them as described in Supplementary Section \ref{supp:methods-existing}. We refer to these as ``DiffPool-based classification''. Finally, we refer to classification using NoN-GCNs as ``NoN-GCN-based'' classification.

For a given data split, for each feature we consider in regular classification, we train the corresponding classifier using the ADAM optimizer on $V^{(2)}_{tr}$. We test the following learning rates $\{0.1, 0.01, 0.001\}$ and choose the best one with respect to performance when predicting on $V^{(2)}_{val}$. Then, we use this best classifier to predict on $V^{(2)}_{te}$.

For a given data split, for DiffPool-based classification, we perform a grid search over the following hyperparameters: \texttt{hidden dimension: $\{32, 64, 128\}$} and \texttt{output dimension: $\{32, 64, 128\}$}. We choose the best combination with respect to performance when prediction on $V^{(2)}_{val}$ and use this best classifier to predict on $V^{(2)}_{te}$.

For a given data split, for NoN-GCN-based classification, we train a neural network that consists of two NoN-GCN layers, each followed by dropout layers, followed by a logistic regression classifier (i.e., one fully connected hidden layer). We specifically add this logistic regression classifier on the end of the neural network, rather than directly performing classification from the final NoN-GCN layer, to make the NoN-GCN-based classification as fairly comparable as possible to the regular classification. Note that for synthetic NoNs with two random graph models, we tested a version of the neural network with three NoN-GCN layers. However, because two NoN-GCN layers was as good as three for the majority of the evaluation tests, and because three layers took much more time to compute, we continued with two layers.
We also use the ADAM optimizer. We perform a grid search over the following hyperparameters: \texttt{learning rate: $\{0.1, 0.01, 0.001\}$}, \texttt{dropout: $\{0.0, 0.1, 0.2, 0.3, 0.4, 0.5\}$}, \texttt{hidden dimension: $\{128, 256, 512\}$} and choose the best combination with respect to performance when predicting on $V^{(2)}_{val}$. Then, we use this best classifier to predict on $V^{(2)}_{te}$.

Both DiffPool and our NoN-GCN require initial features. Ideally, they should use the same type of initial features so that they are as fairly comparable as possible. Our NoN-GCN has stricter limitations for what initial features can be used because it requires level 2 nodes' initial features to be in the same low dimensional space as level 1 nodes' initial features, otherwise it does not make sense to aggregate them. So, we determined initial features for our NoN-GCN first. We tested random features of lengths 128, 256, and 512, and nodes' GDVs (each index in the GDV corresponds to the number of times the node participates in that specific graphlet orbit; hence, GDVs are in the same low dimensional space) and found that GDVs were the best. So, we use nodes' GDV as initial features for our NoN-GCN. Thus, we also use nodes' GDVs as initial features into DiffPool.


For synthetic NoNs, we report classification accuracy (\# of correct predictions / total \# of entities) since class sizes are balanced. For the real-world NoNs, we report area under precision-recall (AUPR), precision@k, recall@k, and F-score@k, since class sizes are not balanced. Here @k refers to the corresponding measure when only considering the top k predictions. That is, for each approach, for each GO term, we rank each protein for which a prediction is made by the probability that it annotated by the given GO term, as determined by the approach's classifier. Then, we compute the corresponding measure on the top k items of the ranked list. To determine k, for each approach, for each GO term, we do the following. 
We choose the $k$ that maximizes the F-score@k where precision@k is greater than recall@k. We impose precision@k to be greater than recall@k because we believe that in the biomedical domain, precision is more important -- fewer but mostly correct predictions (e.g., 9 correct out of 10 made), which corresponds to high precision, is better than a greater number of mostly incorrect predictions (e.g., 300 correct out of 1,000 made), which corresponds to high recall, in terms of potential wet lab validation. By choosing $k$ in this way, we give each classifier the best case advantage. We report precision, recall, and F-score at this $k$.

We also test if each approach's performance is significantly better than random. That is, given an approach, for each measure, for each GO term, we use a one sample one-tailed t-test (recall that each approach is run 3 times, corresponding to 3 different training/validation/testing splits) to see if the approach's performance is significantly greater than the value expected by random. Then, for each measure, for each approach, we perform FDR correction over the 131 GO terms. For each measure, for each GO term, any approach with a corrected p-value $<$ 0.05 is considered significantly better than random for that GO term.

\section{Results} \label{supp:results}

\subsection{Synthetic NoNs} \label{supp:synthetic-results}

We expect an approach only using one level to reach an accuracy of $\frac{\textrm{\# of models}}{\textrm{\# of labels}}$, i.e., 0.5. To see why, consider the following example using the L1 GDVM approach. Here, there are four possible labels corresponding to the four possible combinations of random graph models at each level: GEO-GEO, GEO-SF, SF-GEO, SF-SF. Since L1 GDVM uses level 1 information, it will be able to distinguish between GEO and SF level 1 networks but not between level 2 nodes with GEO- and SF- topology. So, L1 GDVM will only have enough information to predict $\frac{2}{4} = 0.5$ of the labels correctly.

\begin{figure}
     \centering
     \subfloat[]{\includegraphics[width=0.4\textwidth]{l1l2_config_0-05_0-0_acc.png}}
     \subfloat[]{\includegraphics[width=0.4\textwidth]{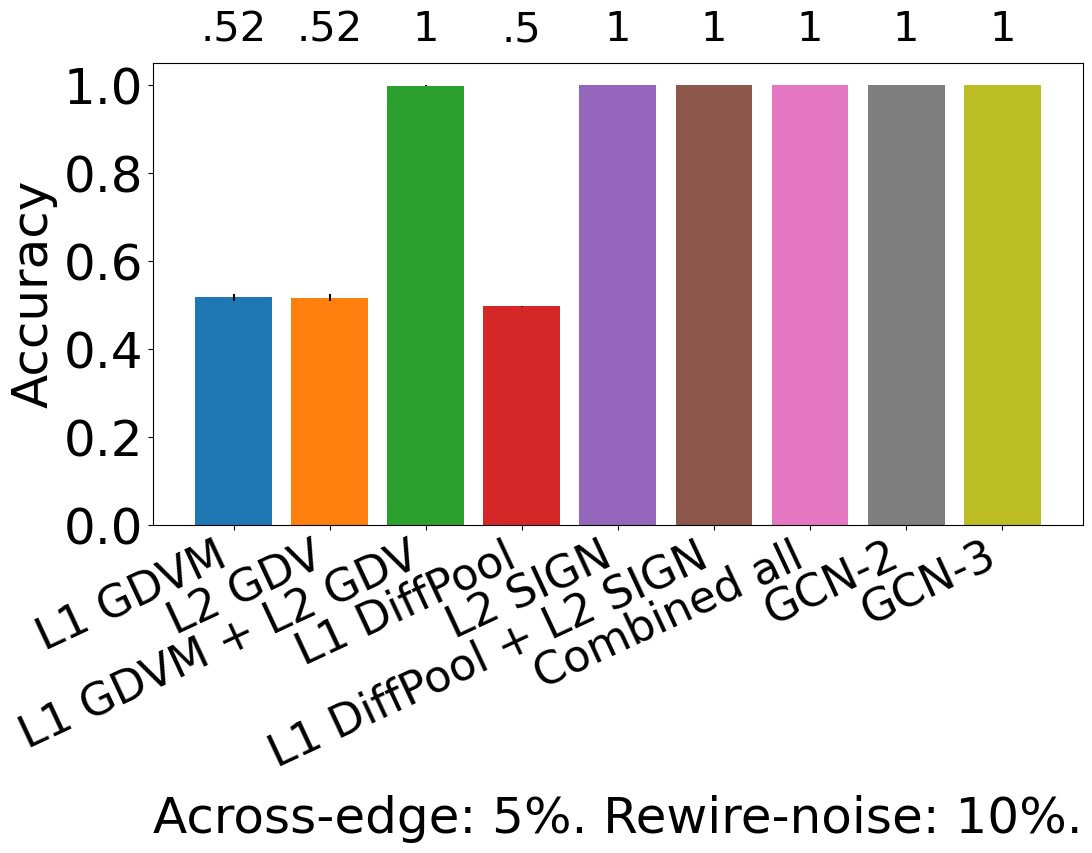}}
     
     \subfloat[]{\includegraphics[width=0.4\textwidth]{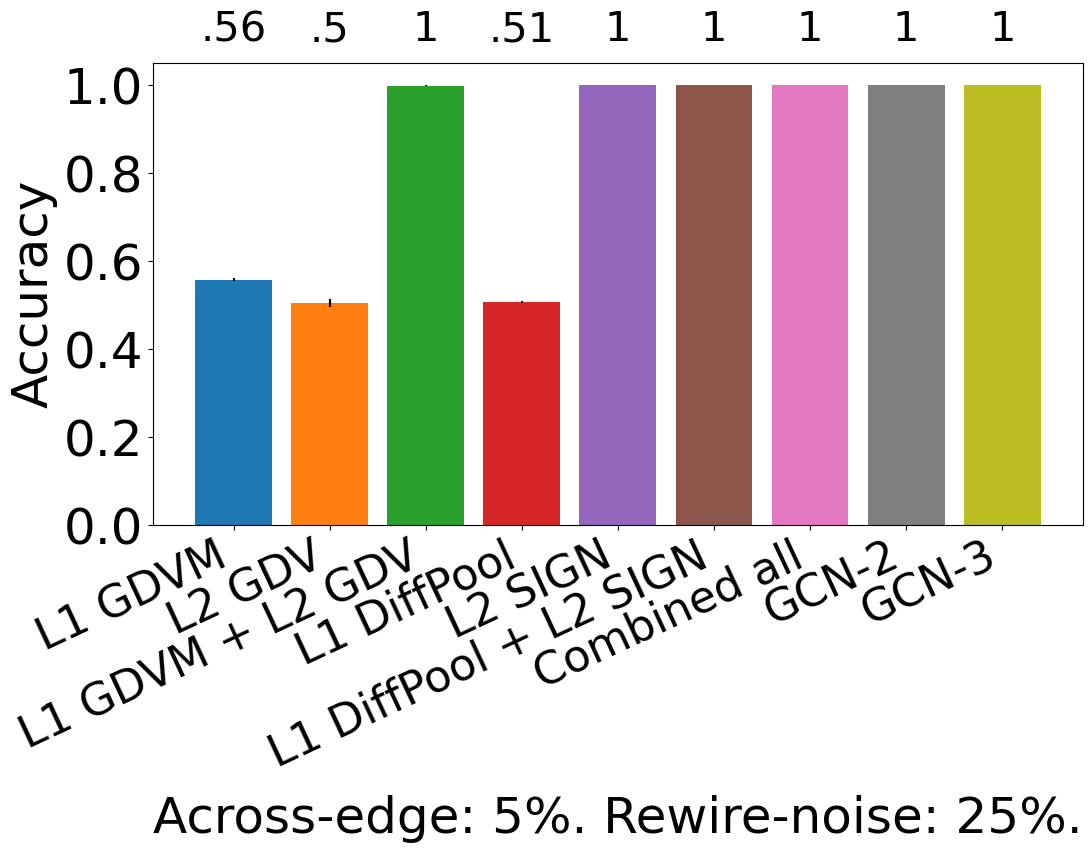}}
     \subfloat[]{\includegraphics[width=0.4\textwidth]{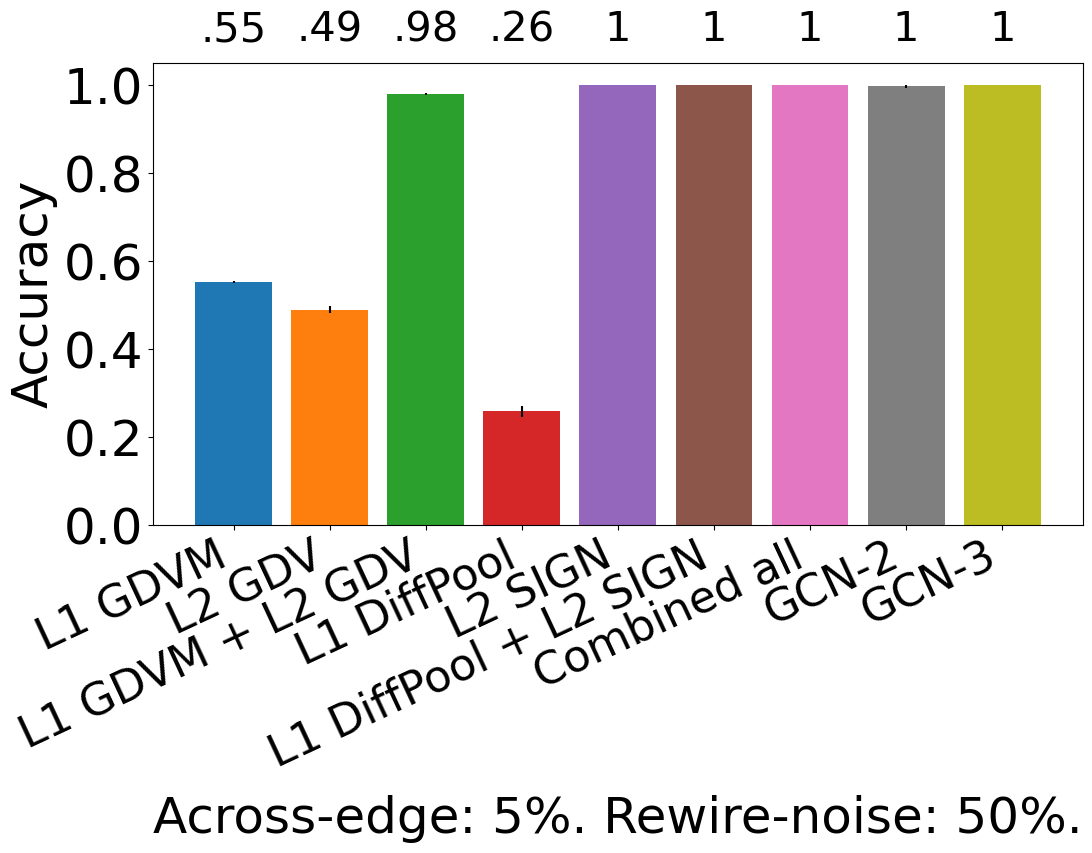}}
     
     \subfloat[]{\includegraphics[width=0.4\textwidth]{l1l2_config_0-05_0-75_acc.png}}
     \subfloat[]{\includegraphics[width=0.4\textwidth]{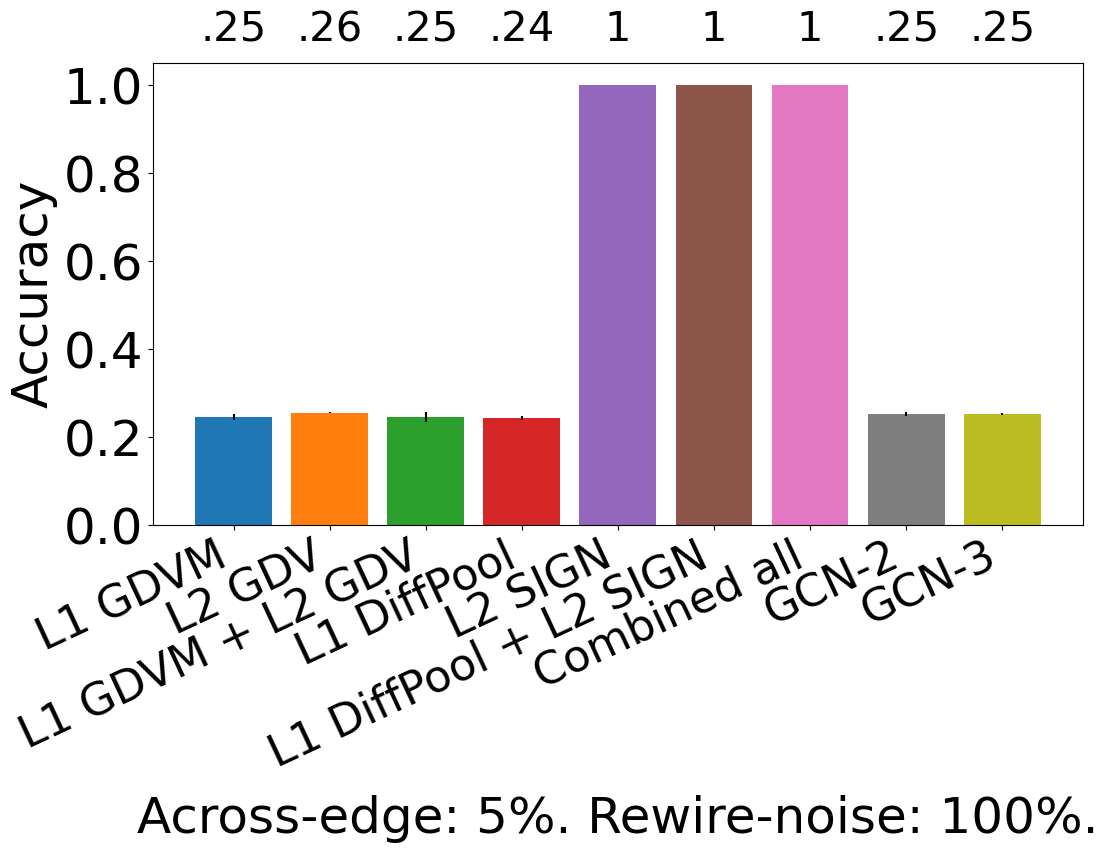}}
     
     
        
    \caption{Comparison of the nine relevant approaches in the task of entity label prediction for synthetic NoNs with 5\% across-edge amount and the following rewire-noise amounts: \textbf{(a)} 0\%, \textbf{(b)} 10\%, \textbf{(c)} 25\%, \textbf{(d)} 50\%, \textbf{(e)} 75\%, and \textbf{(f)} 100\%. ``Combined all'' refers to L1 GDVM + L2 GDV + L1 DiffPool + L2 SIGN. Raw prediction accuracies are shown above. ``Combined all'' refers to L1 GDVM + L2 GDV + L1 DiffPool + L2 SIGN. Accuracy is shown above the bars. 
    }
    \label{suppfig:synthetic-5-across-edge}
\end{figure}

\begin{figure}
     \centering
     \subfloat[]{\includegraphics[width=0.4\textwidth]{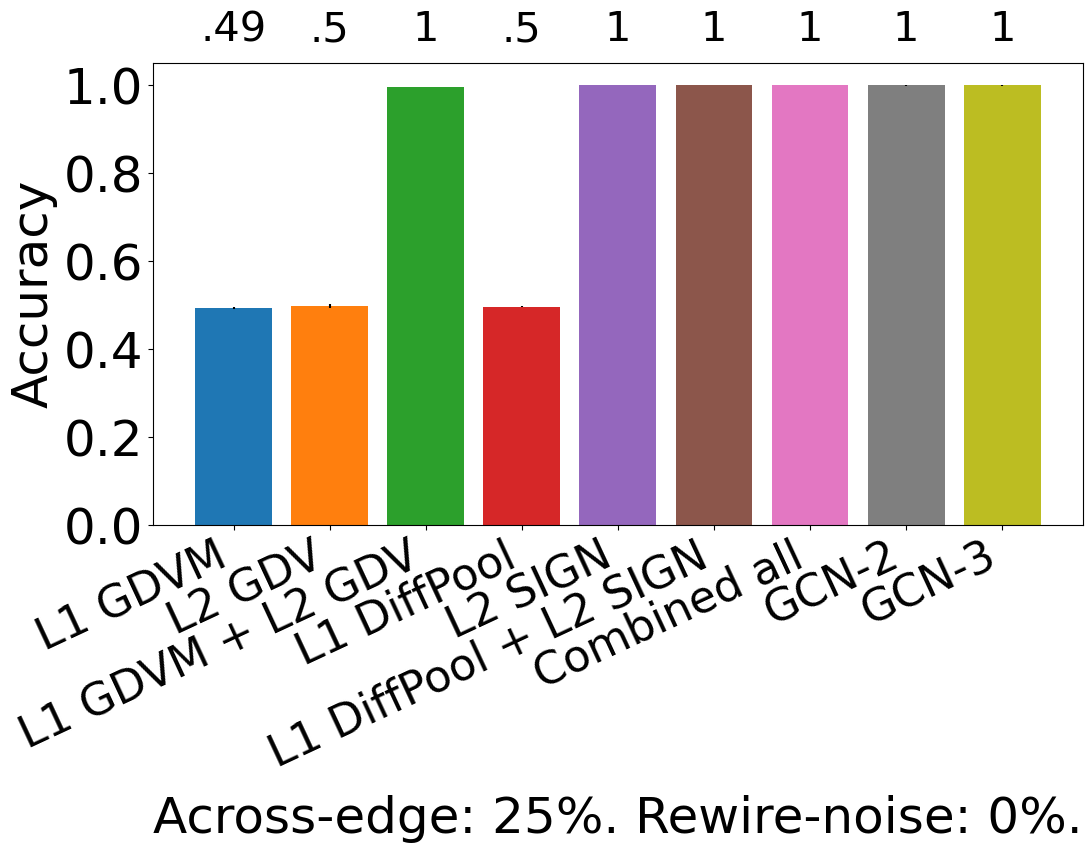}}
     \subfloat[]{\includegraphics[width=0.4\textwidth]{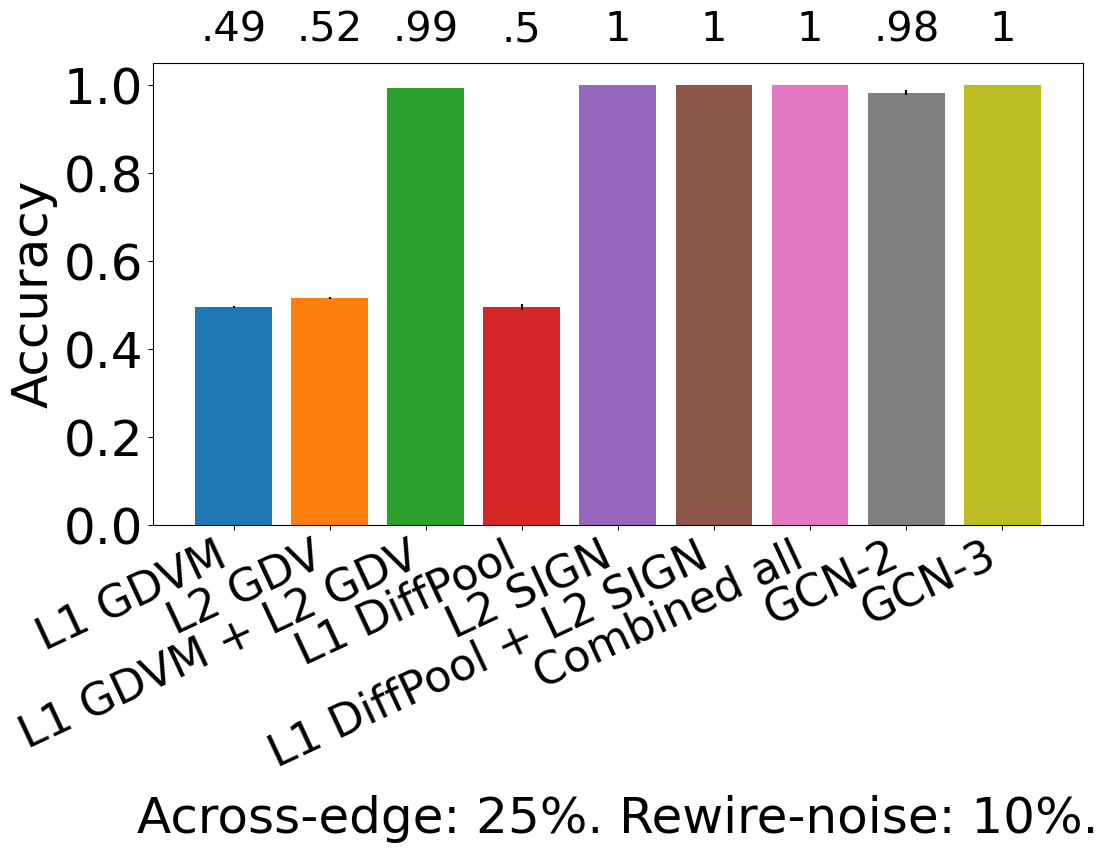}}
     
     \subfloat[]{\includegraphics[width=0.4\textwidth]{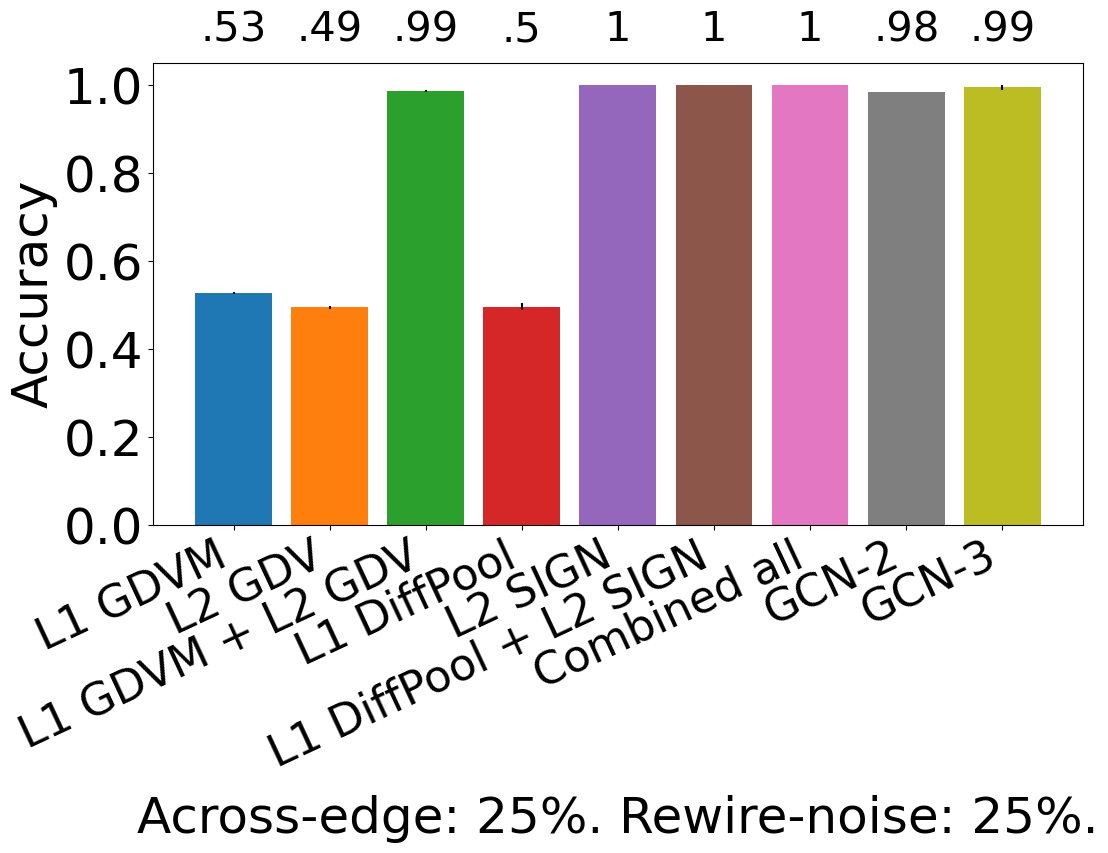}}
     \subfloat[]{\includegraphics[width=0.4\textwidth]{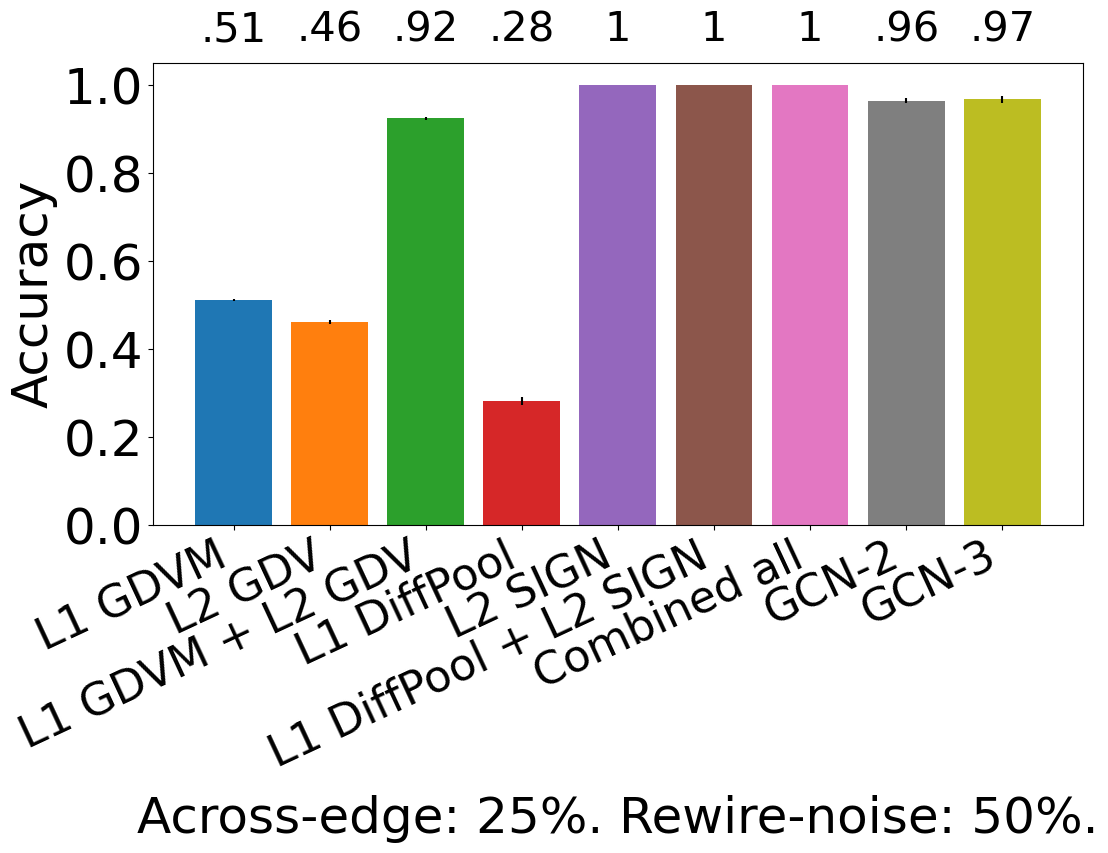}}
     
     \subfloat[]{\includegraphics[width=0.4\textwidth]{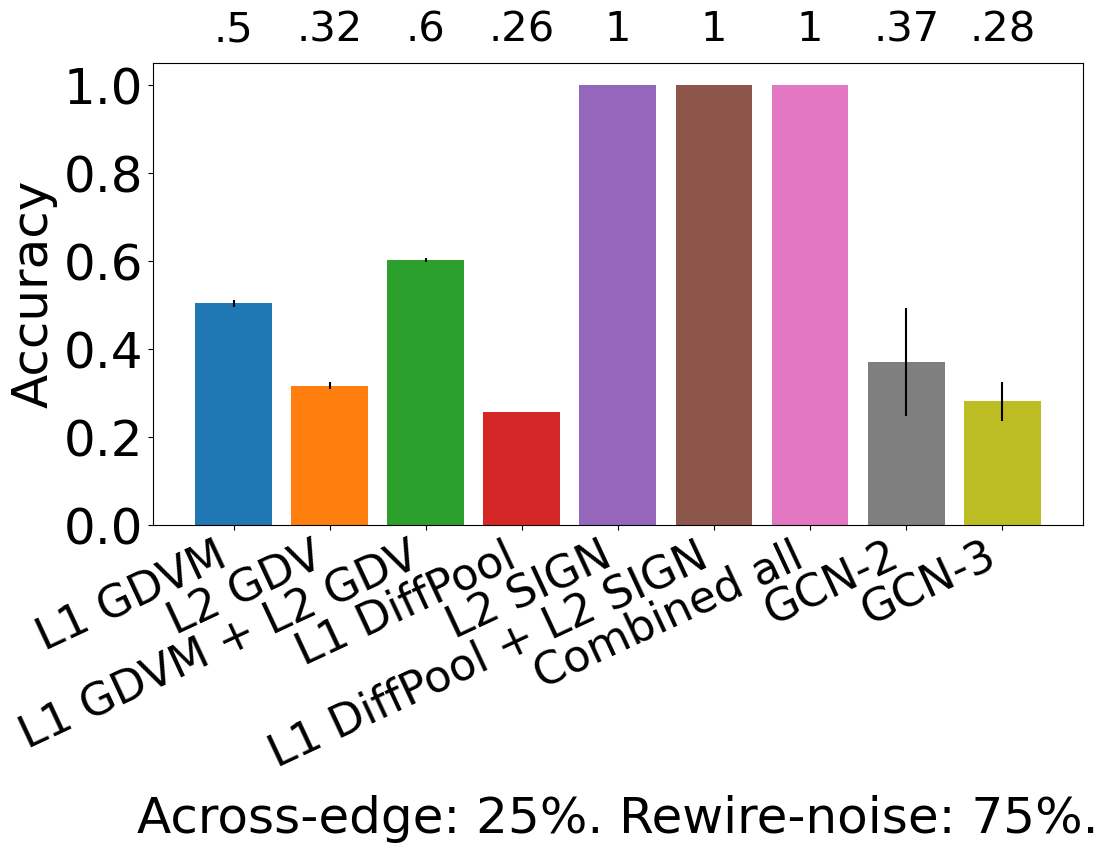}}
     \subfloat[]{\includegraphics[width=0.4\textwidth]{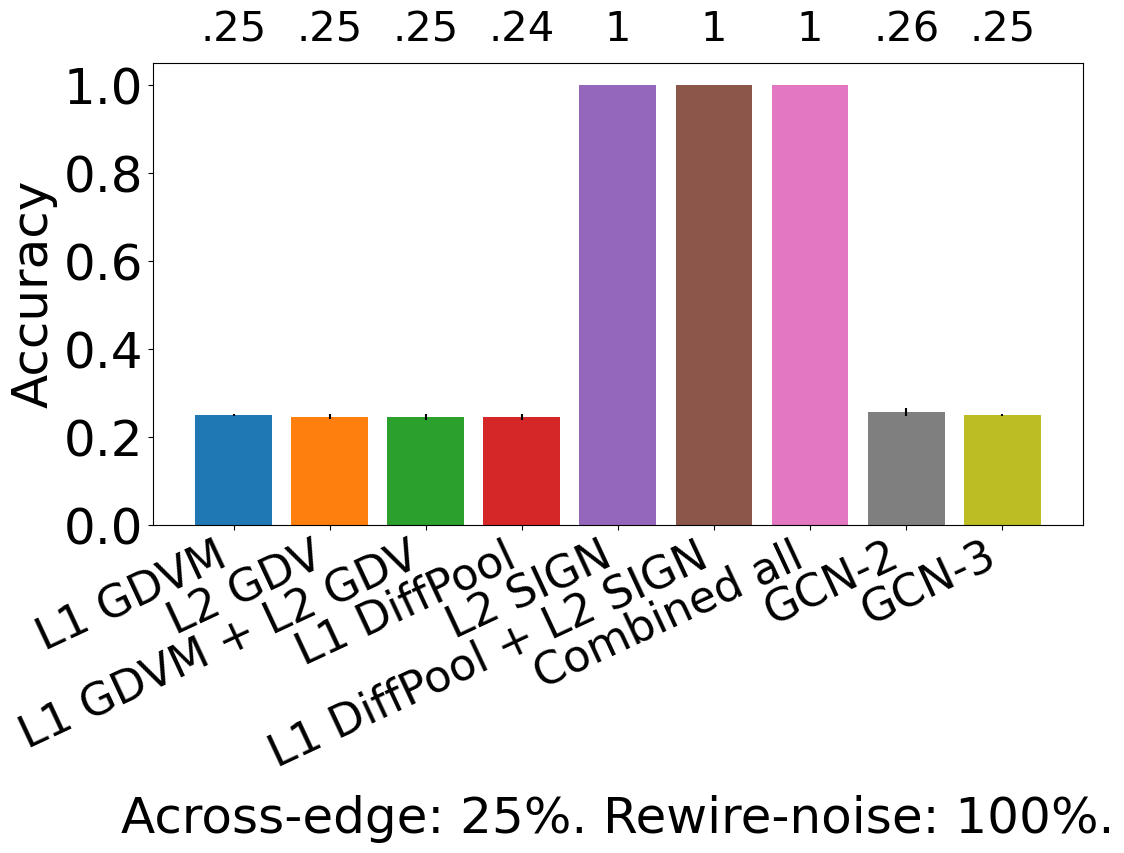}}
     
     
        
    \caption{Comparison of the nine relevant approaches in the task of entity label prediction for synthetic NoNs with 25\% across-edge amount and the following rewire-noise amounts: \textbf{(a)} 0\%, \textbf{(b)} 10\%, \textbf{(c)} 25\%, \textbf{(d)} 50\%, \textbf{(e)} 75\%, and \textbf{(f)} 100\%. ``Combined all'' refers to L1 GDVM + L2 GDV + L1 DiffPool + L2 SIGN. Raw prediction accuracies are shown above. ``Combined all'' refers to L1 GDVM + L2 GDV + L1 DiffPool + L2 SIGN. Accuracy is shown above the bars.}
    \label{suppfig:synthetic-25-across-edge}
\end{figure}

\begin{figure}
     \centering
     \subfloat[]{\includegraphics[width=0.4\textwidth]{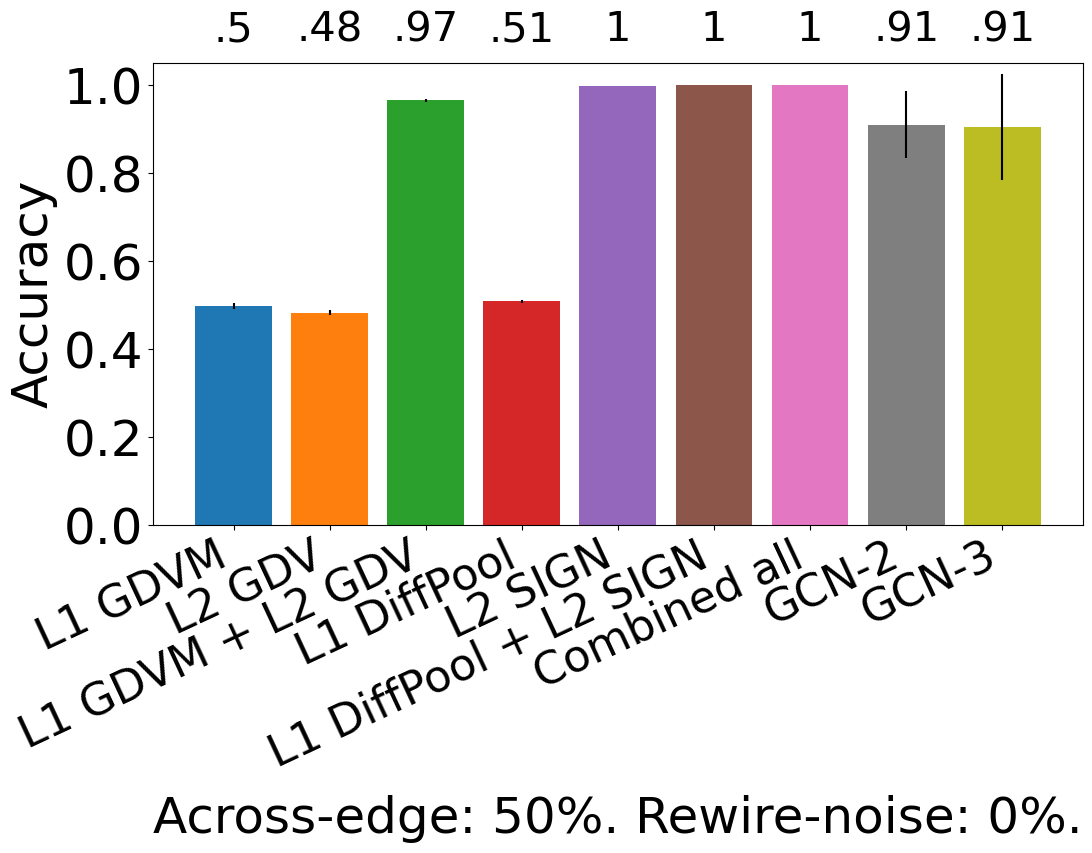}}
     \subfloat[]{\includegraphics[width=0.4\textwidth]{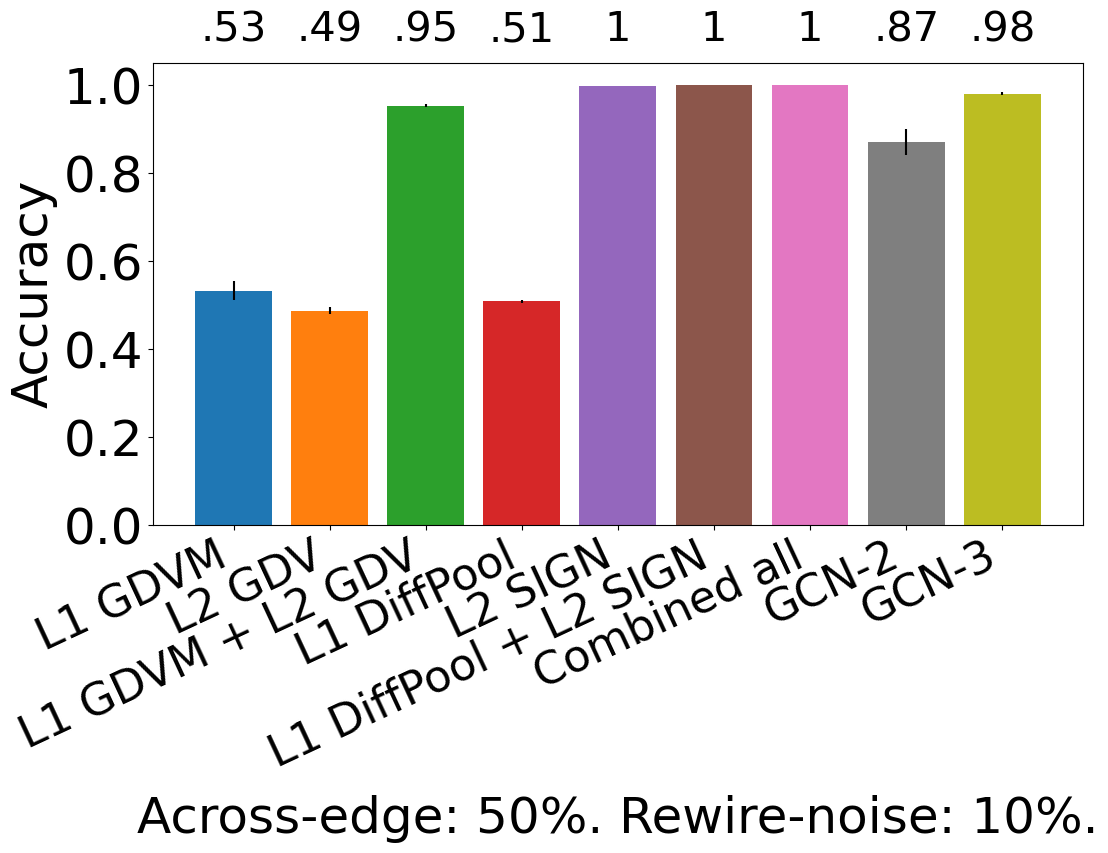}}
     
     \subfloat[]{\includegraphics[width=0.4\textwidth]{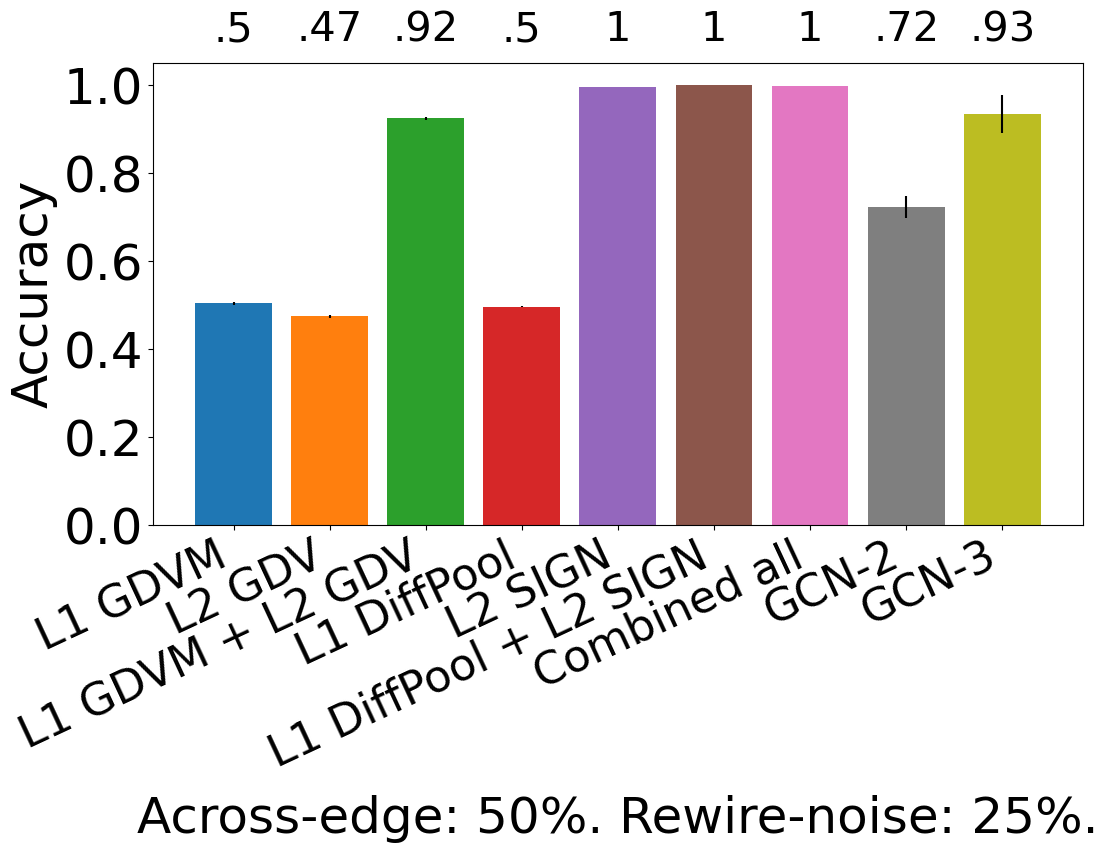}}
     \subfloat[]{\includegraphics[width=0.4\textwidth]{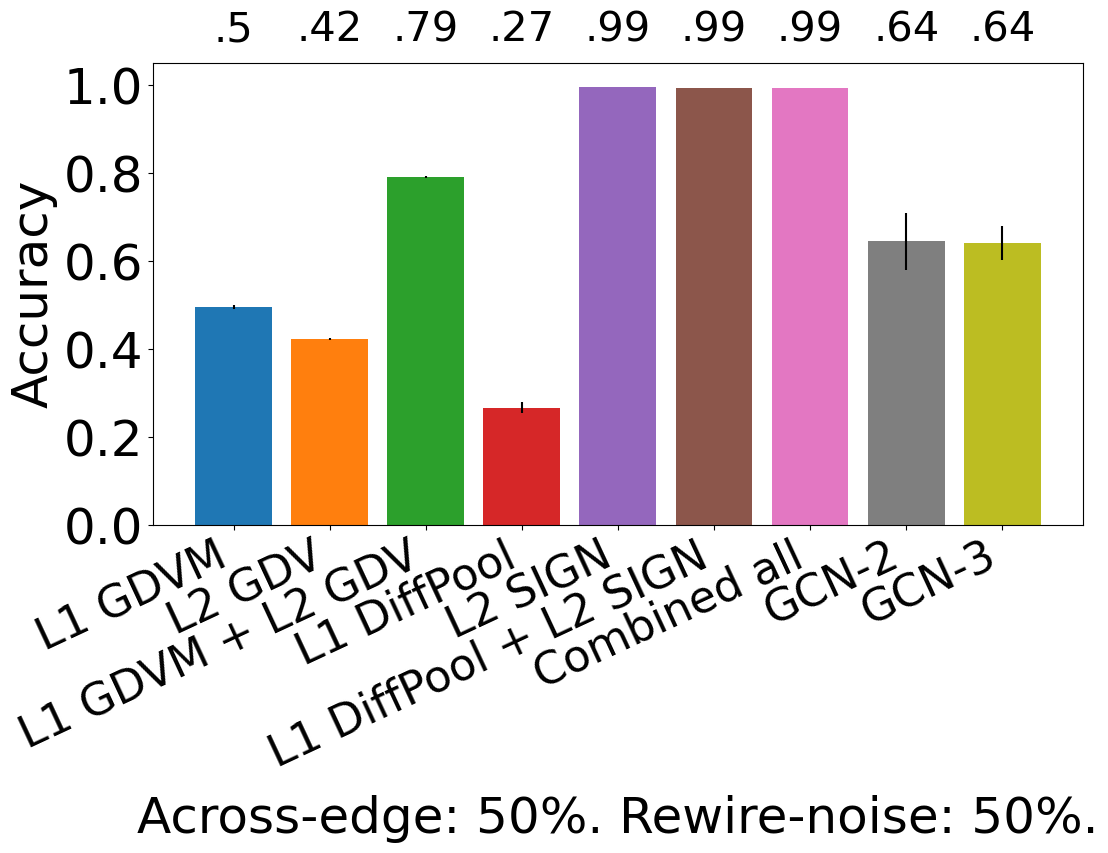}}
     
     \subfloat[]{\includegraphics[width=0.4\textwidth]{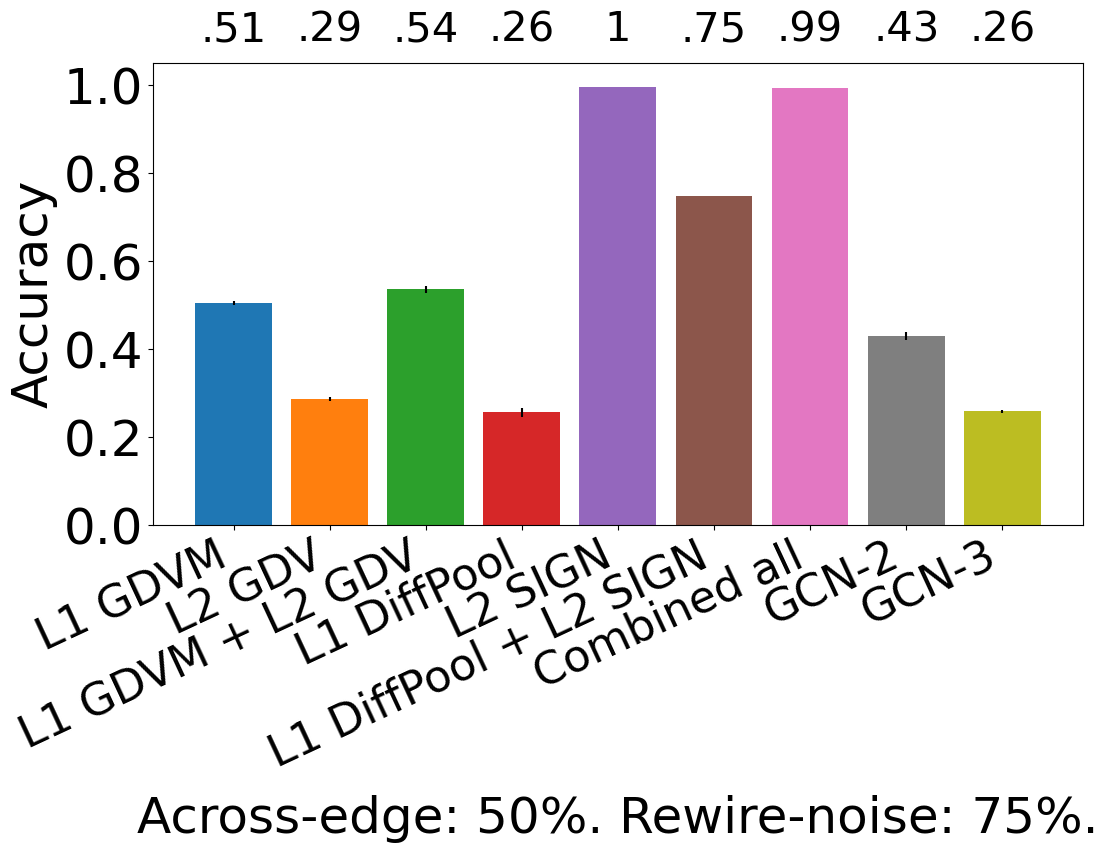}}
     \subfloat[]{\includegraphics[width=0.4\textwidth]{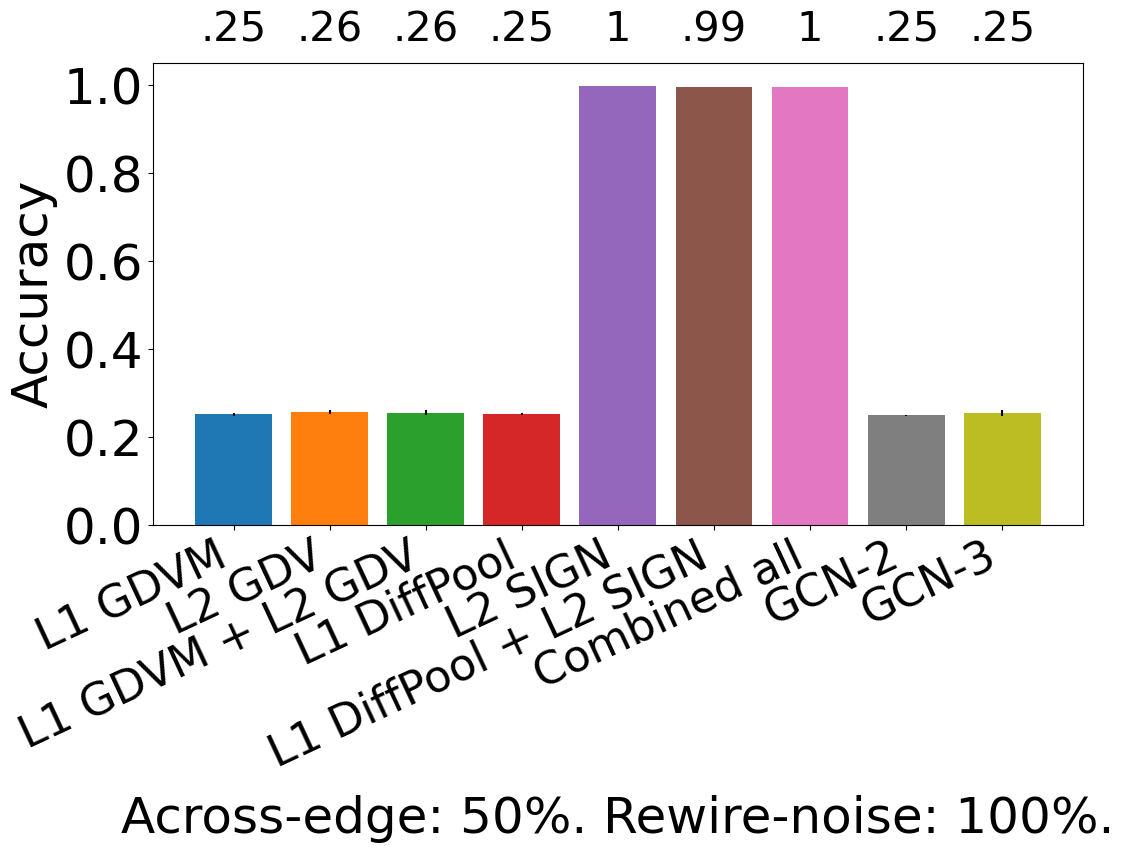}}
     
     
     
        
    \caption{Comparison of the nine relevant approaches in the task of entity label prediction for synthetic NoNs with 50\% across-edge amount and the following rewire-noise amounts: \textbf{(a)} 0\%, \textbf{(b)} 10\%, \textbf{(c)} 25\%, \textbf{(d)} 50\%, \textbf{(e)} 75\%, and \textbf{(f)} 100\%. ``Combined all'' refers to L1 GDVM + L2 GDV + L1 DiffPool + L2 SIGN. Raw prediction accuracies are shown above. ``Combined all'' refers to L1 GDVM + L2 GDV + L1 DiffPool + L2 SIGN. Accuracy is shown above the bars.}
    \label{suppfig:synthetic-50-across-edge}
\end{figure}

\begin{figure}
     \centering
     \subfloat[]{\includegraphics[width=0.4\textwidth]{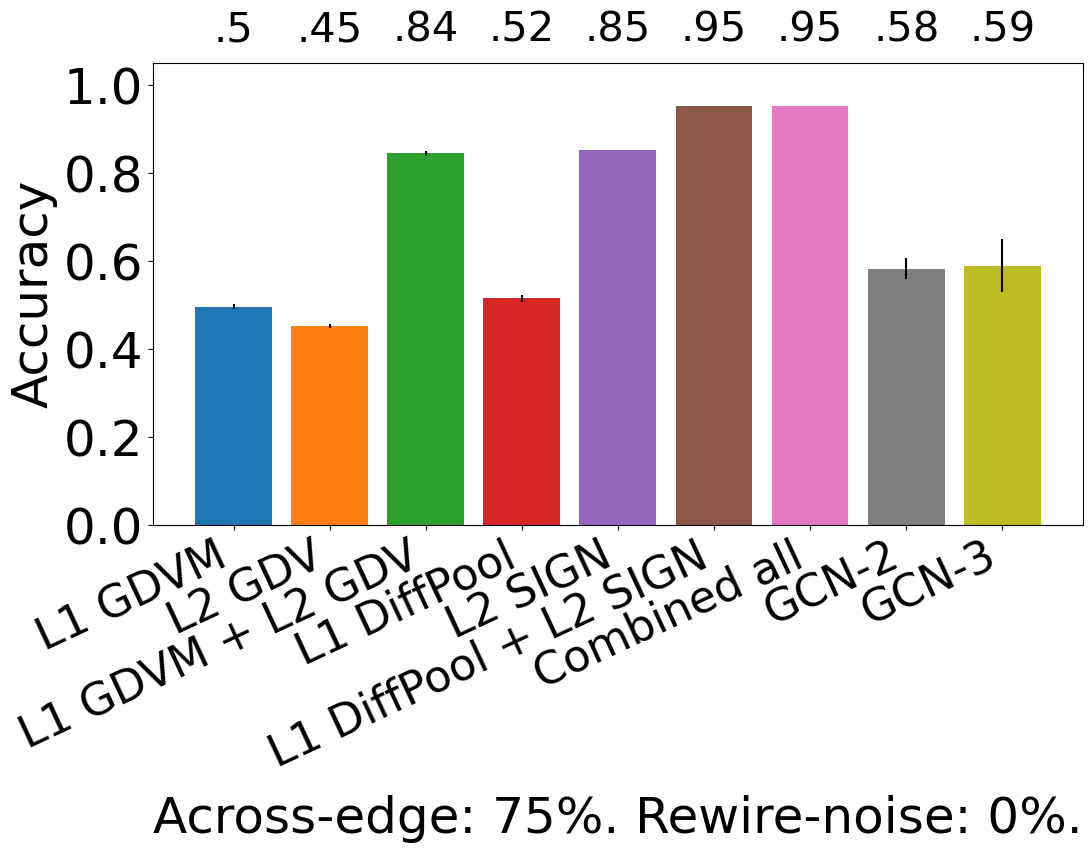}}
     \subfloat[]{\includegraphics[width=0.4\textwidth]{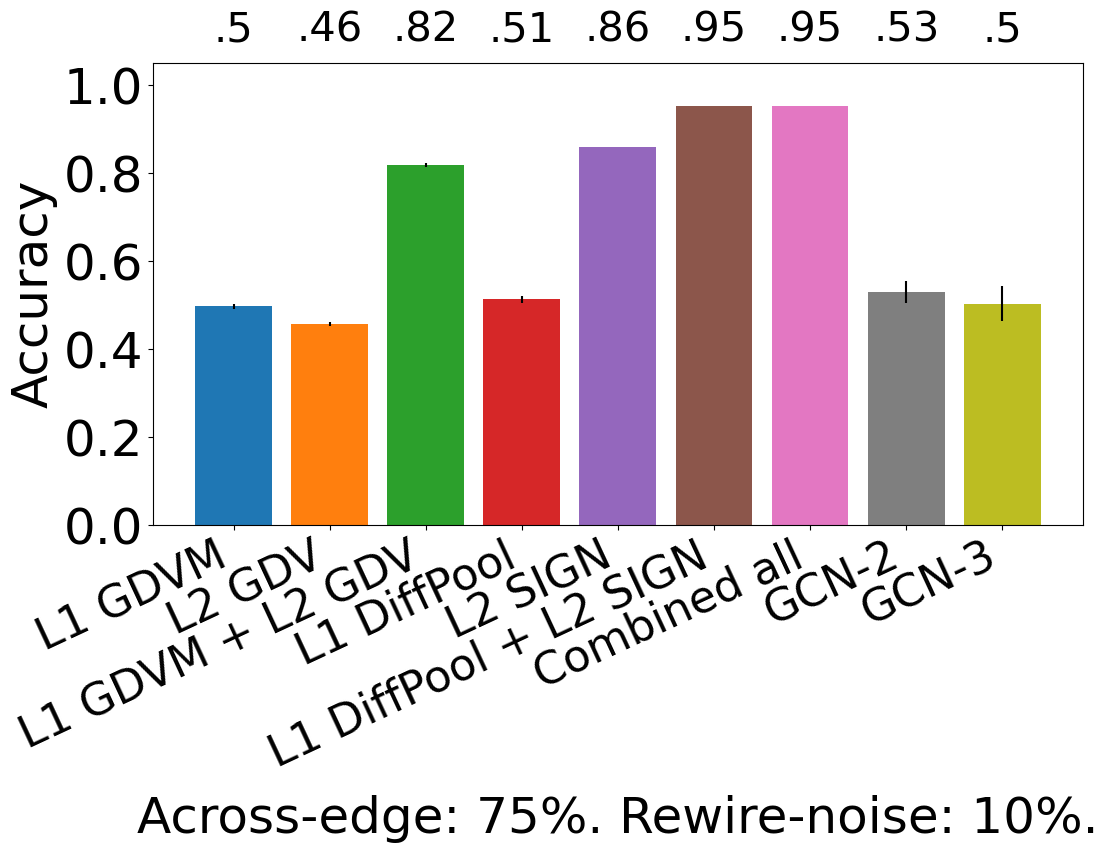}}
     
     \subfloat[]{\includegraphics[width=0.4\textwidth]{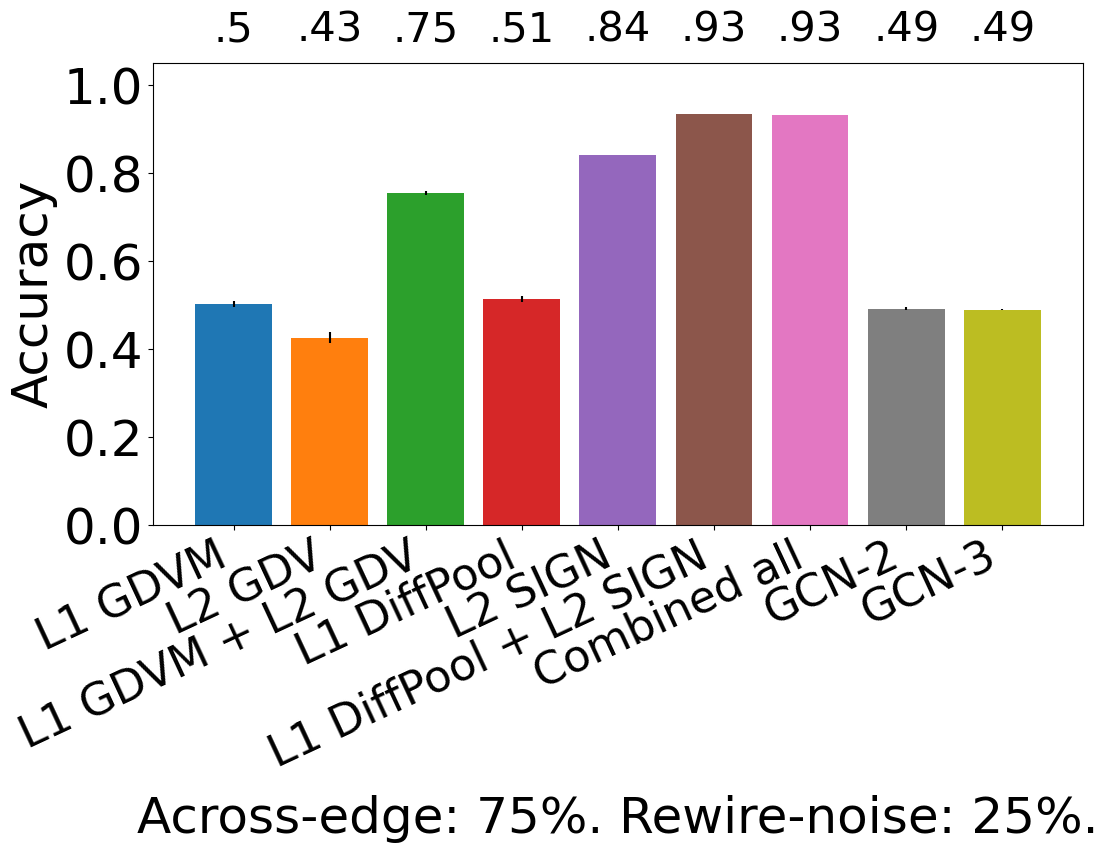}}
     \subfloat[]{\includegraphics[width=0.4\textwidth]{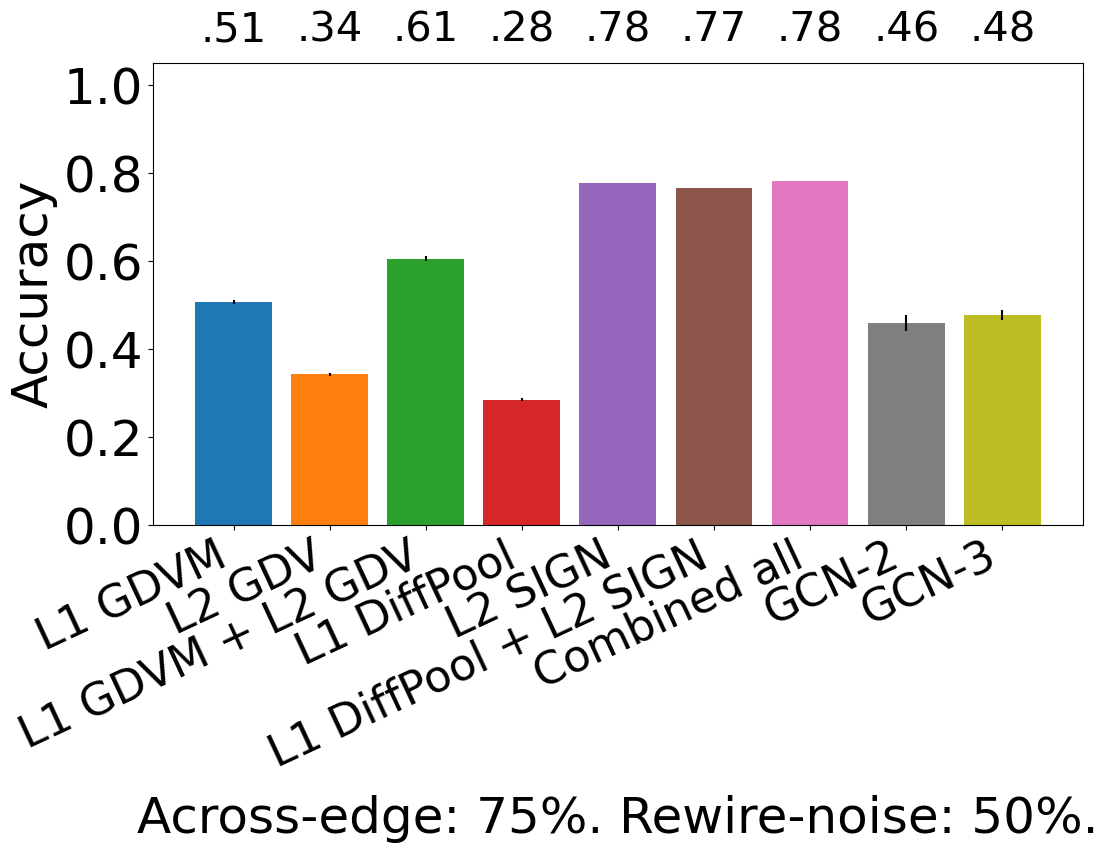}}
     
     \subfloat[]{\includegraphics[width=0.4\textwidth]{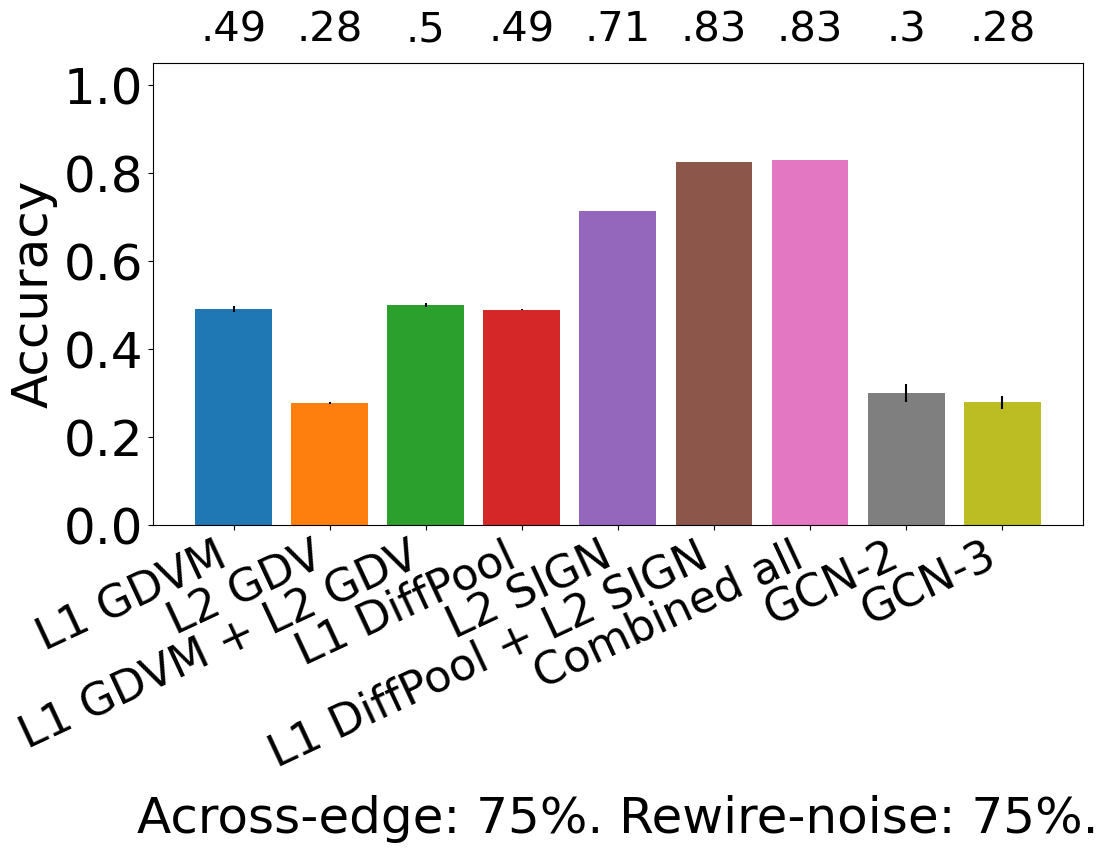}}
     \subfloat[]{\includegraphics[width=0.4\textwidth]{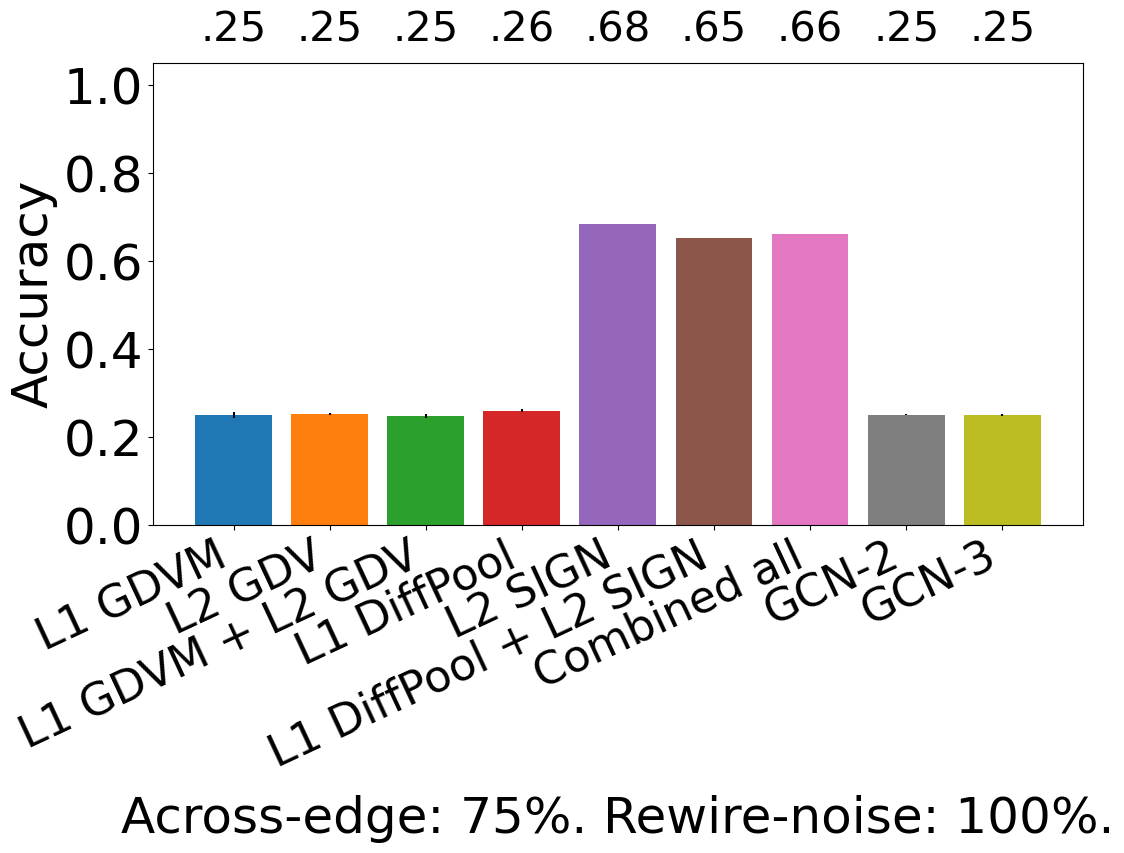}}
     
     
     
        
    \caption{Comparison of the nine relevant approaches in the task of entity label prediction for synthetic NoNs with 75\% across-edge amount and the following rewire-noise amounts: \textbf{(a)} 0\%, \textbf{(b)} 10\%, \textbf{(c)} 25\%, \textbf{(d)} 50\%, \textbf{(e)} 75\%, and \textbf{(f)} 100\%. ``Combined all'' refers to L1 GDVM + L2 GDV + L1 DiffPool + L2 SIGN. Raw prediction accuracies are shown above. ``Combined all'' refers to L1 GDVM + L2 GDV + L1 DiffPool + L2 SIGN. Accuracy is shown above the bars.}
    \label{suppfig:synthetic-75-across-edge}
\end{figure}

\begin{figure}
     \centering
     \subfloat[]{\includegraphics[width=0.4\textwidth]{l1l2_config_0-95_0-0_acc.png}}
     \subfloat[]{\includegraphics[width=0.4\textwidth]{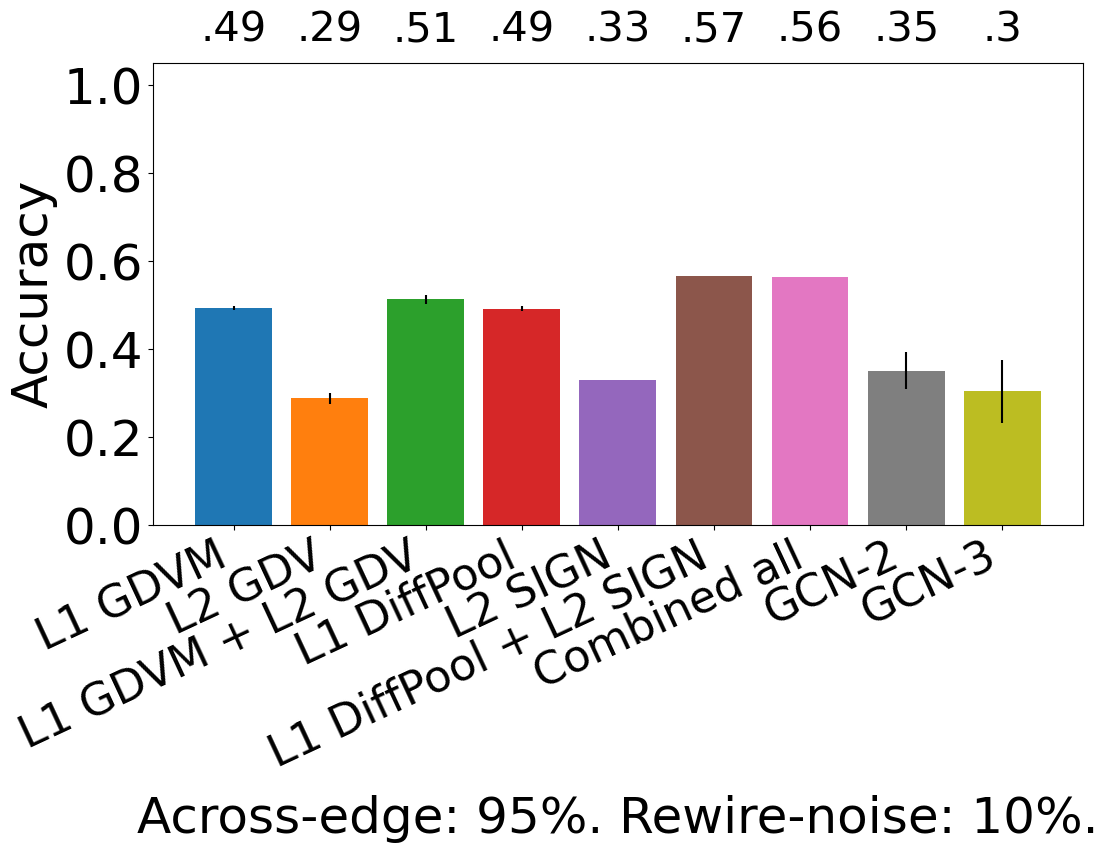}}
     
     \subfloat[]{\includegraphics[width=0.4\textwidth]{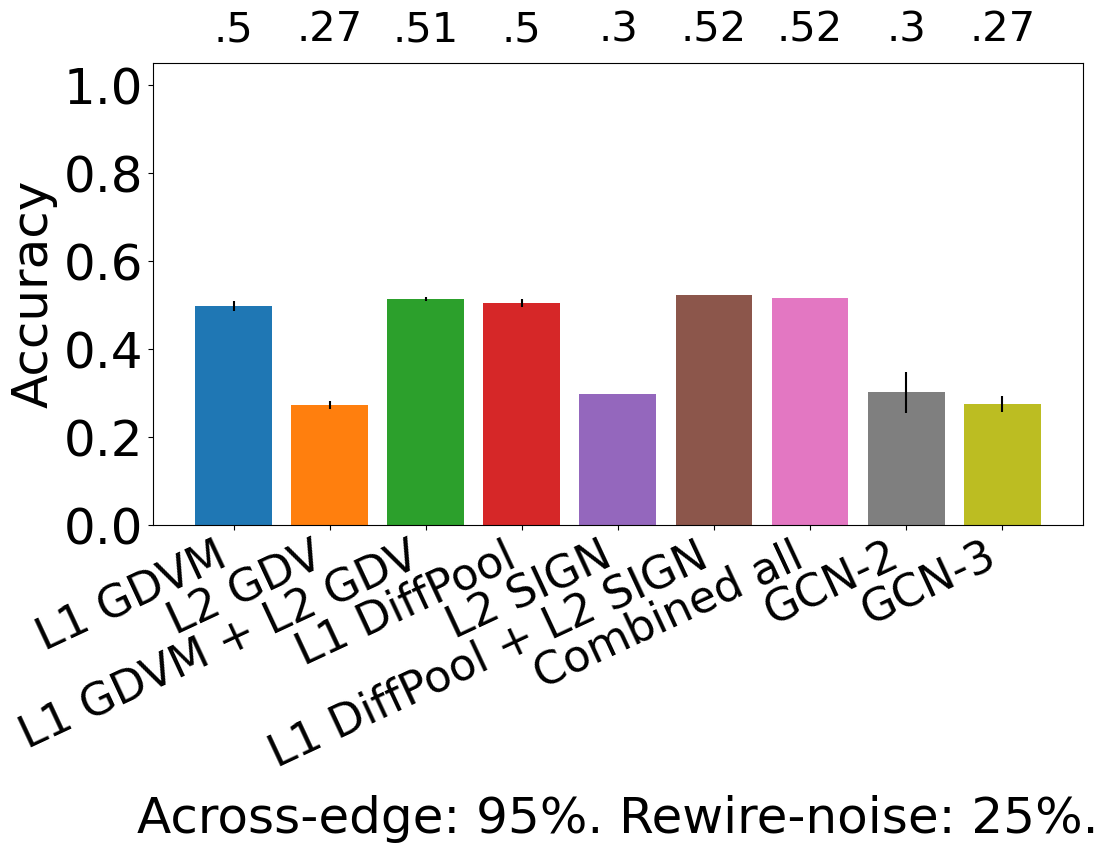}}
     \subfloat[]{\includegraphics[width=0.4\textwidth]{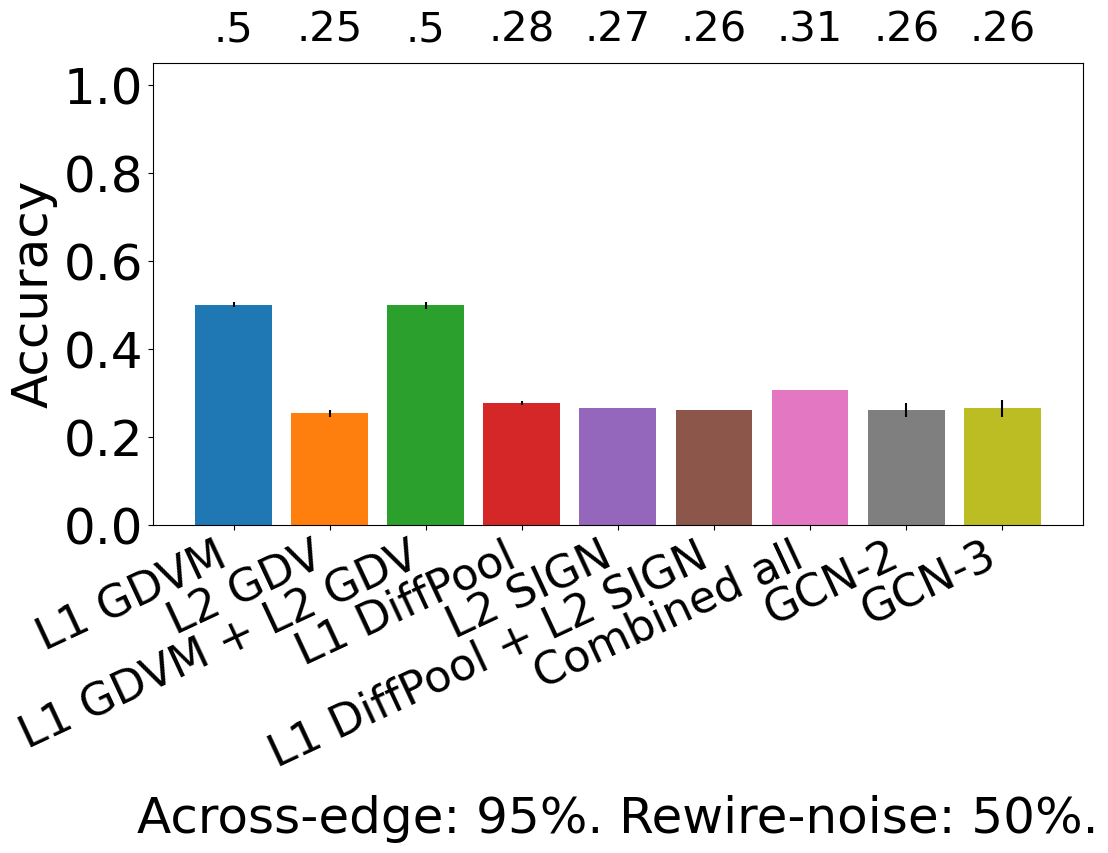}}
     
     \subfloat[]{\includegraphics[width=0.4\textwidth]{l1l2_config_0-95_0-75_acc.png}}
     \subfloat[]{\includegraphics[width=0.4\textwidth]{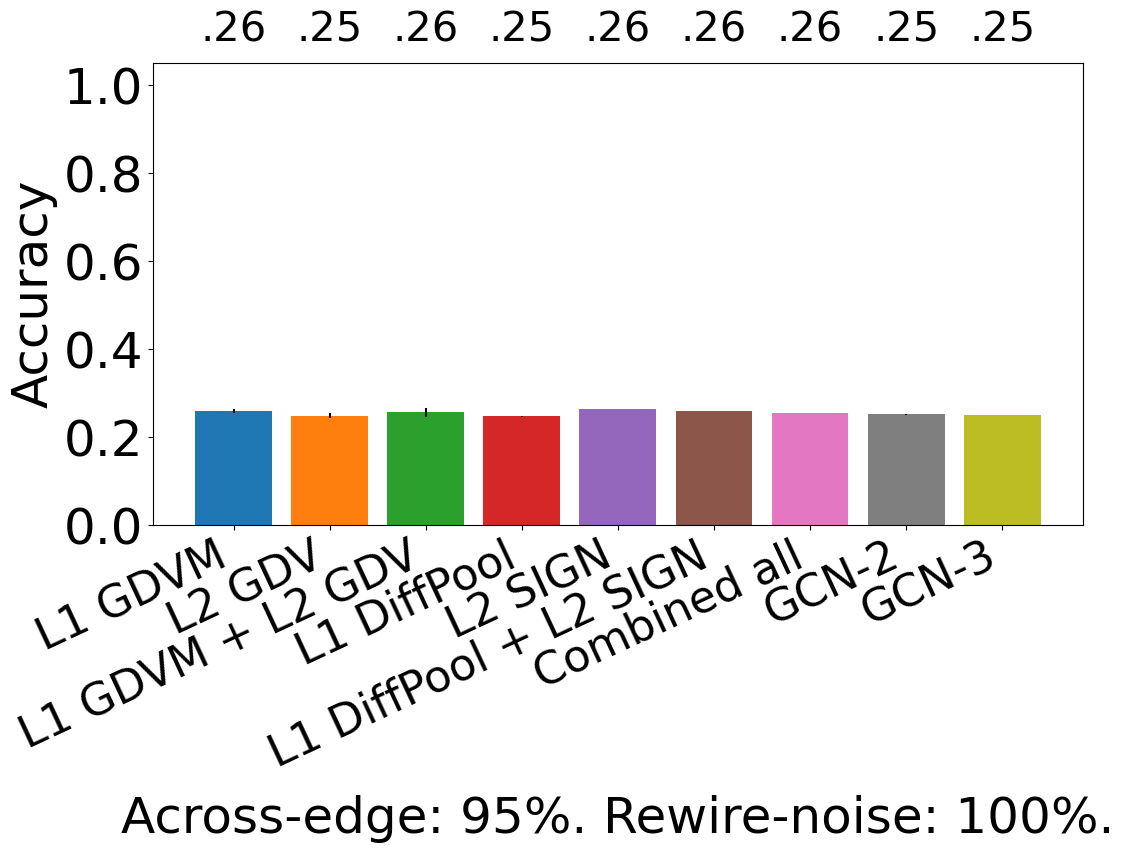}}
     
     
        
    \caption{Comparison of the nine relevant approaches in the task of entity label prediction for synthetic NoNs with 95\% across-edge amount and the following rewire-noise amounts: \textbf{(a)} 0\%, \textbf{(b)} 10\%, \textbf{(c)} 25\%, \textbf{(d)} 50\%, \textbf{(e)} 75\%, and \textbf{(f)} 100\%. ``Combined all'' refers to L1 GDVM + L2 GDV + L1 DiffPool + L2 SIGN. Raw prediction accuracies are shown above. ``Combined all'' refers to L1 GDVM + L2 GDV + L1 DiffPool + L2 SIGN. Accuracy is shown above the bars.}
    \label{suppfig:synthetic-95-across-edge}
\end{figure}

\subsection{Biological NoN} \label{supp:results-figs}

\begin{figure}
     \centering
     \subfloat[]{\includegraphics[width=0.45\textwidth]{overall_rankingsaupr.pval.bh0.01.png}}
     \subfloat[]{\includegraphics[width=0.45\textwidth]{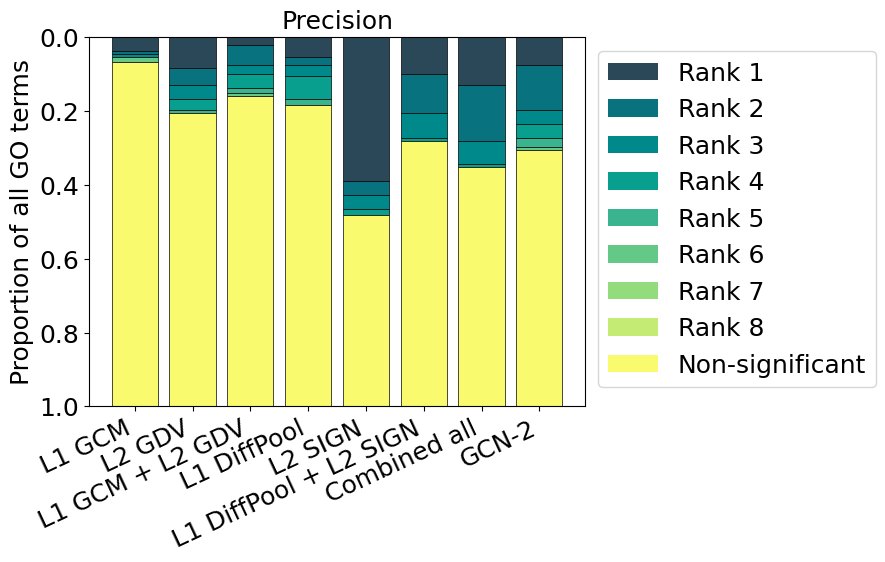}}
     
     \subfloat[]{\includegraphics[width=0.45\textwidth]{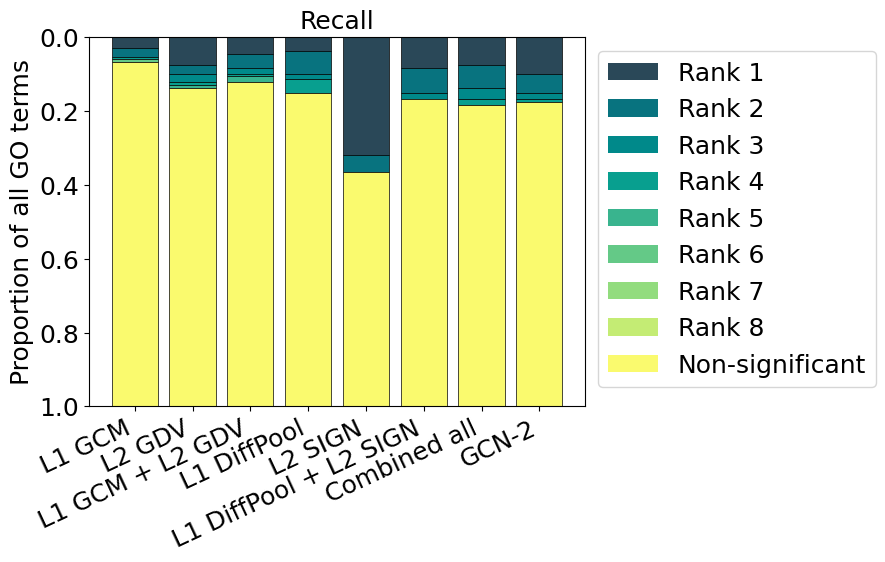}}
     \subfloat[]{\includegraphics[width=0.45\textwidth]{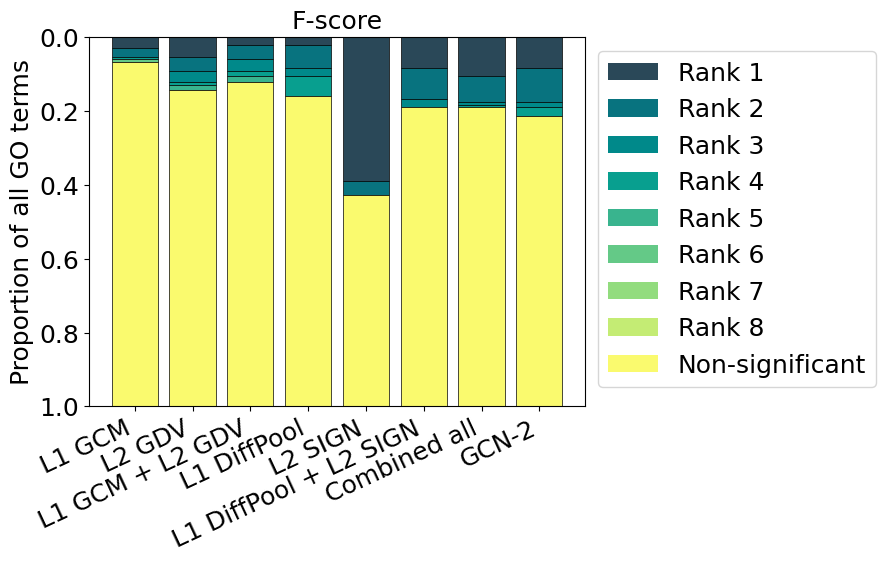}}
     
     

    \caption{Summarized results of the eight relevant approaches in the task of protein functional prediction for evaluation measures \textbf{(a)} AUPR, \textbf{(b)} precision, \textbf{(a)} recall, and \textbf{(a)} F-score. For each GO term (out of the 131 total), we rank the eight approaches' classification performances from best (rank 1) to worst (rank 8). If an approach's performance is not significantly better than expected by random we deem it ``non-significant'' instead. Then for each approach, we calculate the proportion of times it achieves each rank. ``Combined all'' refers to L1 GDVM + L2 GDV + L1 DiffPool + L2 SIGN. }
    \label{suppfig:bio-overall-rankings}
\end{figure}

\begin{figure*}[ht]
     \centering
     \subfloat[]{\includegraphics[width=0.24\textwidth]{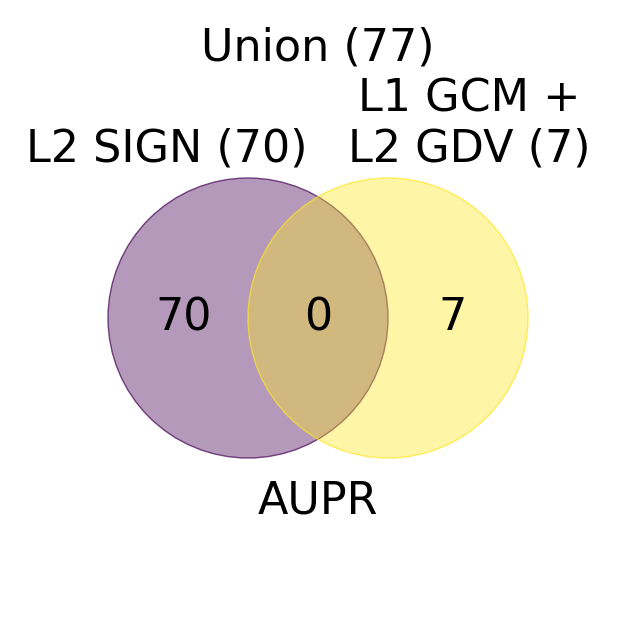}}
     \subfloat[]{\includegraphics[width=0.24\textwidth]{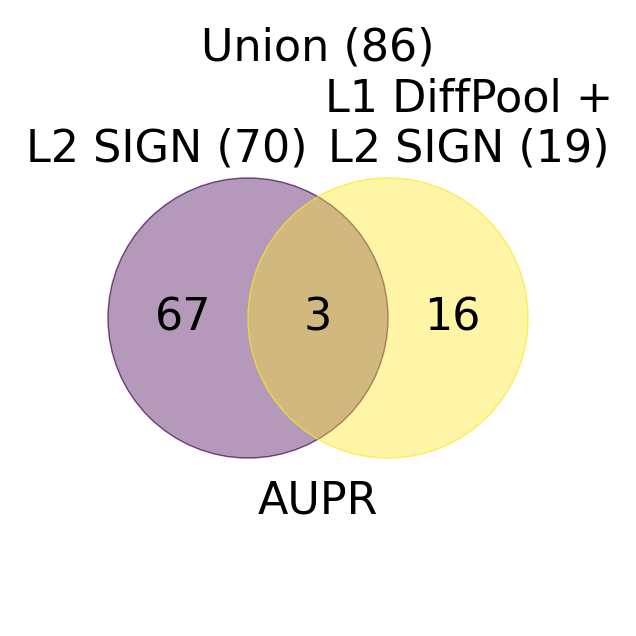}}
     \subfloat[]{\includegraphics[width=0.24\textwidth]{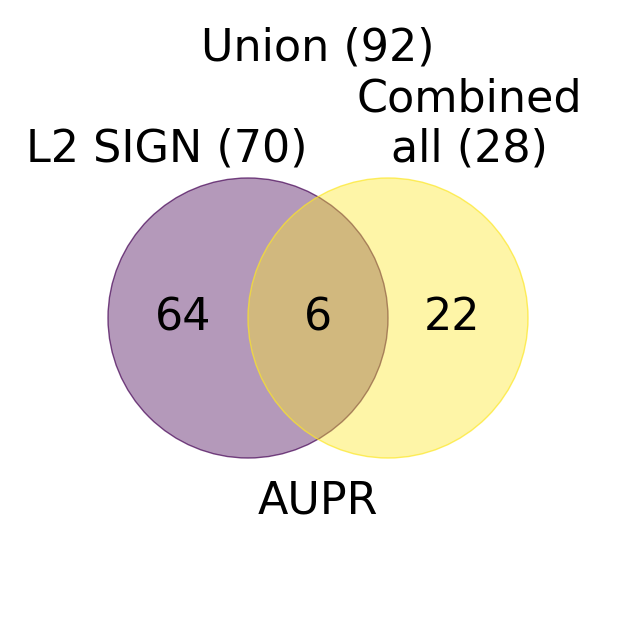}}
     \subfloat[]{\includegraphics[width=0.24\textwidth]{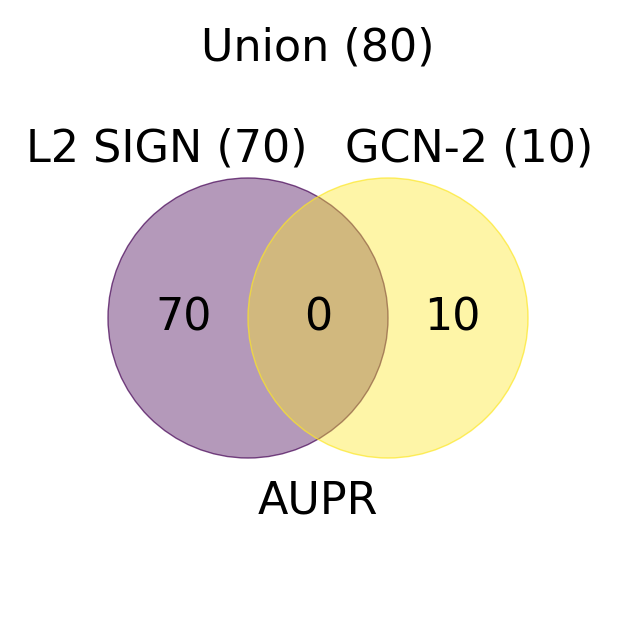}}
     
    \subfloat[]{\includegraphics[width=0.24\textwidth]{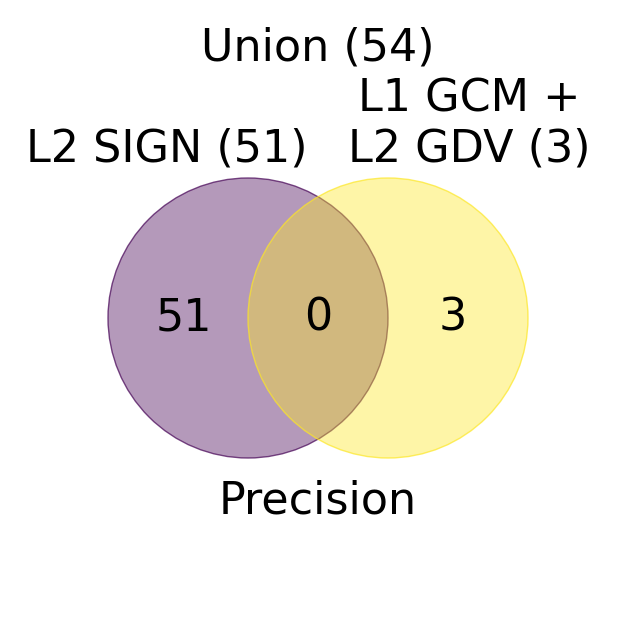}}
     \subfloat[]{\includegraphics[width=0.24\textwidth]{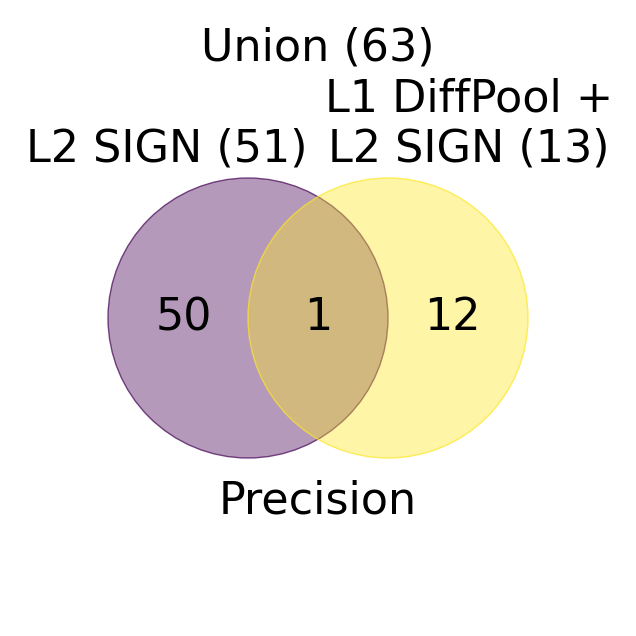}}
     \subfloat[]{\includegraphics[width=0.24\textwidth]{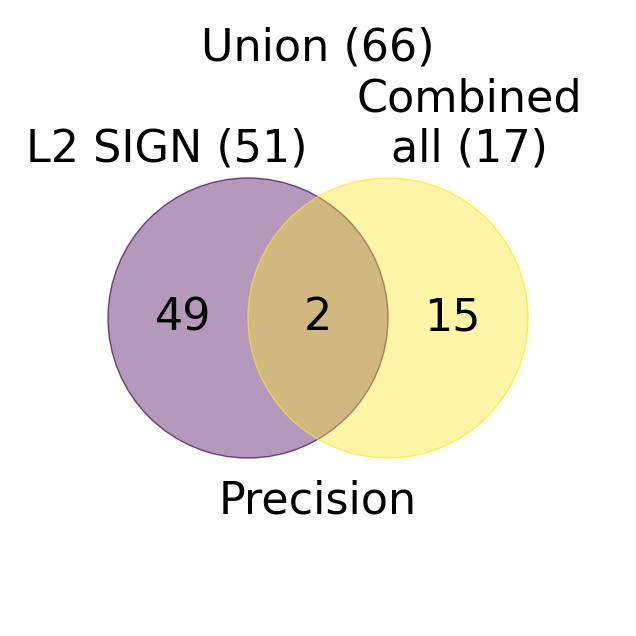}}
     \subfloat[]{\includegraphics[width=0.24\textwidth]{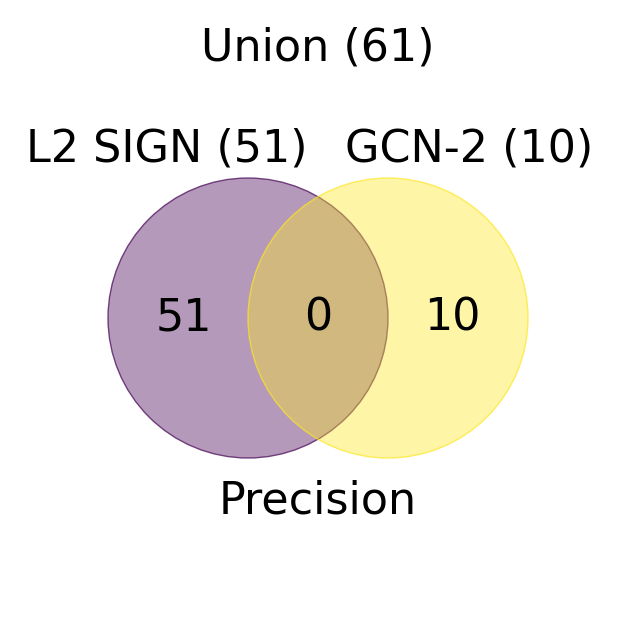}}
     
     \subfloat[]{\includegraphics[width=0.24\textwidth]{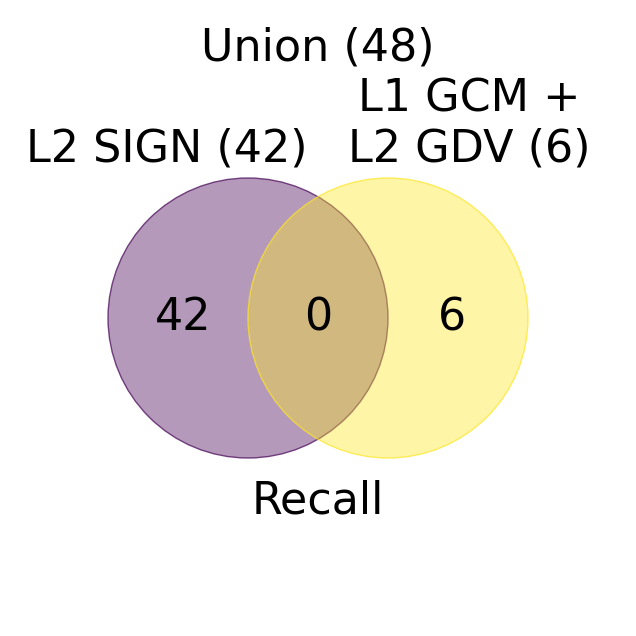}}
    \subfloat[]{\includegraphics[width=0.24\textwidth]{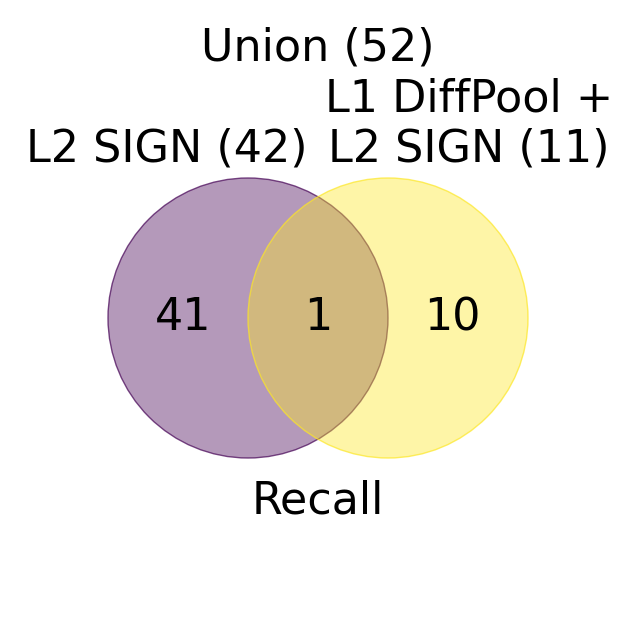}}
     \subfloat[]{\includegraphics[width=0.24\textwidth]{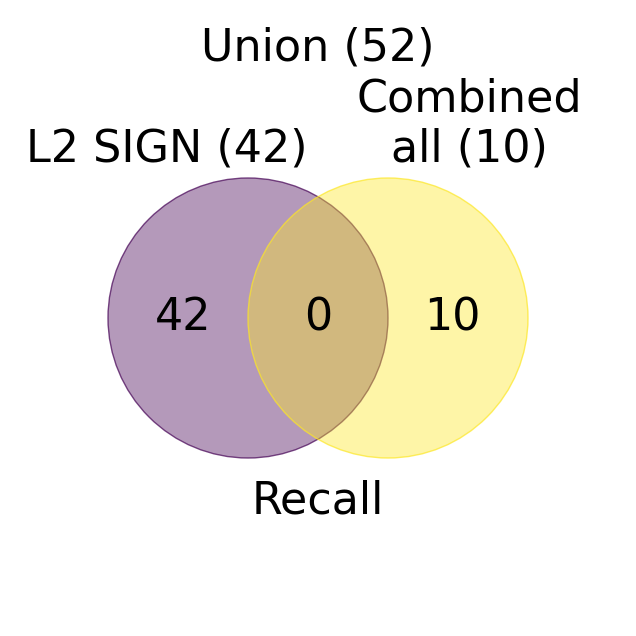}}
     \subfloat[]{\includegraphics[width=0.24\textwidth]{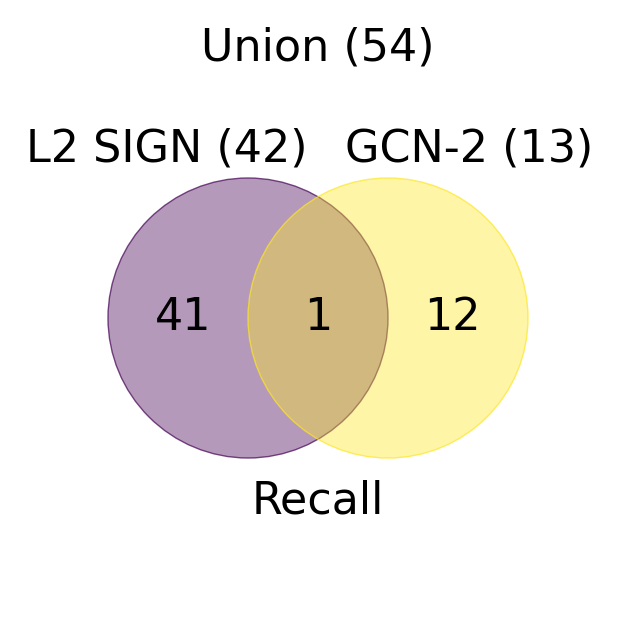}}
     
      \subfloat[]{\includegraphics[width=0.24\textwidth]{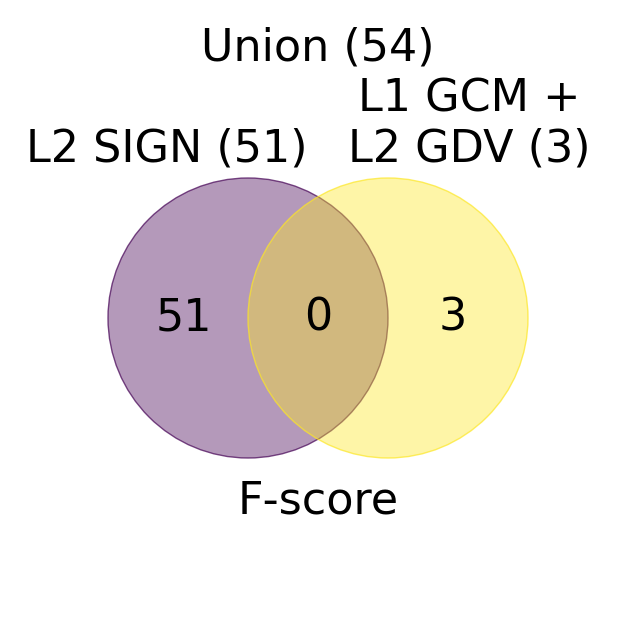}}
     \subfloat[]{\includegraphics[width=0.24\textwidth]{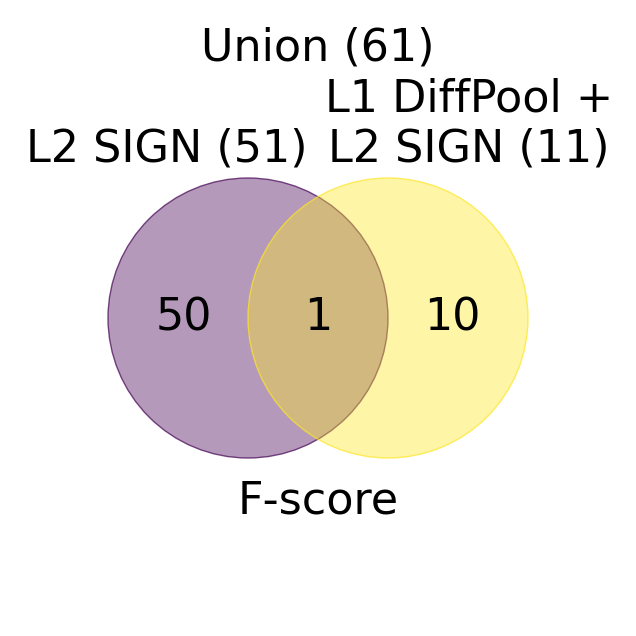}}
    \subfloat[]{\includegraphics[width=0.24\textwidth]{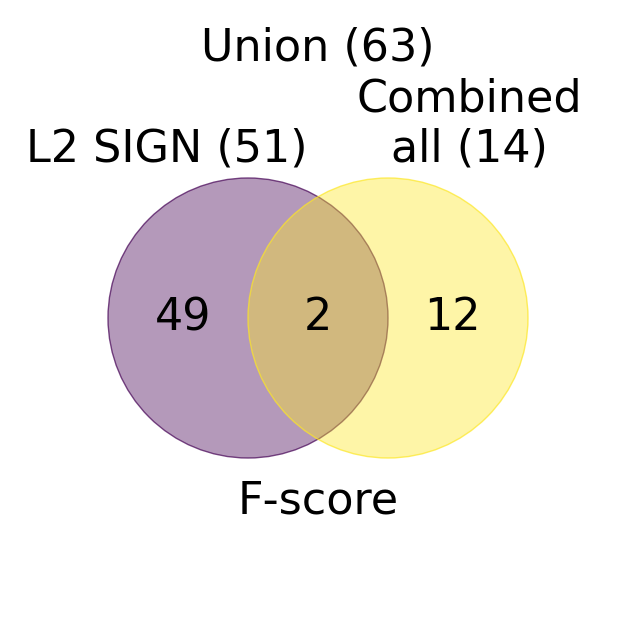}}
     \subfloat[]{\includegraphics[width=0.24\textwidth]{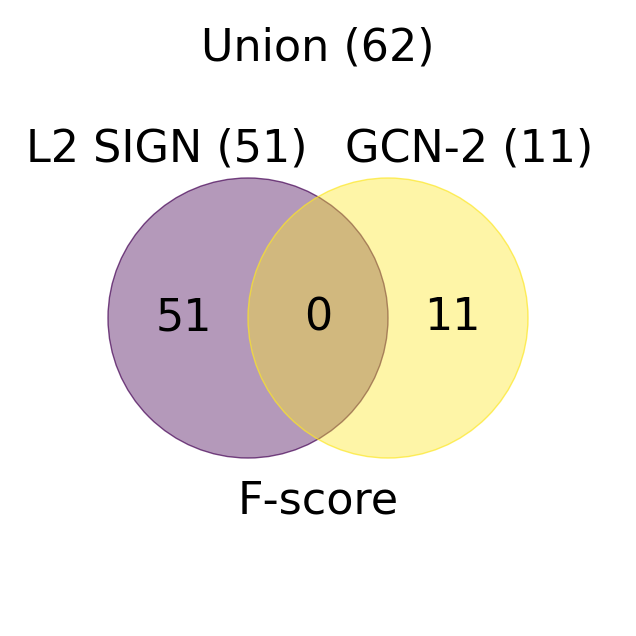}}
    
    \caption{Overlap of GO terms for which L2 SIGN is the best with those for which \textbf{(a, e, i, m)} L1 GCM + L2 GDV, \textbf{(b, f, j, n)} L1 DiffPool + L2 SIGN, \textbf{(c, g, k, o)} Combined all (aka L1 GDVM + L2 GDV + L1 DiffPool + L2 SIGN), and \textbf{(d, h, l, p)} GCN-2 are the best in terms of \textbf{(a, b, c, d)} AUPR, \textbf{(e, f, g, h)} precision, \textbf{(i, j, k, l)} recall, and \textbf{(m, n, o, p)} F-score.}
    \label{suppfig:sign-overlaps}
\end{figure*}

\begin{figure*}[ht]

    \centering
     \subfloat[]{\includegraphics[width=0.24\textwidth]{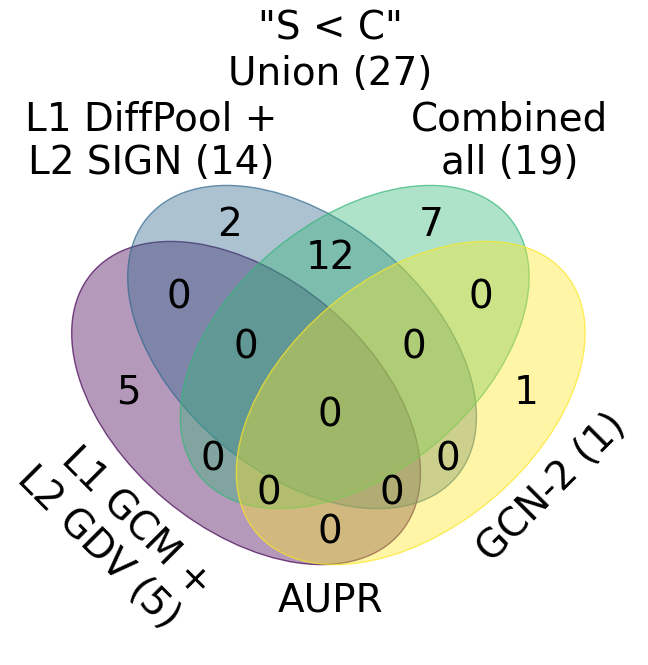}}
     \subfloat[]{\includegraphics[width=0.24\textwidth]{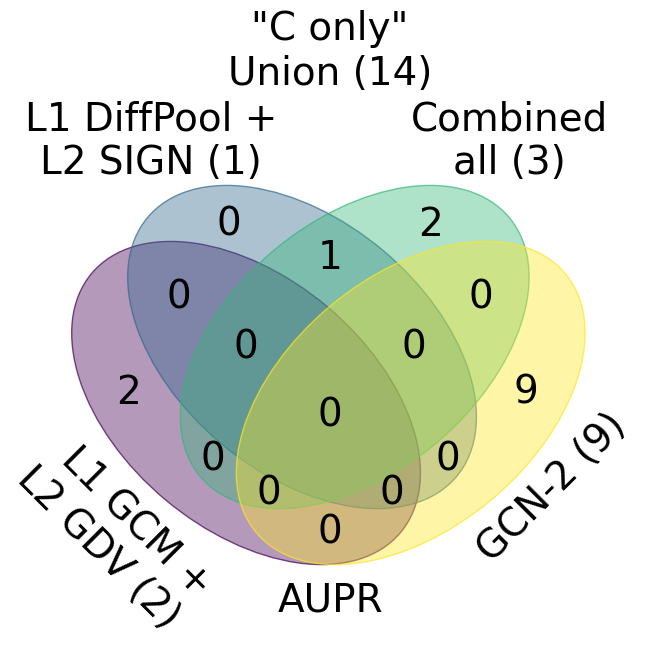}}
     \subfloat[]{\includegraphics[width=0.24\textwidth]{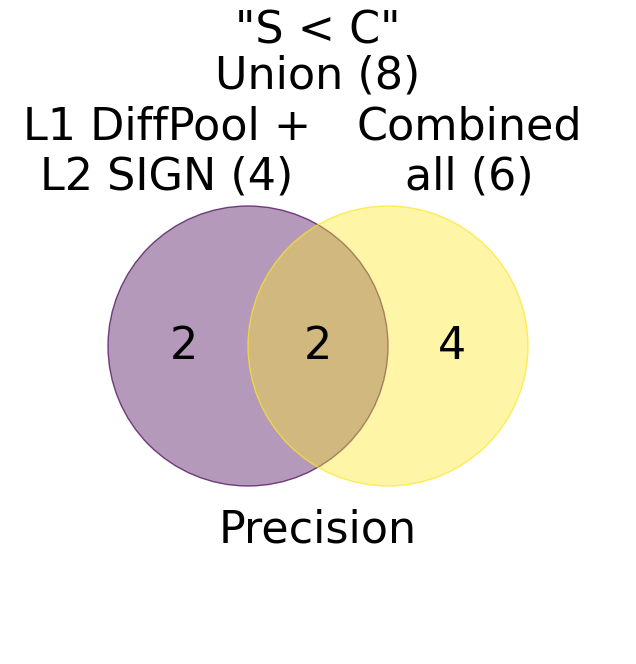}}
     \subfloat[]{\includegraphics[width=0.24\textwidth]{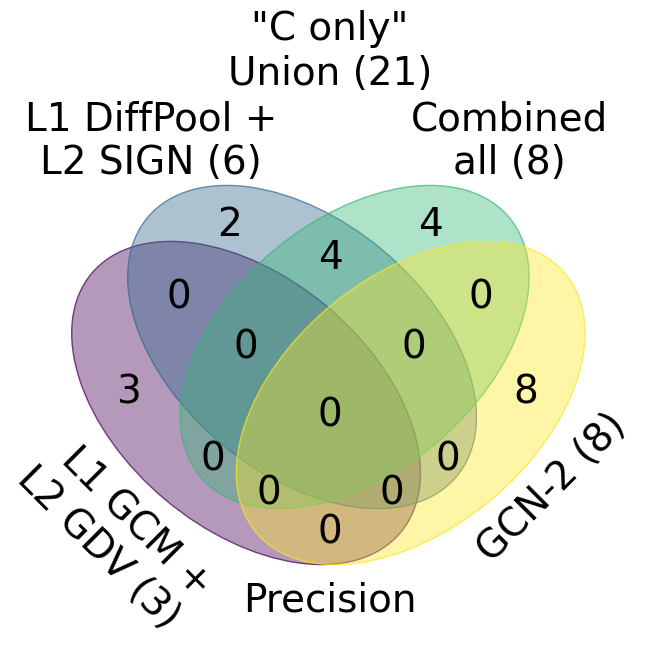}}
     
     \subfloat[]{\includegraphics[width=0.24\textwidth]{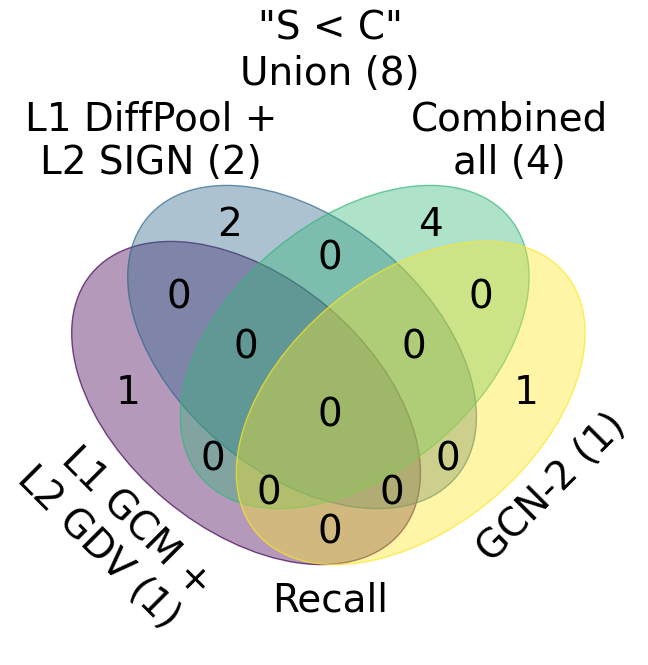}}
     \subfloat[]{\includegraphics[width=0.24\textwidth]{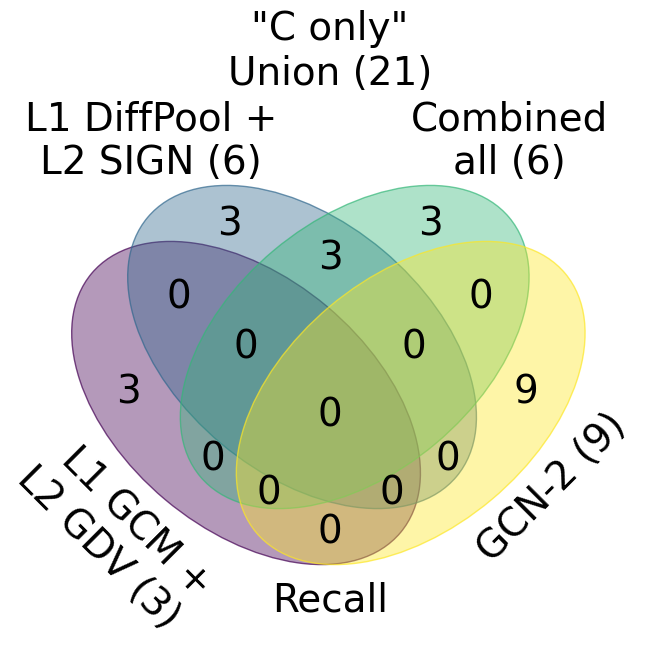}}
     \subfloat[]{\includegraphics[width=0.24\textwidth]{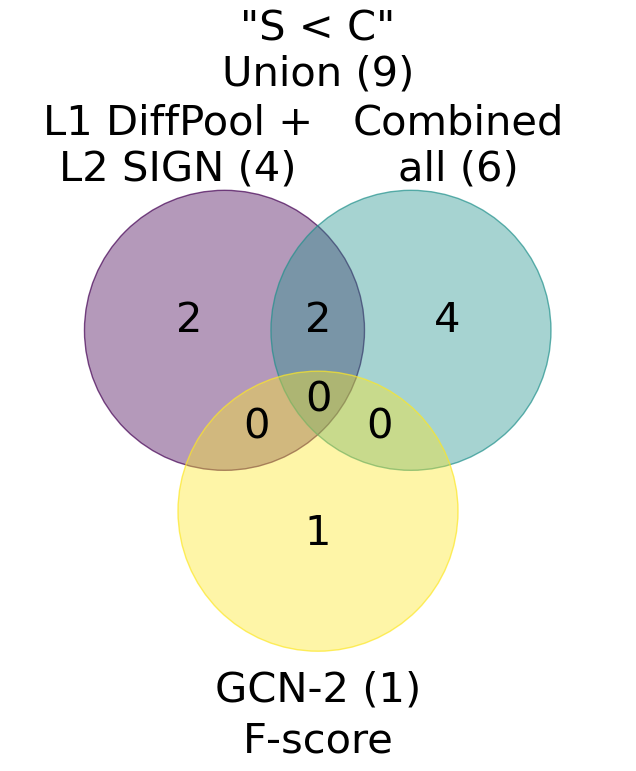}}
    \subfloat[]{\includegraphics[width=0.24\textwidth]{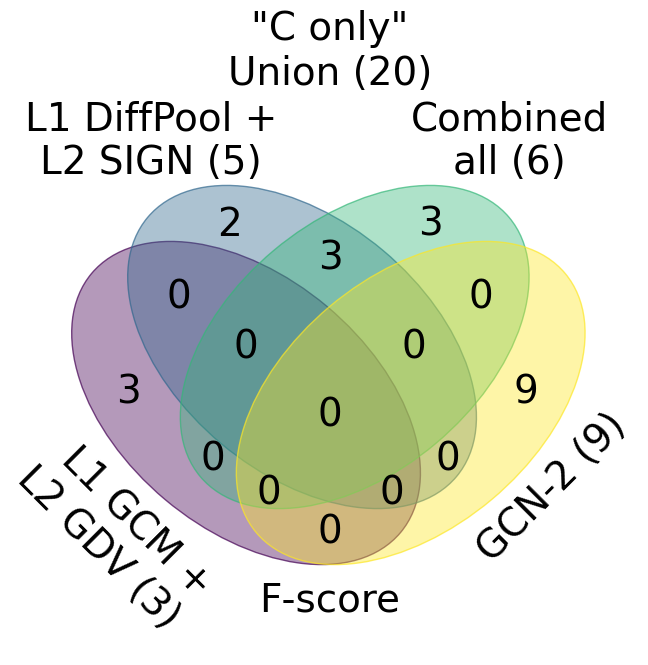}}
     
    \caption{Overlaps of the four combined level approaches for groups \textbf{(a, c, e, g)} ``S $<$ C'' and \textbf{(b, d, f, h)} ``C only'' in terms of \textbf{(a, b)} AUPR, \textbf{(c, d)} precision, \textbf{(e, f)} recall, \textbf{(g, h)} F-score.}
    \label{suppfig:combined-best-overlaps}
\end{figure*}

\begin{figure*}[ht]

    \centering

     \includegraphics[width=0.99\textwidth]{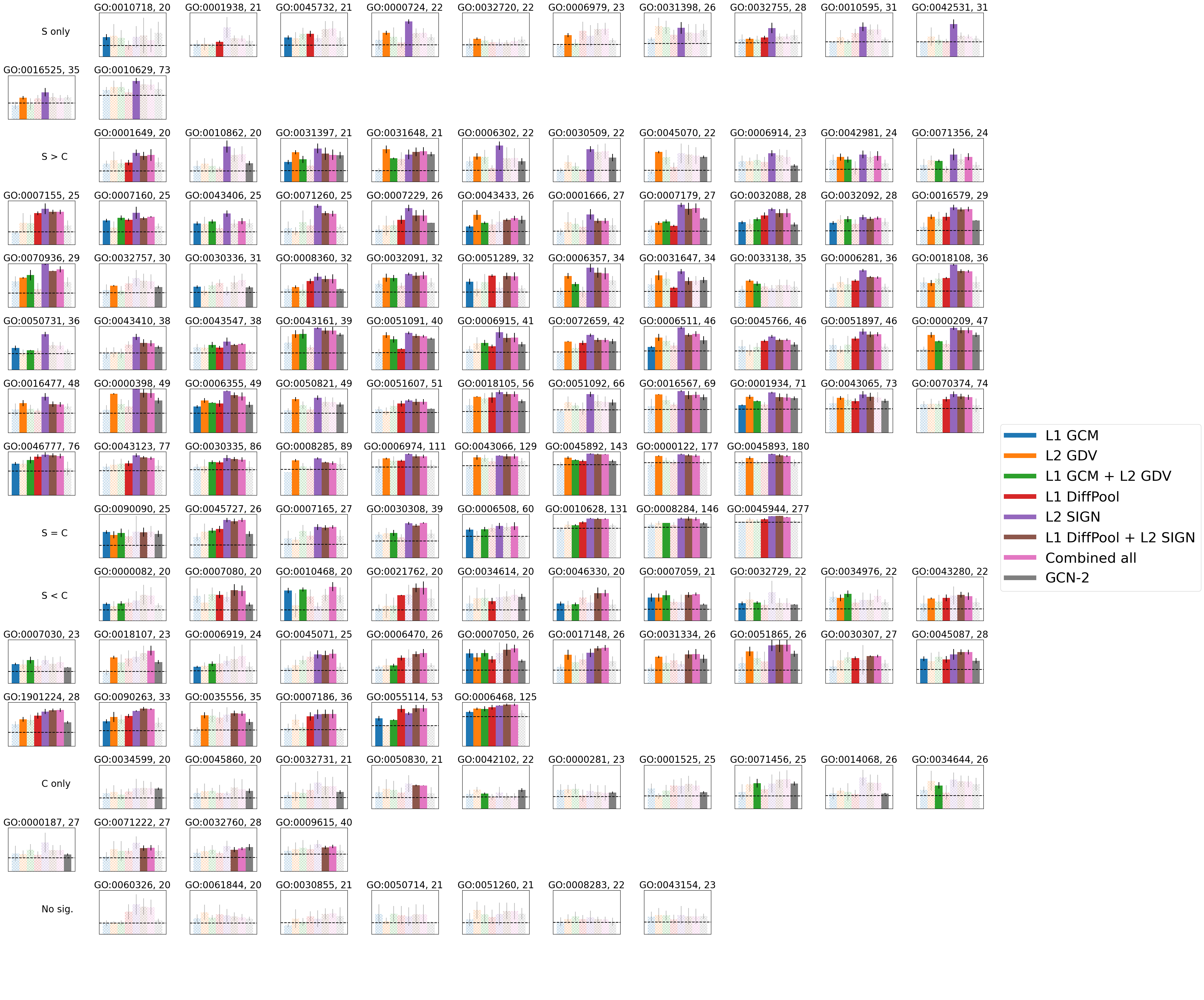}
    \caption{Classification performance of the eight relevant approaches for each GO term in terms of AUPR. GO term IDs and the number of positive instances for that GO term are shown above. Random performance is indicated by the dotted black line. Approaches with performance not significantly greater than random are shown in a lighter shade. GO terms are split into the six groups based on how single versus combined level approaches perform. ``Combined all'' refers to L1 GDVM + L2 GDV + L1 DiffPool + L2 SIGN. Raw scores for each approach for each GO term can be found in Supplementary Table \ref{supptab:goinfo-aupr}.}
    \label{suppfig:aupr-grid}
\end{figure*}

\begin{table}[ht]
    \url{http://nd.edu/~cone/NoNs/goinfo.aupr.csv}
    \caption{Raw AUPR scores of the eight relevant approaches for each GO term in each of the six groups.}
    \label{supptab:goinfo-aupr}
\end{table}

\begin{figure*}[ht]

    \centering

     \includegraphics[width=0.99\textwidth]{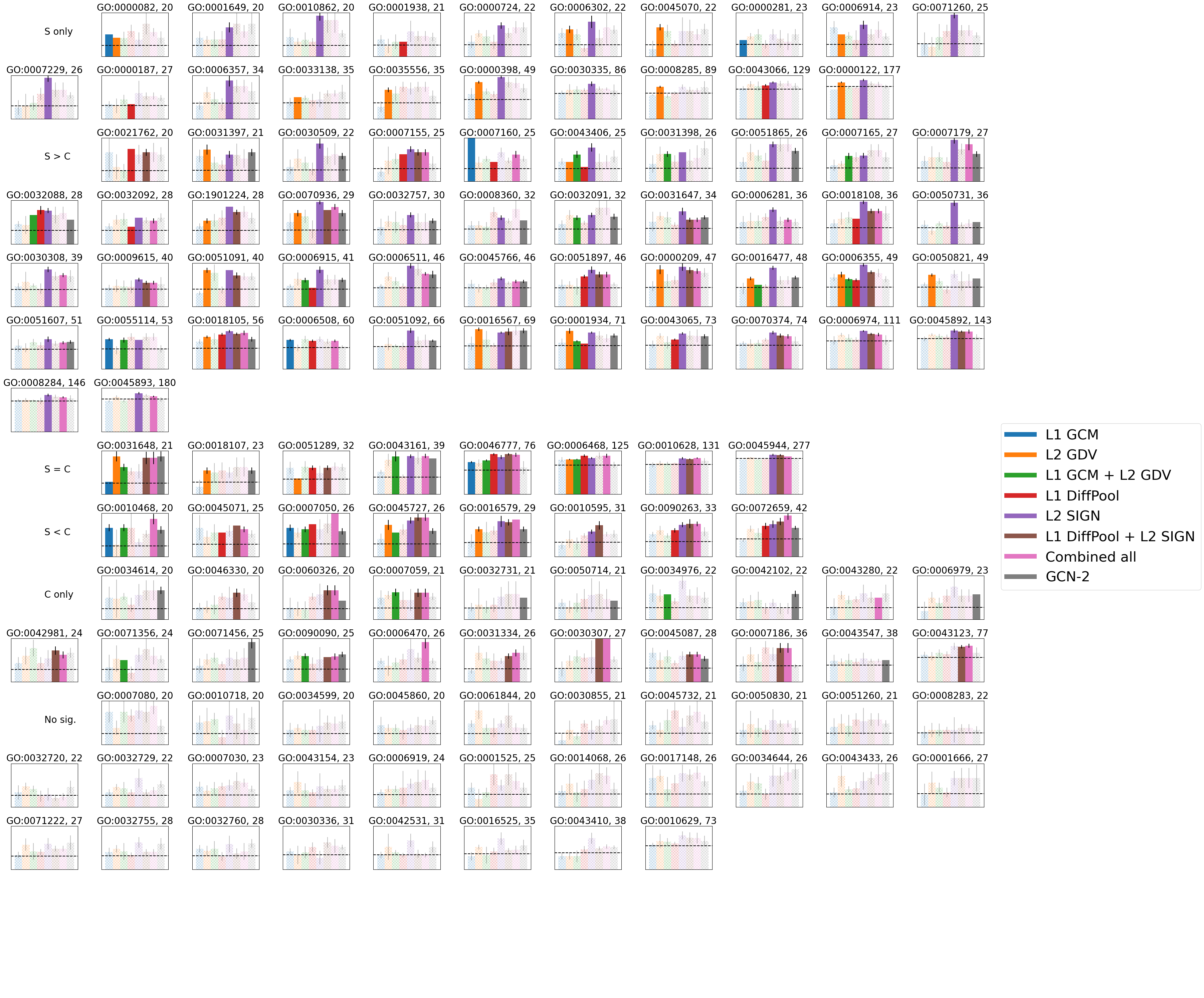}
    \caption{Classification performance of the eight relevant approaches for each GO term in terms of precision. GO term IDs and the number of positive instances for that GO term are shown above. Random performance is indicated by the dotted black line. Approaches with performance not significantly greater than random are shown in a lighter shade. GO terms are split into the six groups based on how single versus combined level approaches perform. ``Combined all'' refers to L1 GDVM + L2 GDV + L1 DiffPool + L2 SIGN. Raw scores for each approach for each GO term can be found in Supplementary Table \ref{supptab:goinfo-precision}.}
    \label{suppfig:precision-grid}
\end{figure*}

\begin{table}[ht]
    \url{http://nd.edu/~cone/NoNs/goinfo.precision.csv}
    \caption{Raw precision scores of the eight relevant approaches for each GO term in each of the six groups.}
    \label{supptab:goinfo-precision}
\end{table}

\begin{figure*}[ht]

    \centering

     \includegraphics[width=0.99\textwidth]{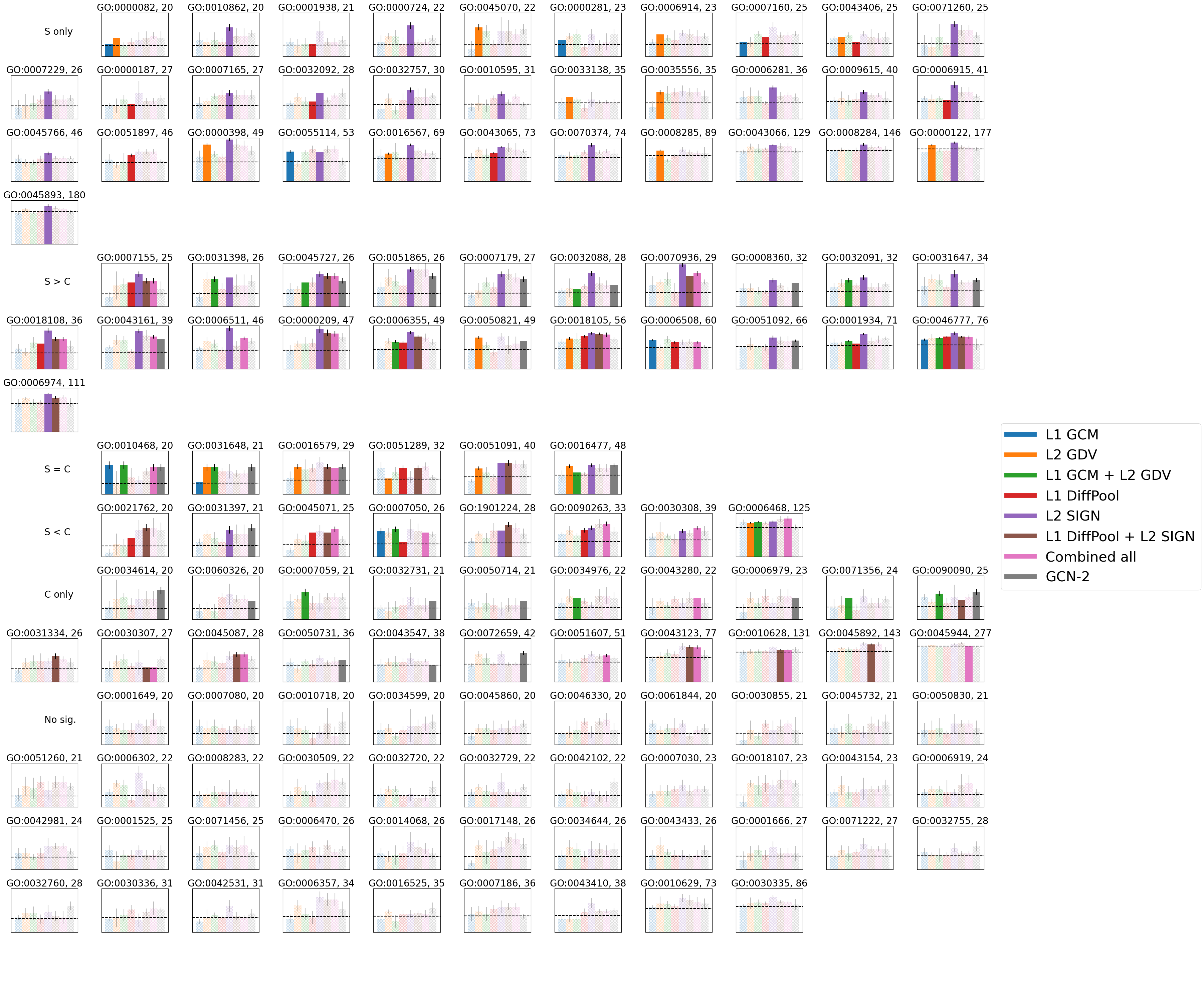}
    \caption{Classification performance of the eight relevant approaches for each GO term in terms of recall. GO term IDs and the number of positive instances for that GO term are shown above. Random performance is indicated by the dotted black line. Approaches with performance not significantly greater than random are shown in a lighter shade. GO terms are split into the six groups based on how single versus combined level approaches perform. ``Combined all'' refers to L1 GDVM + L2 GDV + L1 DiffPool + L2 SIGN. Raw scores for each approach for each GO term can be found in Supplementary Table \ref{supptab:goinfo-recall}.}
    \label{suppfig:recall-grid}
\end{figure*}

\begin{table}[ht]
    \url{http://nd.edu/~cone/NoNs/goinfo.recall.csv}
    \caption{Raw recall scores of the eight relevant approaches for each GO term in each of the six groups.}
    \label{supptab:goinfo-recall}
\end{table}

\begin{figure*}[ht]

    \centering

     \includegraphics[width=0.99\textwidth]{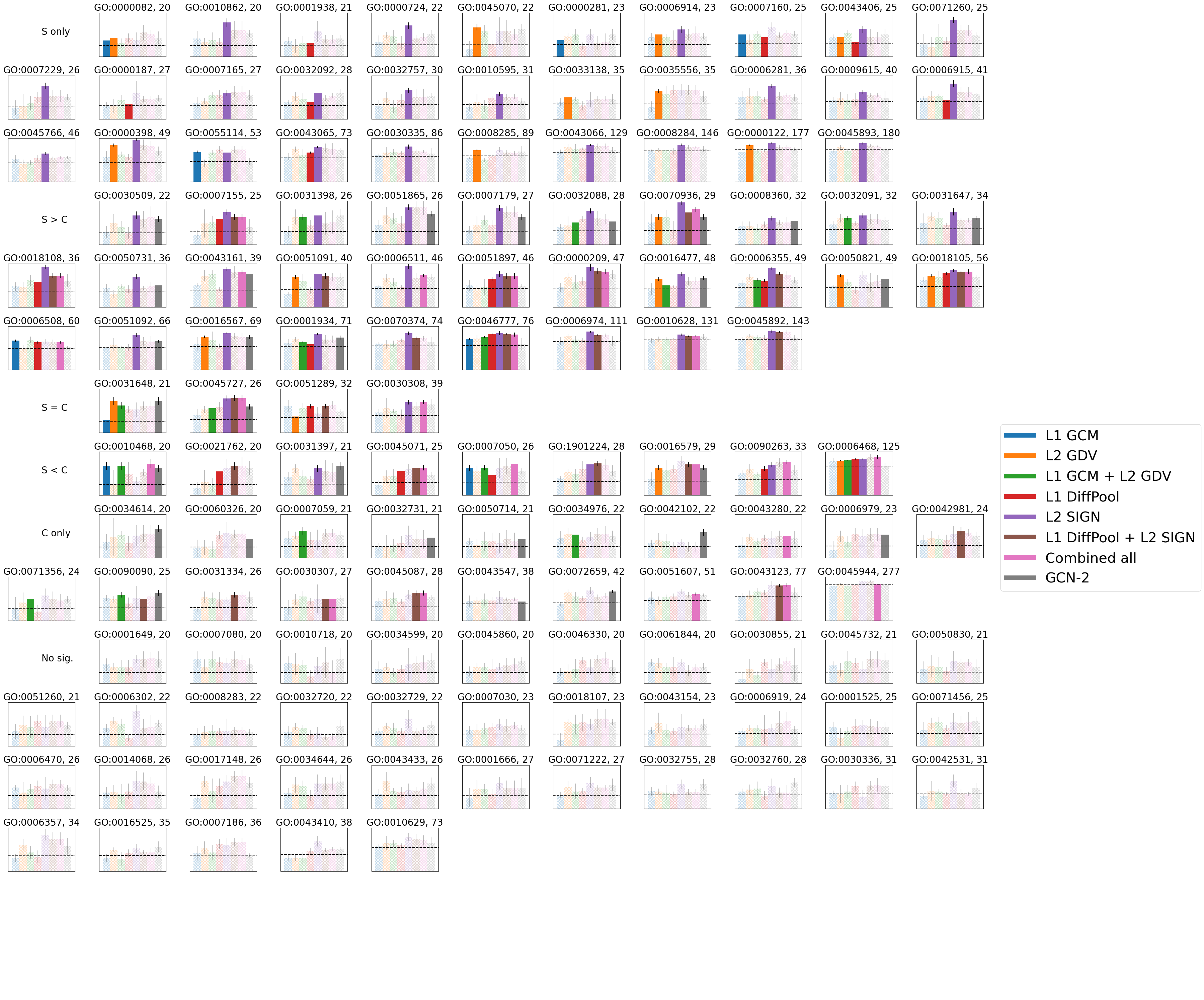}
    \caption{Classification performance of the eight relevant approaches for each GO term in terms of F-score. GO term IDs and the number of positive instances for that GO term are shown above. Random performance is indicated by the dotted black line. Approaches with performance not significantly greater than random are shown in a lighter shade. GO terms are split into the six groups based on how single versus combined level approaches perform. ``Combined all'' refers to L1 GDVM + L2 GDV + L1 DiffPool + L2 SIGN. Raw scores for each approach for each GO term can be found in Supplementary Table \ref{supptab:goinfo-f-score}.}
    \label{suppfig:f-score-grid}
\end{figure*}

\begin{table}[ht]
    \url{http://nd.edu/~cone/NoNs/goinfo.f-score.csv}
    \caption{Raw F-scores of the eight relevant approaches for each GO term in each of the six groups.}
    \label{supptab:goinfo-f-score}
\end{table}

\subsection{Running times} \label{supp:results-time}


Reported times for all approaches except those involving DiffPool (L1 DiffPool, L1 DiffPool + L2 SIGN, and L1 GDVM + L2 GDV + L1 DiffPool + L2 SIGN) are obtained by running on the same machine, fully using one core, for fairness; of course, for practical purposes, some approaches can easily be parallelized given available resources.
Training for DiffPool-based approaches must be done on GPU. We report their training times on a cluster machine, which means that their times for training are affected by resource availability/scheduling. While DiffPool-based approaches are not run under the same conditions as other approaches, we still commented on their running times, as in a realistic scenario, approaches may be run using different resources as we have done here.

We run all approaches except those involving DiffPool using one core on a 64-core AMD Opteron 6376 machine. We run approaches involving DiffPool on a cluster machine with Dual Twelve-core 2.2GHz Intel Xeon processors and 4 NVIDIA GeForce GTX 1080 Ti GPUs, accessed through Notre Dame's Center for Research Computing.

\begin{table}[ht]
\begin{tabular}{l|S[table-format=3.2]S[table-format=3.2]S[table-format=3.2]}
\hline
        & \multicolumn{1}{C{2cm}}{Feature extraction} & \multicolumn{1}{C{2cm}}{Training\newline(1 epoch)} & \multicolumn{1}{C{1cm}}{Total}\\
\hline
L1 GDVM & 485.0 & 2.2 & 487.2  \\
L2 GDV    & 2.1 & 1.5 & 3.6  \\
L1 GDVM + L2 GDV    & 487.1 & 2.1 & 489.2  \\
L1 DiffPool & 485.0 & 140.1 & 625.3   \\
L2 SIGN    & 6.4 & 17.6 & 23.4  \\
L1 DiffPool + L2 SIGN & 491.4 & 25.6 & 516.4  \\
Combined all & 493.5 & 29.4 & 522.5  \\
    GCN-2    & 487.1 & 30.3 & 517.1  \\
    GCN-3    & 487.1 & 128.8 & 615.1  \\
    
\hline
\end{tabular}
\caption{Running times of each approach in seconds. ``Combined all'' refers to L1 GDVM + L2 GDV + L1 DiffPool + L2 SIGN.}
\label{supptab:running-times}
\end{table}

\bibliographystyle{plain}
%




\end{document}